\documentclass[]{aastex62}

\usepackage{graphicx}
\usepackage{float}
\usepackage[caption=false]{subfig}
\usepackage{enumitem}
\usepackage{amsmath}
\usepackage{url}
\newcommand{\spitzer}{{\it Spitzer}}
\newcommand{\kepler}{{\it Kepler}}

\received{July 16, 2020}
\revised{August 28, 2020}
\accepted{September 12, 2020}
\submitjournal{AJ}

\shorttitle{NIRCam Lab Stability Studies}
\shortauthors{Schlawin et al.}
\begin{document}
\turnoffedit1

\title{JWST Noise Floor I: Random Error Sources in JWST NIRCam Time Series}

\correspondingauthor{Everett Schlawin}
\email{eas342 AT EMAIL Dot Arizona .edu}

\author[0000-0001-8291-6490]{Everett Schlawin}
\affiliation{Steward Observatory \\
933 North Cherry Avenue \\
Tucson, AZ 85721, USA}

\author{Jarron Leisenring}
\affiliation{Steward Observatory \\
933 North Cherry Avenue \\
Tucson, AZ 85721, USA}

\author{Karl Misselt}
\affiliation{Steward Observatory \\
933 North Cherry Avenue \\
Tucson, AZ 85721, USA}

\author{Thomas P. Greene}
\affiliation{NASA Ames Research Center \\
Space Science and Astrobiology Division \\
Moffett Field, CA 94035, USA}

\author{Michael W. McElwain}
\affiliation{NASA Goddard Space Flight Center \\
Exoplanets and Stellar Astrophysics Laboratory \\
Greenbelt, MD 20771, USA}

\author{Thomas Beatty}
\affiliation{Steward Observatory \\
933 North Cherry Avenue \\
Tucson, AZ 85721, USA}

\author{Marcia Rieke}
\affiliation{Steward Observatory \\
933 North Cherry Avenue \\
Tucson, AZ 85721, USA}



\begin{abstract}

JWST transmission and emission spectra will provide invaluable glimpses of transiting exoplanet atmospheres,\edit1{ including possible biosignatures.}
This promising science from JWST, however, will require exquisite precision and understanding of systematic errors that can impact the time series of planets crossing in front of and behind their host stars.
Here, we provide estimates of the random noise sources affecting JWST NIRCam time-series data on the integration-to-integration level.
\edit1{We find that 1/f noise} can limit the precision of grism time series for 2 groups (230 ppm to 1000 ppm depending on the extraction method and extraction parameters), but will average down like the \edit1{square root of N frames/reads.}
The current NIRCam grism time series mode is especially affected by 1/f noise because its GRISMR dispersion direction is parallel to the detector fast-read direction, \edit1{ but could be alleviated in the GRISMC direction.}
Care should be taken to include as many frames as possible per visit to reduce this 1/f noise source: thus, we recommend the smallest detector subarray sizes one can tolerate, 4 output channels and readout modes that minimize the number of skipped frames (RAPID or BRIGHT2).
We also describe a covariance weighting scheme that can significantly lower the contributions from 1/f noise as compared to sum extraction.
We evaluate the noise introduced by pre-amplifier offsets, random telegraph noise, and high dark current RC pixels and find that these are correctable below 10 ppm once background subtraction and pixel masking are performed.
We explore systematic error sources in a companion paper.
\end{abstract}

\keywords{Concept: Astronomical detectors --- Concept: Exoplanet atmospheres --- Concept: Near infrared astronomy}



\section{Introduction} \label{sec:intro}

JWST will provide powerful new measurements of exoplanet atmospheres with spectroscopy from 0.6 to 28 $\mu$m
\citep{beichman2014pasp,greene2016jwst_trans,howe2017informationJWST,barstow2015jwstSystematics,schlawin2018JWSTforecasts}.
The sensitivity of JWST combined with its unprecedented wavelength coverage will enable it to measure the abundance of carbon-bearing molecules (CO, CO$_2$ and CH$_4$) in exoplanet atmospheres as well as study cooler, smaller planets than previously characterized.

JWST has also been considered for observations of potentially habitable Earth-like planets that orbit their stars at a distance where water could be in liquid form on their surface.
The TRAPPIST-1 system \citep{gillon2016trappist1Discovery,gillon2017trappist-1sevenp} presents an exciting opportunity to study Earth-sized planets orbiting a 0.1~R$_\odot$ M-type star.
By putting together 10 to 30 transits, it will be possible to collect enough photons to detect CO$_2$ \citep{barstow2016trappist1habitable,krissansen-totton2018trappist1eJWST,lustig-yaeger2019detectabilityTRAPPIST-1}.
Optimistically, 30 transits may contain enough photons to detect O$_3$ in the atmosphere of TRAPPIST-1d with JWST if clouds have a minimal impact on its atmosphere \citep{barstow2016trappist1habitable}.
More detailed modeling of atmospheric evolution indicates that biogenic oxygen may be too difficult to detect with any JWST instrument but desiccated oceans may produce detectable abiogenic oxygen \citep{lustig-yaeger2019detectabilityTRAPPIST-1}.
Alternatively, 10 transits of TRAPPIST-1e could be sufficient if the planet's atmosphere has CO$_2$ and CH$_4$ biosignatures similar to early-Earth \citep{krissansen-totton2018trappist1eJWST}.
A critical question in studying planets like TRAPPIST-1d/e with small spectroscopic signatures, however, is whether photon-limited performance is possible with JWST observations.
For difficult biosignatures, noise levels less than 10 ppm (parts per million) are required and this value shrinks in the presence of clouds.

Experience with the \textit{Hubble Space Telescope} (HST), \spitzer, and \kepler\ shows that many systematics can affect high precision time series and prevent photon-limited performance unless corrected \citep[e.g.][]{beichman2014pasp}.
Many of these effects are detector-related including charge trapping in HST's detector \citep{berta2012flat_gj1214,zhou2017chargeTrap}, intra-pixel sensitivity on the short wavelength bands of \spitzer\ IRAC \citep{moralesCalderon2006LdwarfsWeatherIPC} and sensitivity to pointing jitter with \kepler\footnote{See~\url{https://keplerscience.arc.nasa.gov/K2/Performance.shtml}} \citep[e.g.][]{vanderburg2014twoWheeledKeplerPhot,beichman2014pasp}.

NIRCam \citep{rieke2005nircamSPIE} contains a slitless grism within the instrument's long wavelength (LW) channel \citep{greene2017jatisNIRCam}.
This slitless grism mode is similar to the WFC3 grism on HST that has been successfully employed on many transiting planets \citep[e.g.][]{deming13,kreidberg2014wasp43,sing2016continuum,wakeford2017hatp26}.
NIRCam grism observations will collect moderate resolution R$\sim$1100-1700 spectra over the wavelength range from 2.4 to 5.0~$\mu$m.
A dichroic beamsplitter allows simultaneous operations of the short wavelength (SW) weak lens imaging or (in future cycles) Dispersed Hartmann Sensor spectroscopy \citep[DHS;][]{schlawin2017dhs}. After approval and implementation, the DHS mode will permit spectroscopy from 1.0 to 2.0~$\mu$m using a dispersive device that measures relative mirror phases.
The DHS mode will also permit time series spectroscopy on very bright targets ($K_S \sim 1$ mag in the Vega system) inaccessible by other JWST instrument modes.

Given the finite lifetime of the James Webb mission, it is important to characterize the random and systematic errors before launch to maximize the science return of the observatory.
We focus this paper on the random errors and a companion paper on the systematic errors that can affect time-series observations with the NIRCam instrument.
Parallel efforts to characterize other instrument's noise sources are being carried out, such as with MIRI's detectors \citep{matuso2019siAsDetectorStability}.

Section \ref{sec:knownEffects} lists the known random effects that can degrade the precision of time series measurements compared to the photon noise limit.
Sections \ref{sec:preAmp} through \ref{RTNandRC} detail these noise contributions.
We conclude in Section \ref{sec:Conclusion} that 1/f noise is the largest random error to impact NIRCam grism time series measurements.
Paper~II will address the systematic errors like pointing jitter, \edit1{non-linearities, variable aperture losses} and charge trapping that degrade the precision of time series measurements.

\section{List of Known Detector Effects}\label{sec:knownEffects}
\begin{figure*}[!hbtp]
\centering
\includegraphics[width=.42\columnwidth]{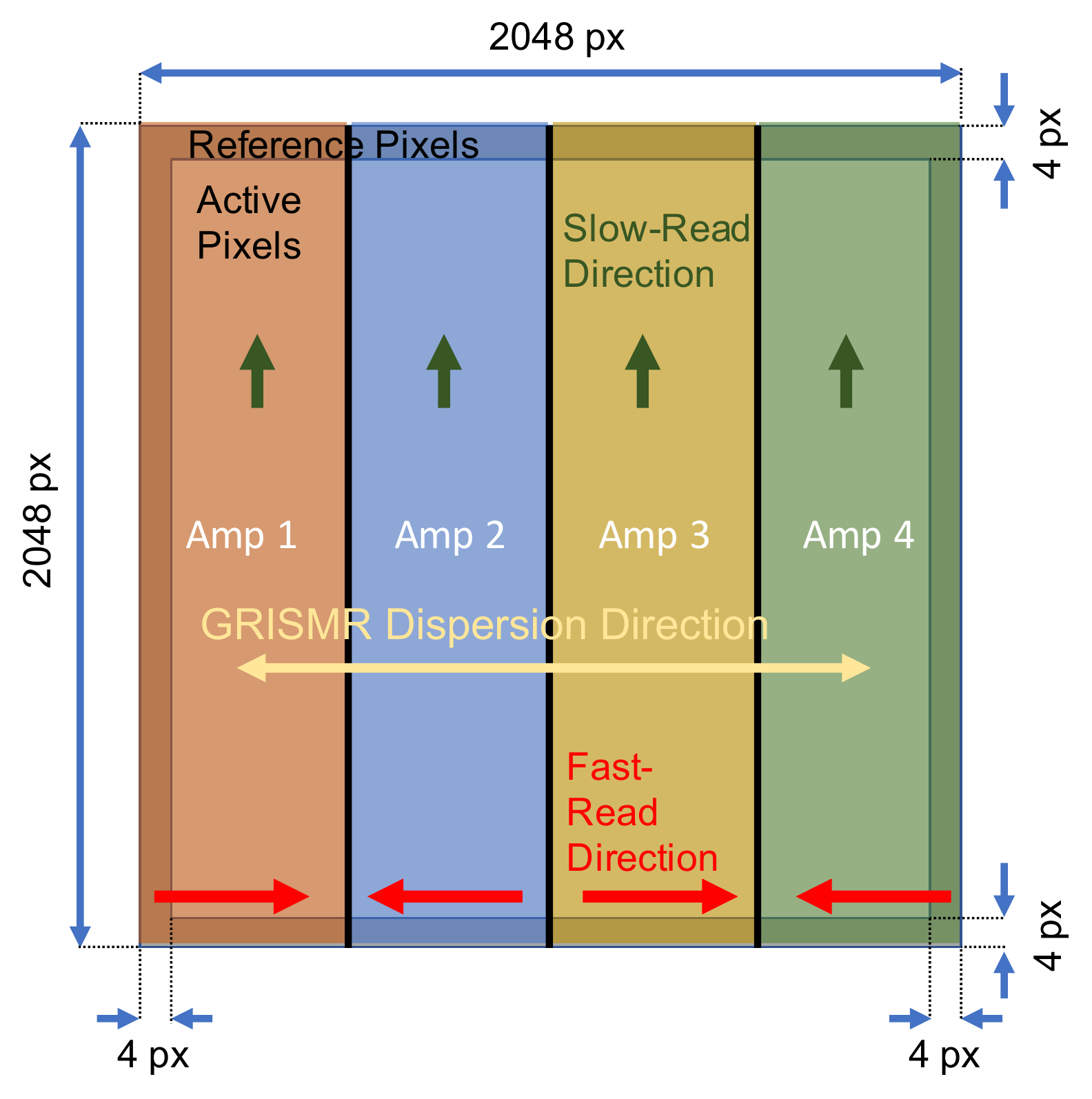}
\caption{In FULL frame imaging and STRIPE mode, four amplifiers are used to simultaneously read out four vertical stripes of pixels.
A 4-pixel-wide boundary of reference pixels can be used to subtract pre-amplifier offsets and some 1/f noise.
The fast-read directions are marked by red arrows, while the slow-read directions are marked by green arrows.
Sources dispersed by the GRISMR element cross the amplifier boundaries and are parallel to the fast-read direction, making spectra susceptible to residual 1/f noise not removed through reference pixel correction.
Note: image is not to scale - the reference pixels have been exaggerated for visibility.}\label{fig:detectorLayout}
\end{figure*}

An ideal HgCdTe photodiode array would simply count the photons incident on its surface (as described in Appendix \ref{sec:detectorPrimer}).
The NIRCam, NIRSpec, and NIRISS instruments on JWST use Teledyne HAWAII-2RG (H2RG) HgCdTe detectors \citep[e.g.][]{gardner2006SSRv,rauscher2012degradation}.
For NIRCam, the eight short wavelength detectors have a 2.5-$\mu$m cutoff while the two long wavelength detectors have a 5.2-$\mu$m cutoff \citep{jdoxNIRCamDetectors}.
Each of the detectors is connected to a System Image Digitizing Enhancing Controlling And Retrieving Application Specific Integrated Circuit (SIDECAR ASIC) that controls how the pixels are clocked, reset, and read \citep{loose2006sidecarAsic}.
Each detector has a default full-frame configuration with four amplifier outputs that are read out simultaneously on four vertical subsections of the detectors (depicted in Figure \ref{fig:detectorLayout}), or alternatively is read out with a single amplifier at about one fourth the data rate.

Each active pixel's NP semiconductor junction may be measured non-destructively multiple times (referred to as ``reads'' or ``samples'') before the pixel is reset, ending an integration.
\edit1{The NP semiconductors are discussed in more detail in Appendix \ref{sec:detectorPrimer}}.
An analog-to-digital converter (ADC) turns the number of electrons collected in the detector pixel into a count of Data Numbers (DN) from 0 to 65,535 (ie. a 16-bit number) with a conversion gain of $\sim$1.8~e$^{-}$/DN for the long-wavelength detectors and $\sim$2.0~e$^{-}$/DN for the short-wavelength detectors.
One pass through all the pixels constitutes a ``frame.''
An integration consists of a collection of frames that ends when the pixels are reset.
The frames may be co-added into groups of 1 or more frames to reduce data bandwidth, and the integrations are packaged together into a single ``exposure.''
In this paper, we discuss the RAPID read mode (with one frame per group), BRIGHT1 (with one frame per group and a skipped frame between groups) and BRIGHT2 (with two frames per group).
\citet{jdoxNIRCamDetectorPatterns} contains more information on the readout patterns.
In addition to active pixels, there is a boundary of 4 pixels that are electrically identical to the active pixels but are not sensitive to light \citep{loose2003H2RGs,rieke2007irDetectorReview}.
These can be used to correct for amplifier voltage drifts.

In the ideal case, the number of DN measured by the detector will be proportional to incident flux from an astrophysical source such as a star-planet system.
If there are no astrophysical variations, the only variability of a light curve will be the Poisson photon shot noise.
There are many non-ideal effects, however, that can affect the stability of a light curve measurement of a science target.
Here, we review many of the known noise sources that could impact the stability of a time series beyond photon counting statistics.
\edit1{The detector effects are} listed briefly and then discussed in more detail as related to the JWST NIRCam detectors below that.

The following sources of noise introduce random variations to time series (on top of the photon noise) on an integration-by-integration level:
\begin{enumerate}[noitemsep]
	\item \textbf{Pre-amplifier reset offsets:} (a SIDECAR ASIC-related effect) \label{it:preAmpOffsets}
	When the electronic pre-amplifiers in NIRCam's SIDECAR ASICs are reset, there is a noise source during the reset process.
	This noise will add a random offset to the pedestal level to all pixels corresponding to one output amplifier.
	Pre-amplifier resets and voltage drifts can also appear in slope images \edit1{(DN/s)}.
	The noise introduced by pre-amplifier offsets cannot be decreased by averaging more pixels within an amplifier region of a frame, but reference pixel or background subtraction can efficiently remove the pre-amp offsets.
	For scientists familiar with CCDs, the pre-amplifier offsets look similar to a bias frame that changes in time.
	These are discussed further in Section \ref{sec:preAmp}.
	\item \textbf{Amplifier Boundary Discontinuities:} The NIRCam detectors can be read out in FULL frame, STRIPE and WINDOW modes. The FULL frame and STRIPE modes make use of 4 amplifiers to simultaneously record or reset pixels. The parallel operation of the amplifiers reduces the frame time and reset time by a factor of $\sim$4 but can introduce subtle voltage biases between the readout channels. If these voltage biases are not corrected, they can potentially introduce discontinuities in a spectrum which crosses amplifier boundaries.
	The amplifier boundary discontinuities are a consequence of the interplay between the detector video output lines with pre-amplifier reset offsets described above as well as the ADCs.
	Slight gain differences can exist between the four vertical stripes read out by the four amplifiers, but these are stable in time and will come out with flat fielding.
	\item \textbf{1/f noise:} The JWST HgCdTe detector readout system adds correlated read noise to the digital images.
	This correlated noise is caused by (1) the read out integrated circuits (ROICs), which have a p-type metal-oxide semiconductor field-effect transistor (PFET) source follower, as well as (2) DC biases in the SIDECAR ASIC electronics \citep{rauscher2011irsSquared}.
	This electronic read noise  has the property that most of the noise power is concentrated at low frequencies.
	1/f noise causes spatial correlations primarily in the fast-read direction (along detector rows).
	The 1/f noise from the SIDECAR ASICs is highly correlated between amplifiers because it is caused by a common reference voltage.
	However, the 1/f noise from the PFETs can vary between amplifiers because each video output has its own PFET \citep{rauscher2011irsSquared}.
	The 1/f noise can be reduced by subtracting the values from background pixels or reference pixels that are read closely in time.
	The NIRSpec instrument on JWST has a mode called IRS$^2$ specifically designed to reduce 1/f noise \citep{rauscher2011irsSquared}, but it is not available on NIRCam.
	\item \textbf{Even/odd offsets:} Alternating columns have different bias levels. Additionally, there are offsets between columns that can change from frame to frame, especially on NIRCam's LW detector. \label{it:evenOddOffsets}
	These can be reduced with reference pixels or background subtraction.
	\item \textbf{kTC noise:} Each reset at the beginning of an integration introduces thermal noise associated with the unknown amount of charge stored in a pixel at the reset voltage level.
	Fortunately, the bias offset produced by kTC noise will completely subtract out when fitting a slope image from multiple samples up the ramp.
	This noise would  only become relevant when attempting to make use of the first group of a ramp, such as when only 1 group is commanded.
	\item \textbf{Elevated Columns} During an integration, entire columns appear to spontaneously increase by a few to 10s of DN. The behavior of these elevated columns can be seen to move to adjacent columns in subsequent integrations. The root cause for the column transitions and its movement is still being investigated.
	\item \textbf{Random Telegraph Noise (RTN)} Some of the pixels will exhibit spontaneous jumps in signal even with no illumination \citep{bacon2005burstNoise}. Fortunately, RTN appears in only $\sim$1000 pixels out of 4$\times 10^6$ on the array.
	The RTN pixels in the ALONG detector used for grism time series include many jumps, of order $\sim$50 DN, up and down along a 20 minute exposure.
\end{enumerate}
We discuss the above noise contributions in more detail in Sections \ref{sec:preAmp} through Sections \ref{RTNandRC}.

In addition to these random effects, there are also systematic effects that do not average out when more integrations are collected.
These can also be functions of other variables such as temperature, telescope pointing or the number of filled charge traps in pixels' photodiodes.
Paper II will discuss these systematic effects, including intrapixel sensitivity, detector temperature fluctuations, charge trapping, \edit1{non-linearities, variable aperture losses} and reciprocity failure.
Here, we will focus on the random effects that are largely independent from one integration to another.

\section{Pre-amplifier Offsets and Amplifier Boundaries}\label{sec:preAmp}

\subsection{Pre-amplifier Offsets}
The NIRCam detectors' pre-amplifiers (pre-amps) in the SIDECAR ASIC have a reference voltage that will drift with time, 
which will affect the pedestal level of all pixels referenced to that voltage. 
This was introduced as error source \ref{it:preAmpOffsets} in Section \ref{sec:knownEffects}.
To address the reference voltage drift, the pre-amps can be reset to the reference voltage periodically, which will introduce a random uncertainty 
into the pedestal level. The SIDECAR ASICs support resetting to the reference voltage either at frame boundaries (between frames) or at integration boundaries (between integrations). 
In early ASICs the pre-amplifier drift between resets was seen to be high enough to allow the pedestal to drift off the range of the 
ADC during longer integrations; hence, \edit1{the NIRCam team} elected to reset the pre-amps on frame boundaries rather then integration boundaries. 
Subsequent delivery of flight ASICs showed much lower drift rates. To characterize the impact of reset-per-frame (RPF) versus reset-per-integration (RPI), 
a set of darks was collected during NIRCam cryogenic vacuum (CV) pre-delivery testing at Lockheed-Martin in both RPI and RPF mode.
In these tests, the pedestal level of all pixels could be seen to drift over 19.3 minute dark integrations by 5 to 100 DN in RPI mode.
The drift was usually linear over these 19 minute exposures, but in one case was observed to drift downward from 100 DN to -50 DN and back to 100 DN \citep{robberto2014refPixPreAmp} while RPF introduces a random uncertainty of the order of 30~DN RMS at each frame boundary in the 
sample up the ramp.  Reference pixel correction effectively removes both the random RPF jumps and the systematic RPI drift; analysis \citep{robberto2014refPixPreAmp} showed that either approach was roughly equivalent.  Operationally, NIRCam elected to continue using RPF since there was no benefit to switching to RPI and significant legacy characterization data had been obtained in RPF mode.
Reference pixel correction or careful
background subtraction can efficiently remove the signature of the pre-amp reset or drifts.
The NIRCam grism subarrays and dispersed source location are placed at the edge of the NIRCam Long Wavelength detector to ensure reference pixels are nearby and saved with active pixels.

\begin{figure*}[!hbtp]
\centering
\includegraphics[width=.49\columnwidth]{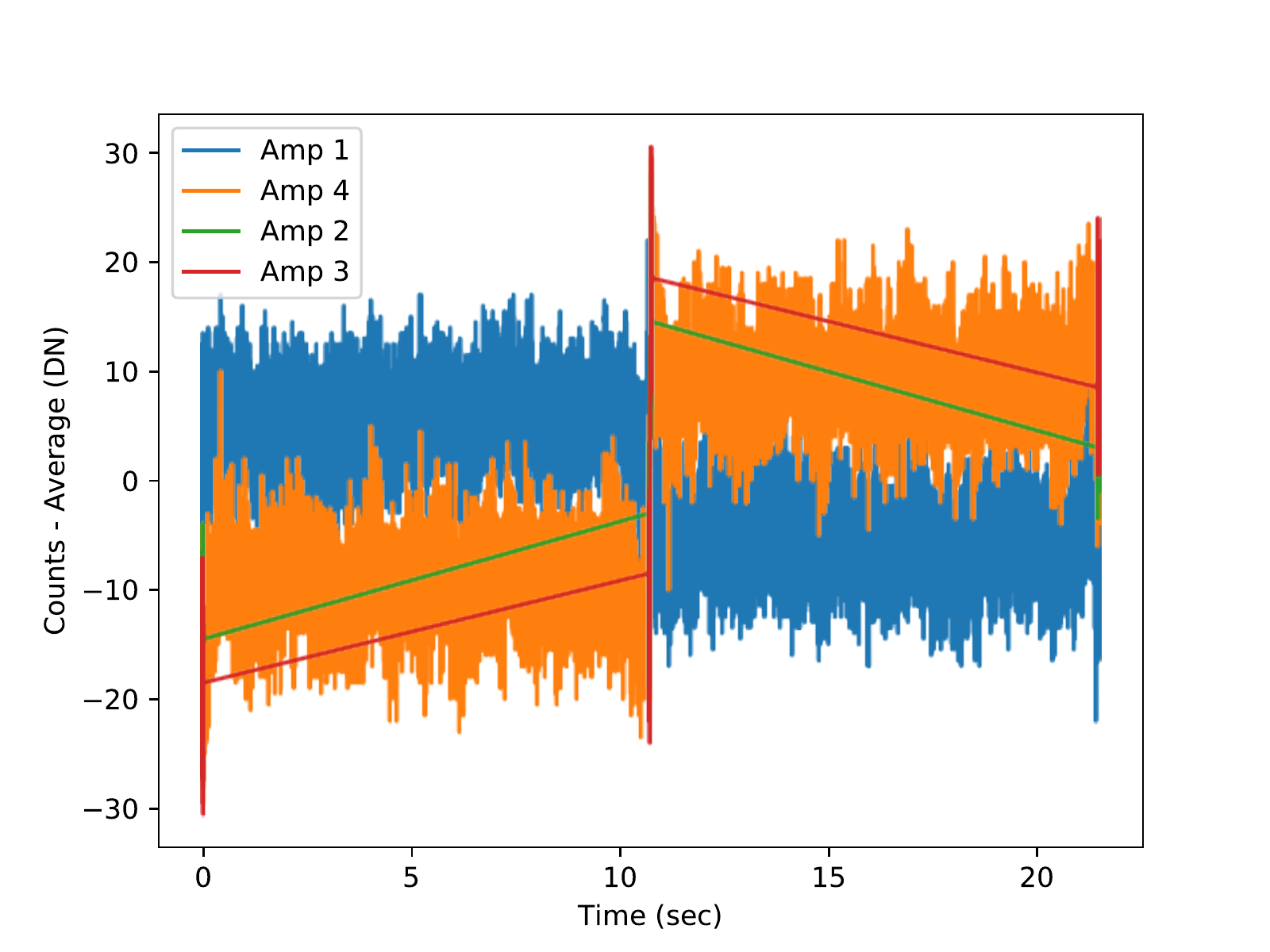}
\includegraphics[width=.49\columnwidth]{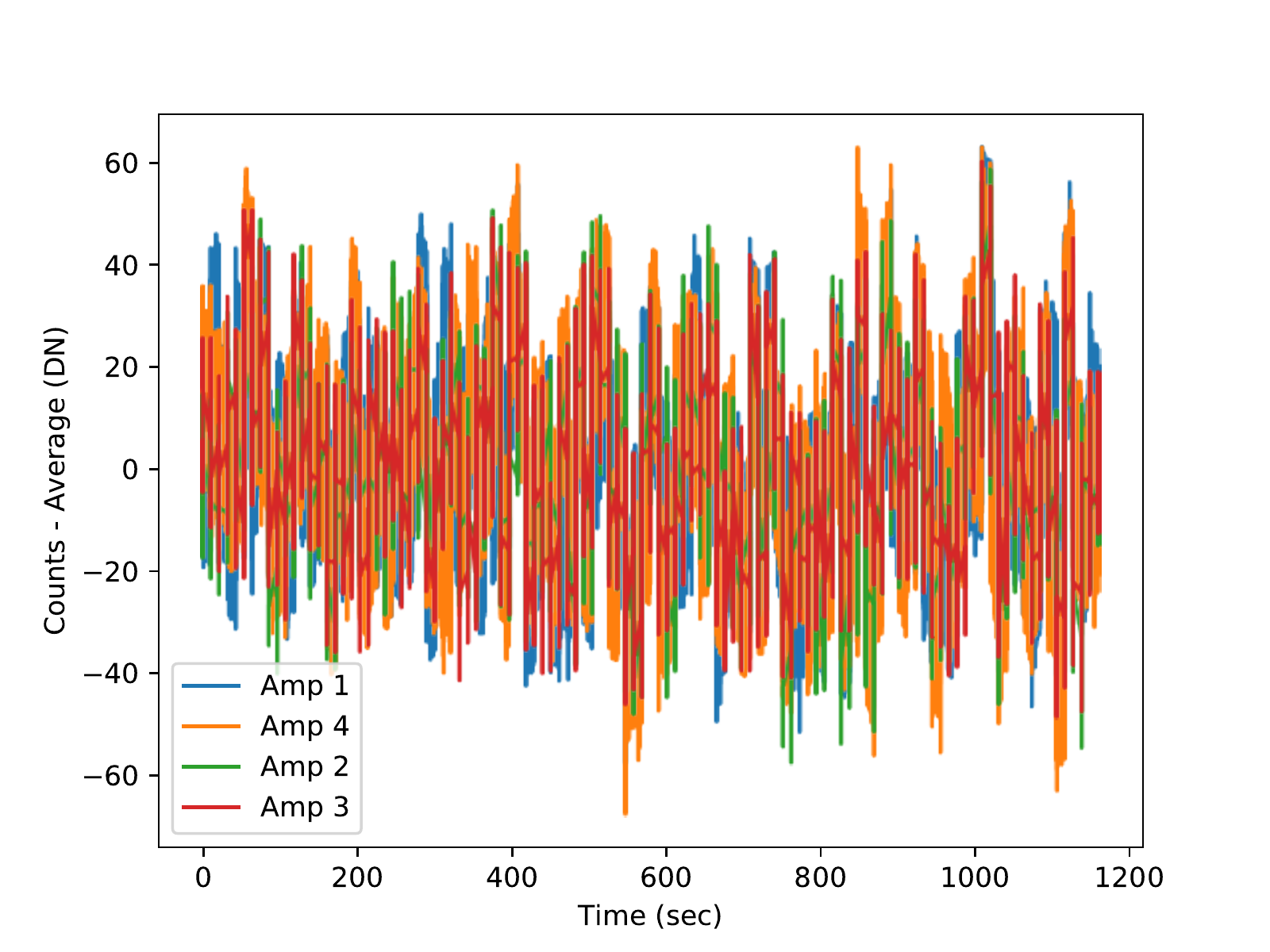}
\caption{Reference pixel time series for a two frame integration (Left) and a long dark integration of 108 frames (Right). Each of the plots show that the SIDECAR ASIC pre-amplifier resets between frames in an integration cause discontinuities in the reference signal. Amplifiers 1 and 4 track the time all through a frame with the side reference pixels, whereas amplifiers 2 and 3 in the middle only contain information on the bottom and top reference pixels at the beginning and ending of a frame.
See Figure \ref{fig:detectorLayout} for a layout of the detector.
The long dark exposure example of 108 frames shows the behavior over long timescales.}\label{fig:ampResetDark}
\end{figure*}

Figure \ref{fig:ampResetDark} shows an example time series of the reference pixels for a dark exposure.
The pedestal/bias level of the reference pixels undergoes a sharp jump between each frame (10.7~s in duration).
Fortunately, the pixels within an amplifier move together, so reference pixel subtraction or background subtraction will efficiently remove the offsets.

\subsection{Amplifier Boundary Discontinuities}

\begin{figure*}[!hbtp]
\centering
\includegraphics[width=.79\columnwidth]{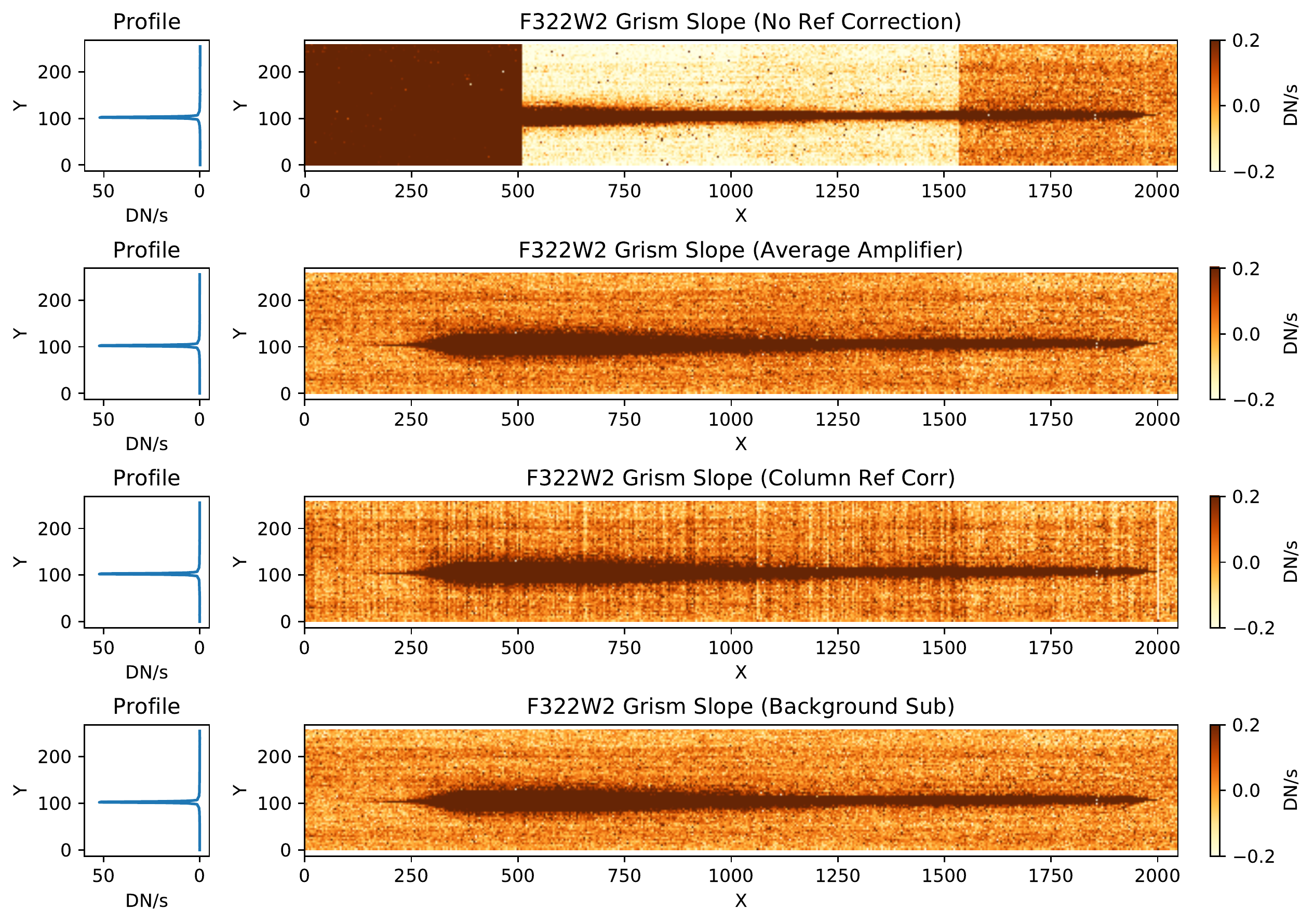}
\caption{An example slope image from \edit1{a CV3 Grism Test}, which consisted of \edit1{10 groups} in RAPID read mode for an integration time of \edit1{107 seconds, and trimmed to a 2048$\times$256 subarray}.
\edit1{The median profile of the image at $\sim$3.5~$\mu$m is shown to the left of each image.}
{\it First Panel:} No reference pixel correction is applied, leaving offsets at the amplifier boundaries.
{\it Second Panel:} Averaging all the reference pixels within an amplifier and subtracting the result reduces the offsets between amplifiers but there is still a \edit1{subtle boundary at X=1536.}.
{\it Third Panel:} Reference pixel correction using the 4 non-illuminated pixels in each column \edit1{largely removes amplifier offsets, but introduces read noise to each column.}
{\it Fourth Panel:} Background subtraction with the mean value of the pixels below \edit1{Y=30} and above \edit1{Y=180} eliminates offsets between amplifiers without introducing any substantial new noise source.
}\label{fig:ampOffsetsCV3GrismSlope}
\end{figure*}

NIRCam's STRIPE mode and full frame mode both employ 4 amplifiers to simultaneously read out or reset 4 pixels at once.
This simultaneous use of 4 amplifiers reduces the frame time (to scan through all pixels in a 2048 px-wide subarray or full frame) by a factor of about 4.
Thus, the STRIPE and full frame modes increase the brightness of objects that can be detected before saturation as well as increases the number of reads for a detector pixel before saturation to average down readout noise.
Users have the option to select 1 or 4 amplifiers with the ``No. of output channels'' in the Astronomer's Proposal Tool \citep[APT][]{apt2020p2}.
On the flip side, the four amplifiers will have different behaviors such as separate gains and offsets in the bias level.
The differential biases between the amplifiers will show up as step function changes in the slope of the image at the amplifier boundaries (512, 1024 and 1536), as visible in Figure \ref{fig:ampOffsetsCV3GrismSlope} (top panel).
If a spectrum of a source is extracted without any vertical background extraction or reference pixel correction, the amplifier offsets cause sharp discontinuities in the summed spectrum.
\edit1{The image shown here is from a grism exposure during Cryogenic Vacuum Testing 3 (CV3) test at NASA Goddard}.
The locations of amplifier boundaries and reference pixels are visible in Figure \ref{fig:detectorLayout}.

Fortunately, the amplifier offsets are efficiently removed by reference pixels and background subtraction.
For example, subtracting the average reference pixel in each amplifier will reduce the offsets between amplifiers, shown in Figure \ref{fig:ampOffsetsCV3GrismSlope}, (second from top panel).
One could subtract the mean reference pixel in each column, shown in Figure \ref{fig:ampOffsetsCV3GrismSlope}, (third from top panel) but given that there are only 4 reference pixels at the bottom of each column, this introduces extra noise in the subtraction.
Instead, one can use the background region for each column ($\sim$105 pixels used here) to improve removal of gradients and the amplifier offsets.
In Figure \ref{fig:ampOffsetsCV3GrismSlope} (bottom panel), we determine the median of all pixels from \edit1{Y=0 to 30 and Y=180 to 255} in each column and subtract this value from the whole column.
This is repeated for all columns.
This background subtraction method can efficiently remove the pre-amplifier offsets to the level where amplifier discontinuities become smaller than the combined read and photon noise in the image.
This background subtraction method can be applied to the STRIPE subarrays that are 64 px, 128, px, 256 px or the full 2048 px tall.
\edit1{We note that some amount of self-subtraction occurs from the extended wings of the PSF for a 64 px tall subarray.
A 20 px aperture contains $\sim$6\% less flux than a 128 px aperture at 3.5~$\mu$m, but will not produce a significant time dependence to the flux.
We discuss aperture losses more in Paper II.}

\section{1/f Noise}
The readout circuitry for NIRCam and other JWST NIR instruments introduces correlated noise from pixel-to-pixel.
This is visible when examining the time series of pixels as they are read out within a frame in the order they are addressed, as shown in Figure \ref{fig:detectorLayout}.\footnote{See a diagram of the readout directions at \url{https://jwst-docs.stsci.edu/near-infrared-camera/nircam-instrumentation/nircam-detector-overview/nircam-detector-readout}}
The spatial coordinates are re-labeled as time coordinates assuming a 10~$\mu$s pixel address time plus overheads between rows, so there is 1 pixel per 10~$\mu$s within a row with 120~$\mu$s gaps between rows.
The power spectrum of this pixel-by-pixel time series reveals high power at low frequencies that drops approximately like a 1/f power law.

\begin{figure*}[!hbtp]
\centering
\includegraphics[width=.4\columnwidth]{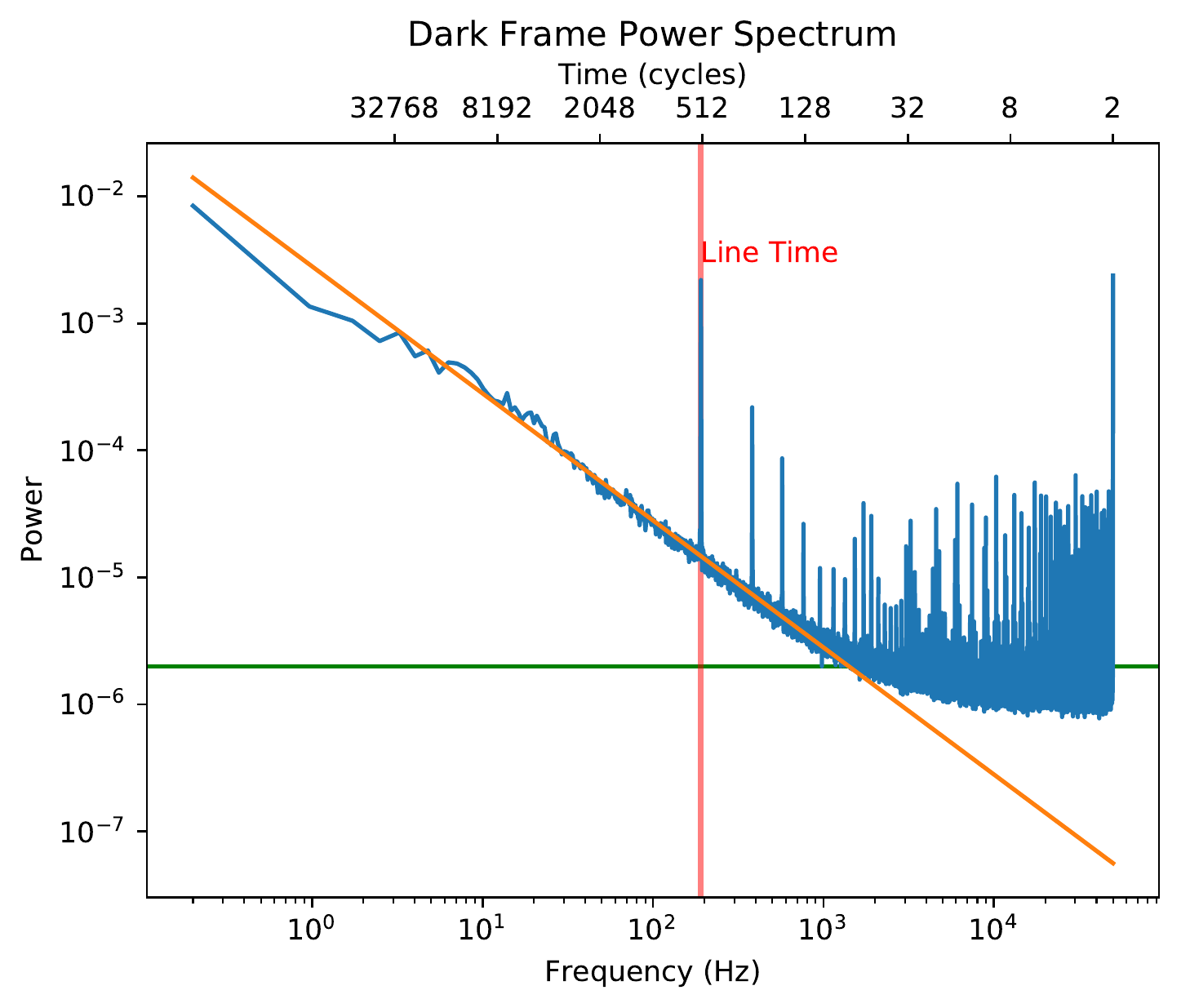}
\includegraphics[width=.4\columnwidth]{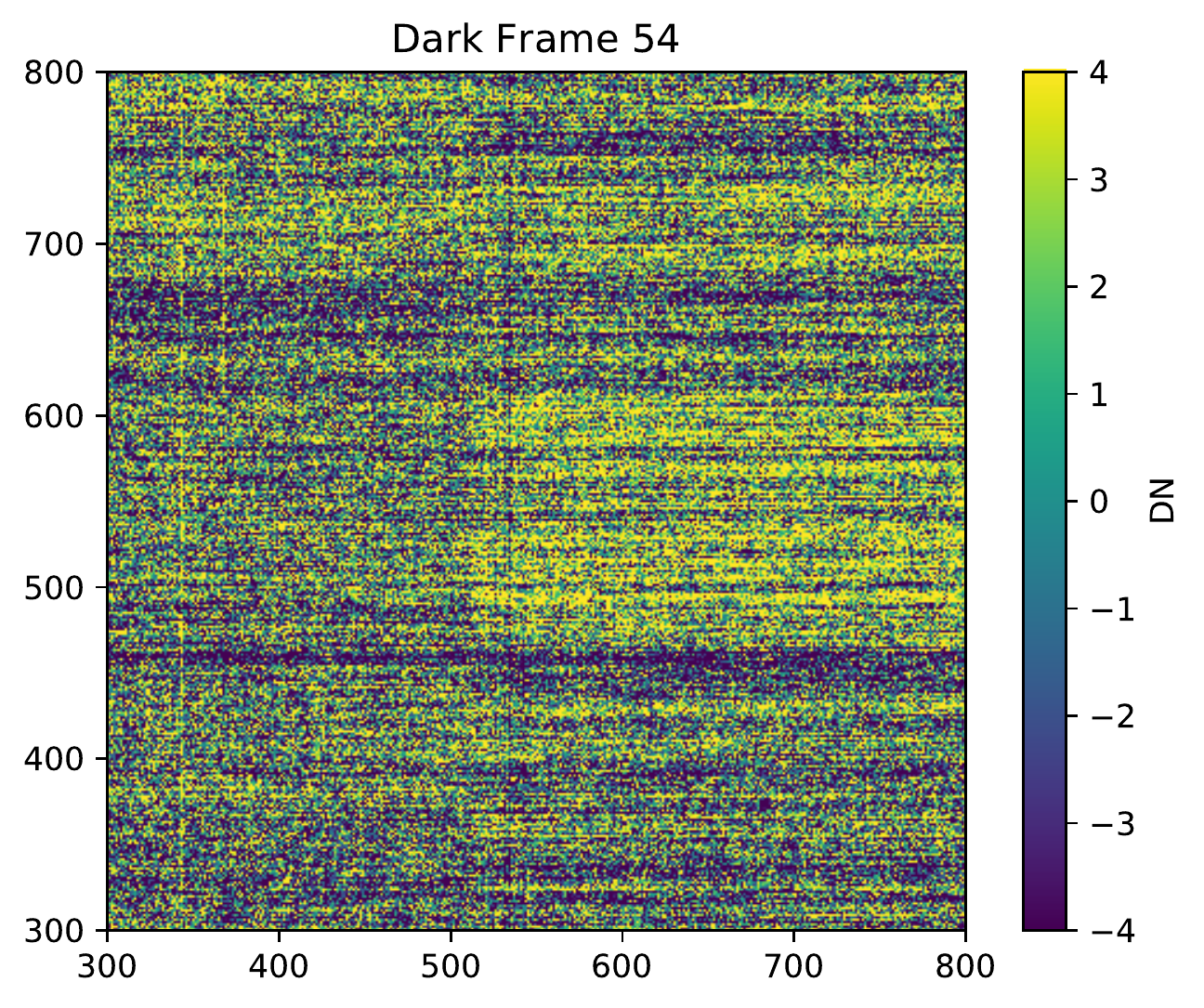}
\caption{{\it Left:} The average Lomb-Scargle periodogram from Amplifier 2's time series over 108 RAPID groups.
The noise in the spectrum is close to a 1/f power law (orange line) from 10 to 1000 Hz.
Above 1000 Hz, it is closer to white noise (green line) with spikes at specific frequencies.
The top x-axis shows the inverse frequency as measured in number of clock cycles, which corresponds to the number of pixels read within a row.
{\it Right:} This 1/f noise introduced in the readout circuit causes prominent correlated noise along the fast-read direction (horizontal) on a dark frame.
The amplifier offsets have been removed, but the 1/f noise from these adjacent amplifiers (separated at X=512) shows different behavior.
}\label{fig:pixelTSeriesPSpec}
\end{figure*}

Figure \ref{fig:pixelTSeriesPSpec} shows the average Lomb Scargle periodogram from output amplifier 2 (covering X=512 px to X=1023 px) over 108 dark groups in RAPID read mode.
The dark frames were part of a single integration taken at NASA Johnson during the OTIS (Optical Telescope element + Integrated Science instrument module) cryogenic test.
To create the periodogram, each group is subtracted by the average of all groups to remove the bias and all pixels that deviate by more than 80 DN from the average are masked to eliminate bad pixels.

Figure \ref{fig:pixelTSeriesPSpec} shows that the noise power drops nearly like 1/f from the lowest frequency in a frame to $\sim$1000 Hz, where the curve flattens out.
This indicates that the noise is closer to white noise, where each pixel's noise would be statistically independent from the others, at high frequencies.
The periodogram also shows many spikes at high frequency on top of the 1/f noise.
One of these prominent spikes occurs at the line reset time of 524 clock cycles, which is the time to read out all the pixels in a row for the amplifier and line overheads before moving to the next row.

The 1/f noise is compared to a simulated time series with the same robust standard deviation but with Gaussian white noise where each pixel is independent and identically distributed as a Gaussian.
It is clear from Figure \ref{fig:pixelTSeriesPSpec} that the 1/f noise and the periodic signatures significantly increase the power over uncorrelated white noise (green line) at low frequencies.
1/f rises above the white noise level at 70 cycles / 1400 Hz for amplifier 2 and the intersection ranges from 70 to 135 across the 4 amplifiers.
This power spectrum shows that the read noise is significantly correlated from pixel to pixel and this will affect both high precision spectroscopy and high precision photometry.

The 1/f noise can be reduced by subtracting reference pixels and/or active pixels surrounding the aperture of interest.
As the power spectrum shows, the 1/f noise begins to contribute significantly for frequencies below 1000 Hz (i.e., timescales longer than 100 cycles, where a cycle is the 10$\mu$s time to address a pixel).
100 cycles corresponds to 100 pixels, except between rows and frames due to overheads.
Ideally, source apertures would be less than 100 pixels wide (equivalent to 1~ms of time) so that background subtraction along rows will remove the 1/f noise but this is impossible for grism spectra dispersed along the rows.
The NIRCam grism time series uses the GRISMR element in the pupil wheel to disperse along the rows.
NIRCam has a GRISMC \edit1{element} that will disperse along the column direction.
We discuss the GRISMC element further in Section \ref{sec:grismcChange}.

It is also possible to modify the detector readout to reduce the 1/f noise by reading reference pixels more regularly.
This is accomplished by the NIRSpec IRS$^2$ mode, which reduces the read noise from 25~e$^-$ to 10~e$^-$ and also reduces the correlations between pixels \citep{rauscher2011irsSquared}.
However, the way that the NIRCam focal plane electronics transmits telemetry packets will not work for IRS$^2$.

\subsection{Long Darks}\label{sec:longDarks}

It is instructive to look at dark frames to isolate the contributions of the read-out electronics, including the SIDECAR ASIC, and how much they contribute to the error budget.
The dark frames do not have issues with the photometric stability of illumination lamps, which is a persistent challenge with ground-based tests.
They also do not contain photon noise in the background, which can make electronic noise sources harder to see.
We study here the contributions from read noise, including the 1/f noise from the electronics.
This experiment with dark frames does not test subtle changes in behavior of the electronic read noise under illumination, such as crosstalk where pixels that have no direct flux from a source in an image correlate with other regions of the detector with high flux rates.

A standard long dark exposure used for detector characterization is a single integration that is 108 groups of 1 frame each (RAPID mode).
The detectors are read in full frame mode so that each frame is 10.73677~s in duration with 4 output amplifiers in use, thus the total time for the exposure is 19.3 min.
We use one of these long dark integrations to create simulated time series as if it were observing many consecutive integrations in full frame mode.
The long darks are broken into 54 sequential pairs of images that are subtracted from each other (group 1 minus group 0, 3 minus 2, 5 minus 4 etc. in a zero-based counting scheme).
This method of breaking up the sequence precludes the ability to study the effects of detector resets, which only occur at the beginning of an integration.
\edit1{We find that the detector reset does affect the dark current in an integration, but we do not expect detector resets to significantly impact the time series.
This is discussed further in Appendix \ref{sec:darkCurrentVariations}.}
We also re-label the times to include the reset time that would normally occur after each integration, so the duration of the exposure from start to finish is 29.0 min.
\edit1{The re-labeling of the times makes the simulated exposure similar to a real time series observation but does not impact the analysis of the time series, which focuses on the scatter in the flux between read pairs.}

\begin{figure*}[!hbtp]
\centering
\includegraphics[width=.32\columnwidth]{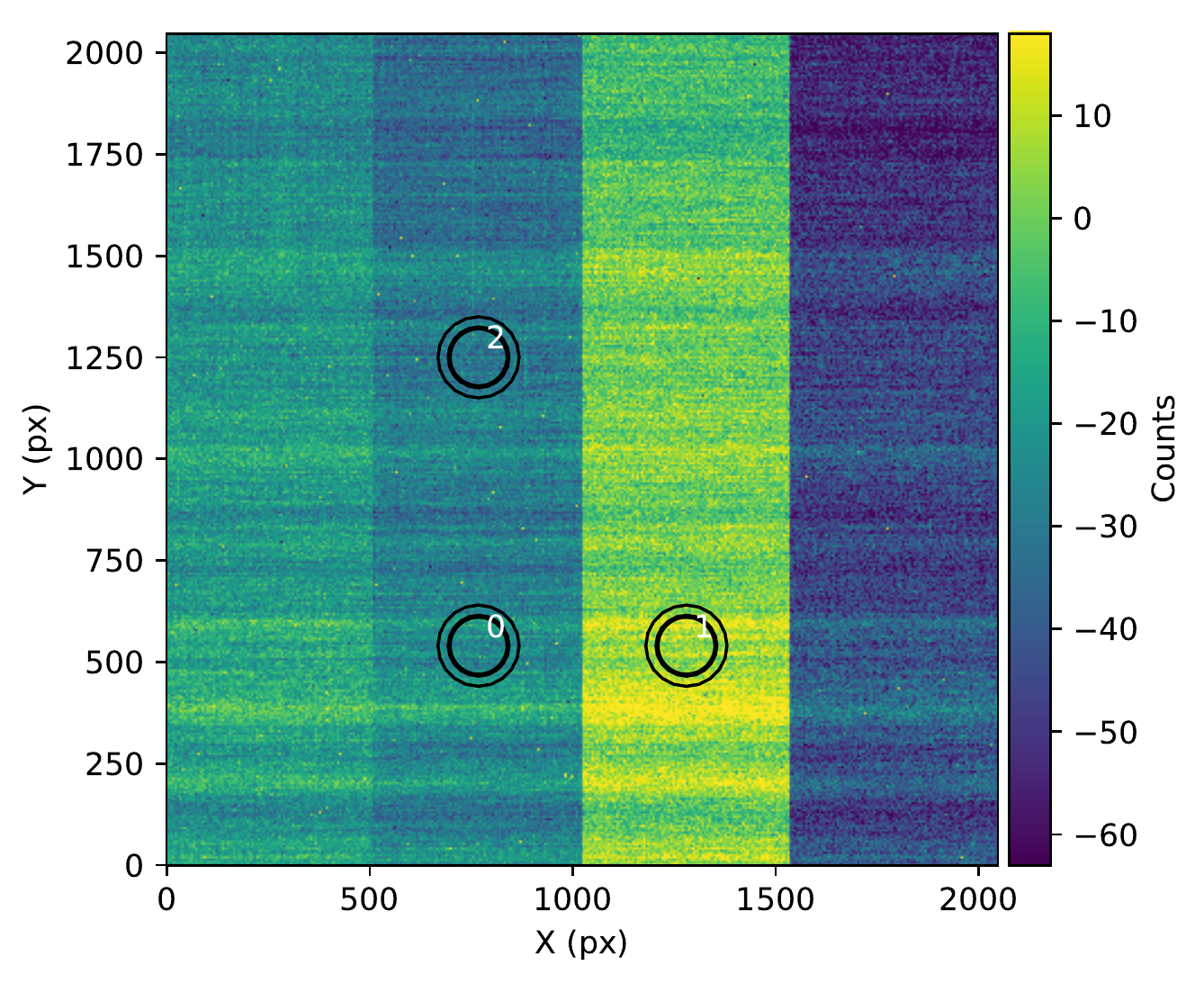}
\includegraphics[width=.32\columnwidth]{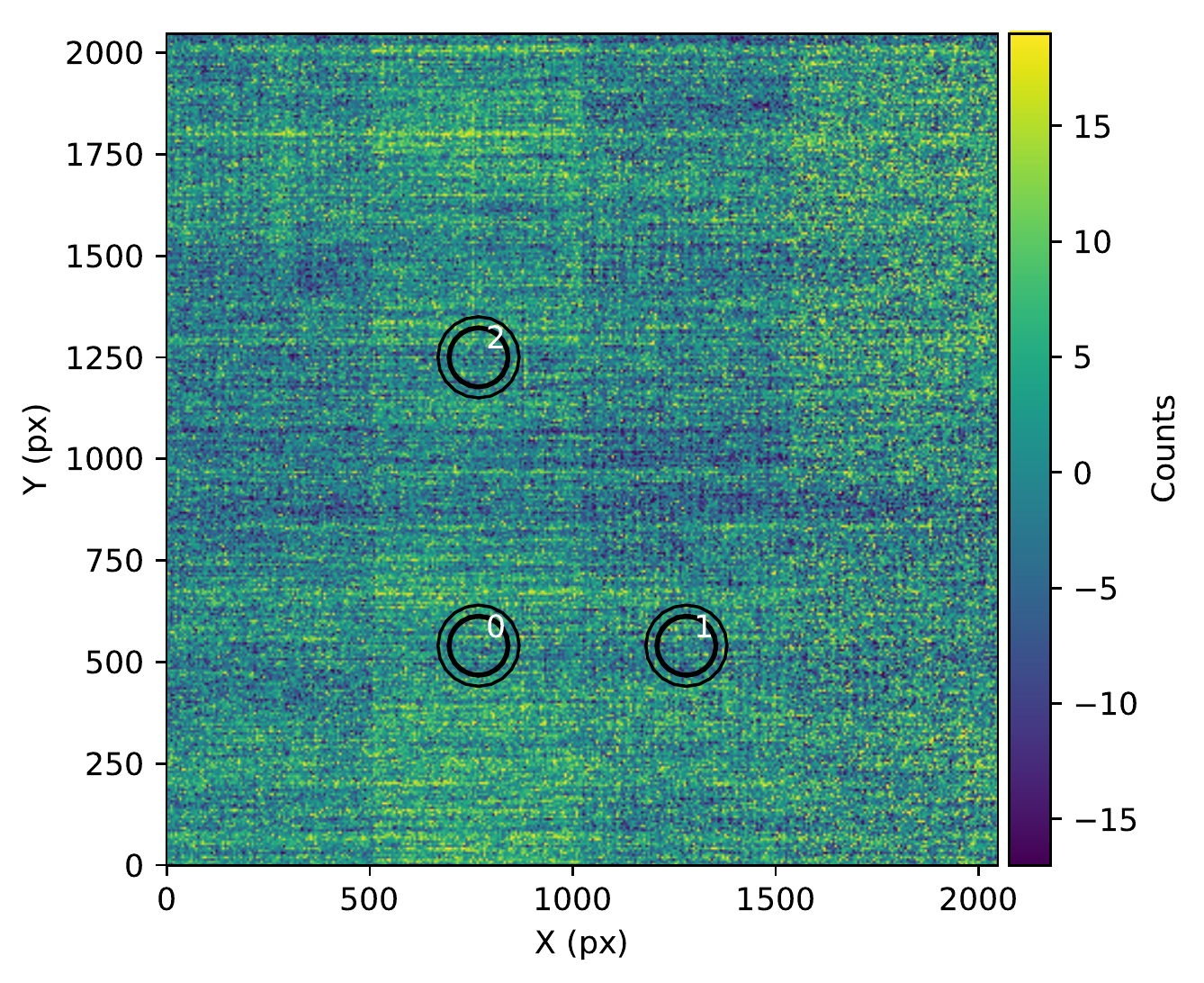}
\includegraphics[width=.32\columnwidth]{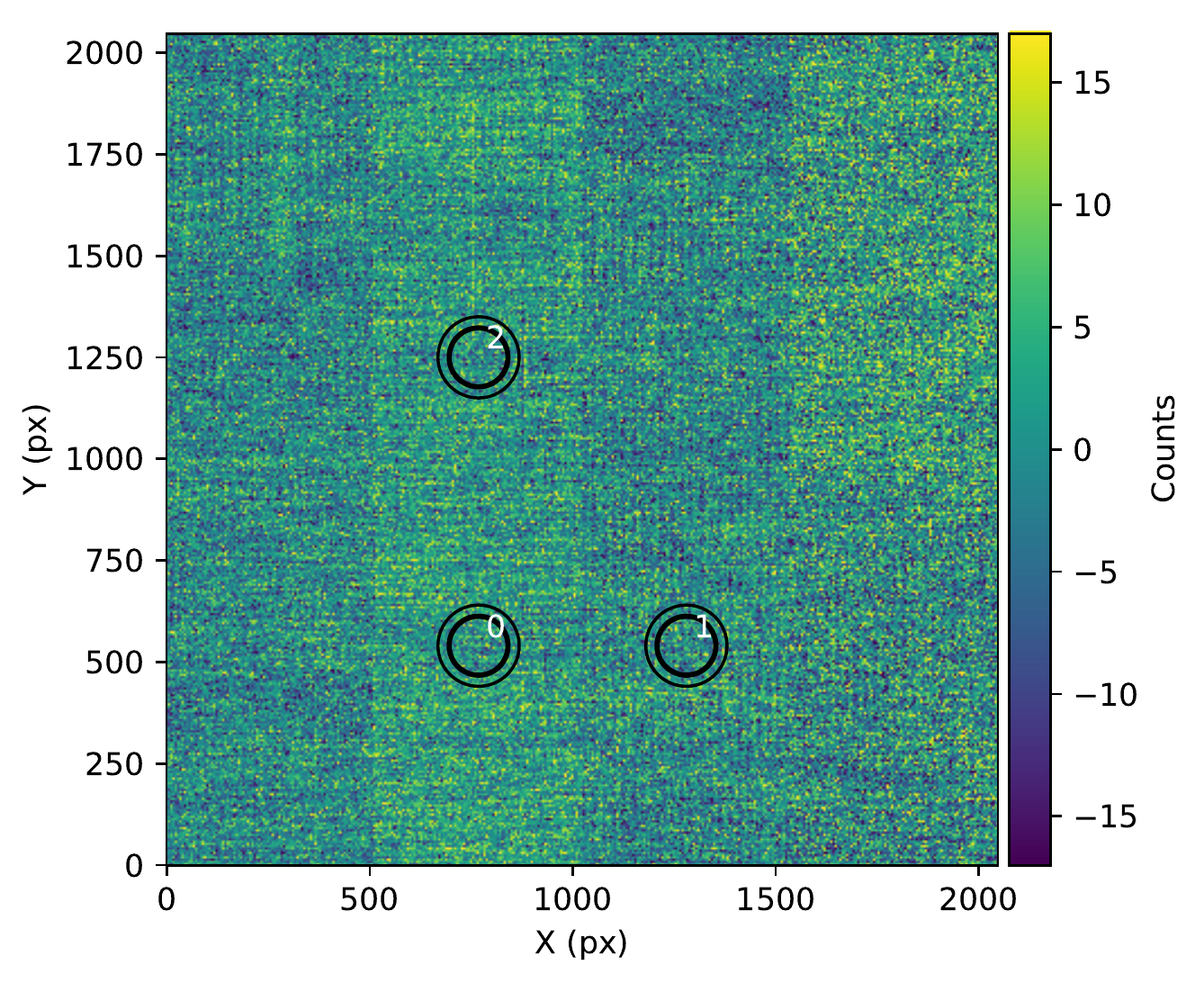}
\includegraphics[width=.32\columnwidth]{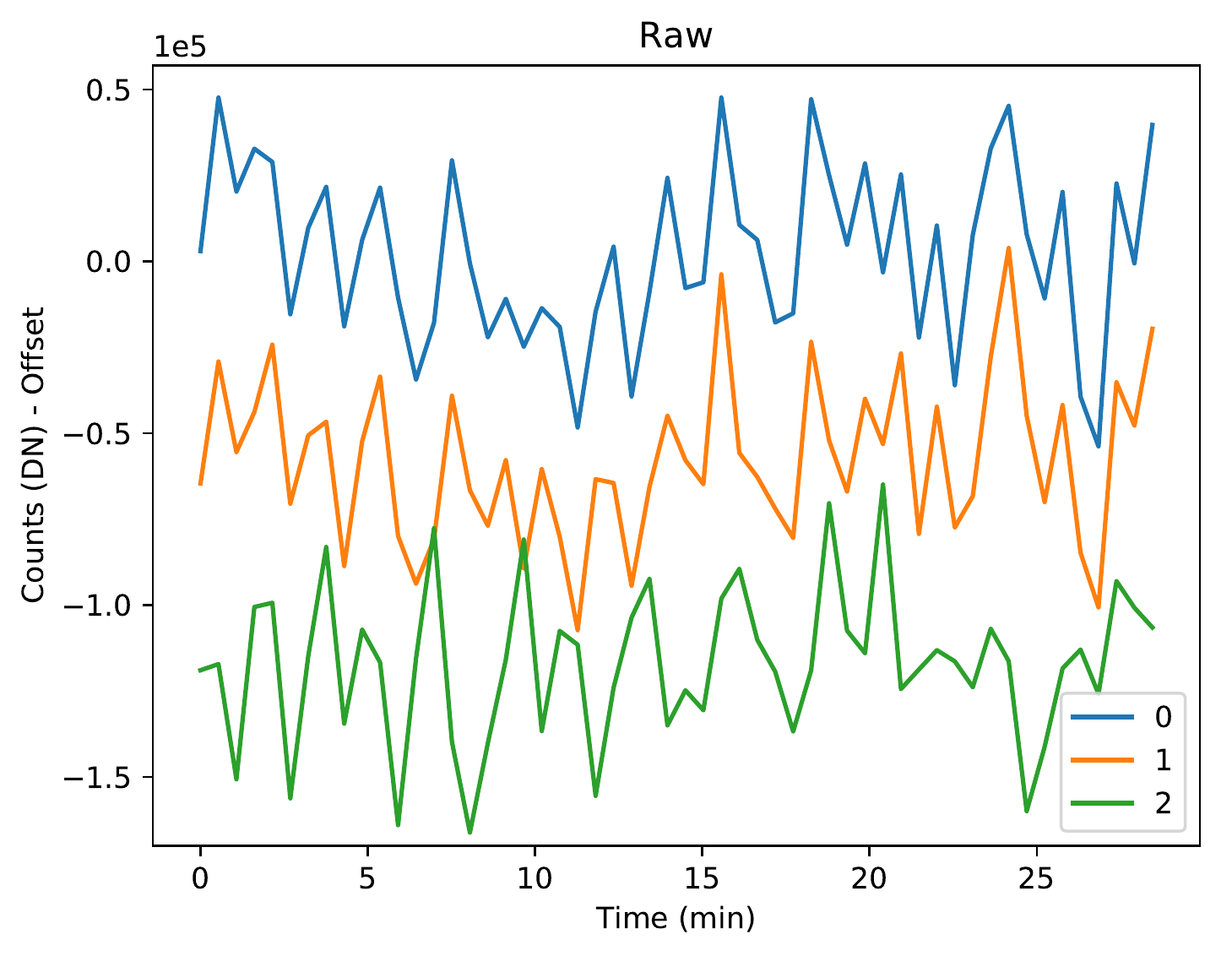}
\includegraphics[width=.32\columnwidth]{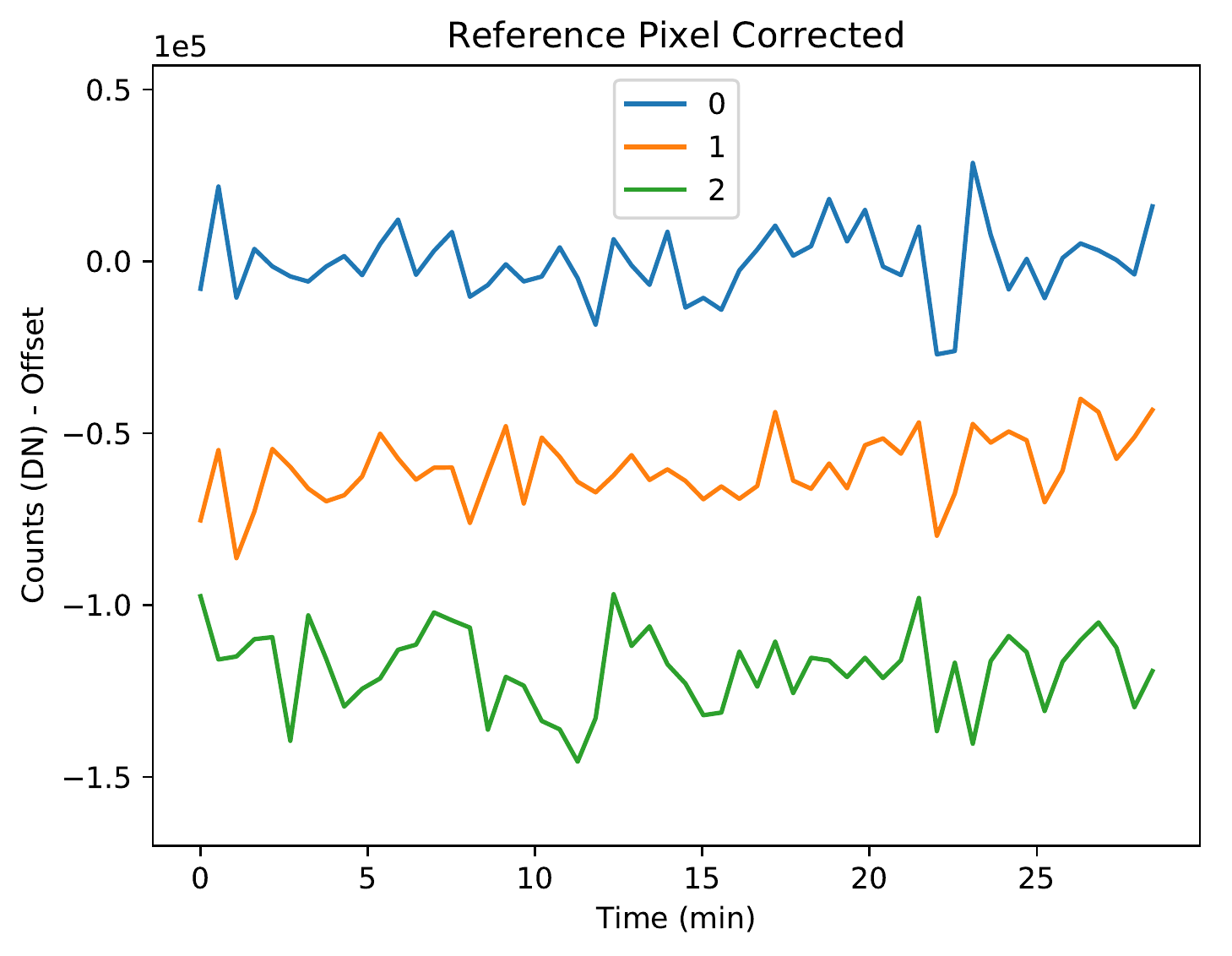}
\includegraphics[width=.32\columnwidth]{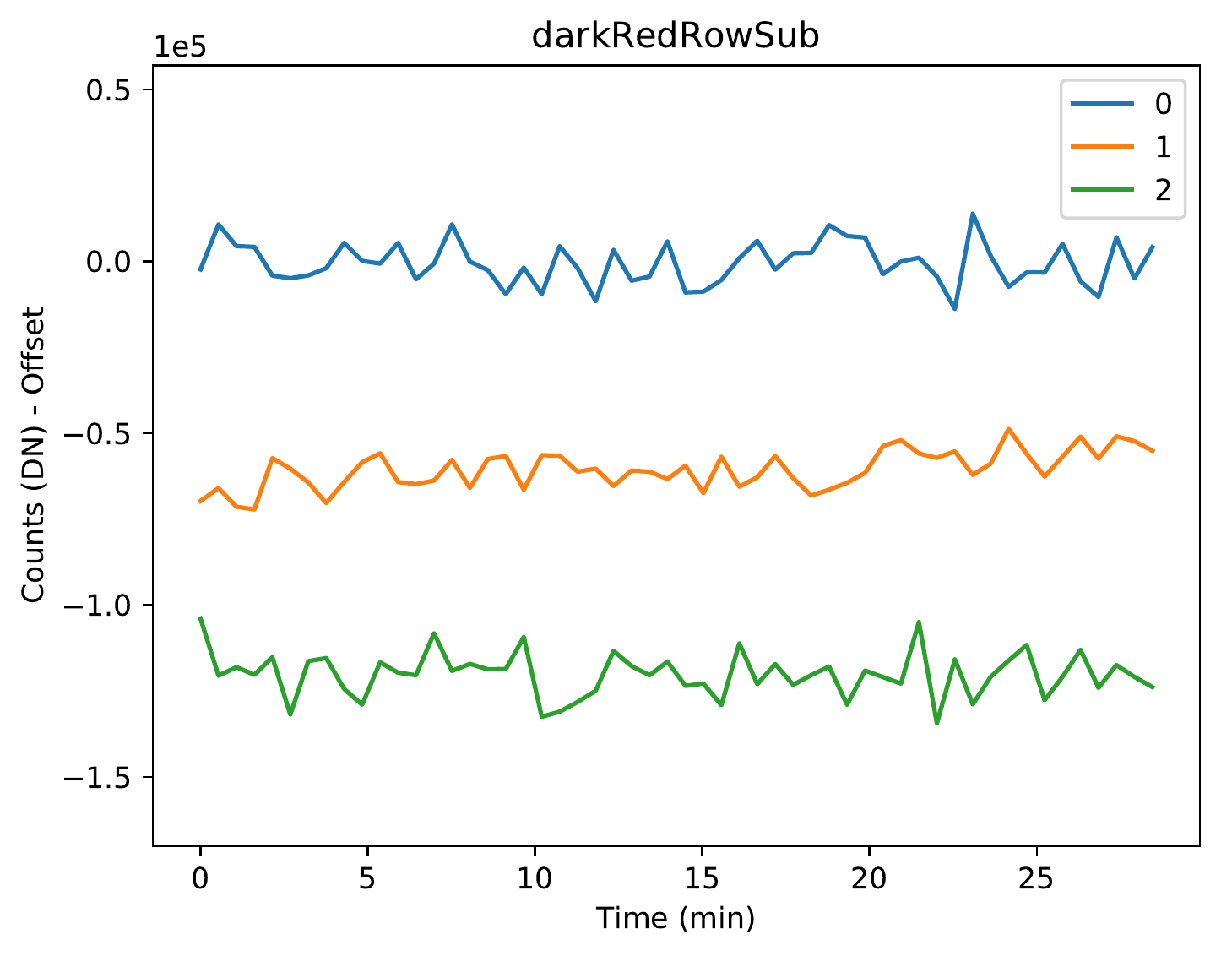}
\caption{Time series data of a 70 pixel radius photometric aperture obtained during a long dark exposure for detector ALONG=A5. The long dark exposure of 108 frames is used to construct 54 pairs of reads that are subtracted, in analogy with a series of integrations.
{\it Top:}  Example read pair subtraction with an overlay of the apertures for 3 different reduction steps from ``Raw" data using no reference pixel correction (left), reference pixel-correction that uses the top, bottom and side reference pixels (middle) and finally a method using a row-by-row median subtraction (right).
{\it Bottom} The time series for each circular aperture is calculated from the above images with \texttt{photutils} using the source circular aperture and a circular background annulus.
The ``raw'' frame photometry (left) has a standard deviation of 23,500 DN, which is reduced to 10,800 DN with reference pixel correction (middle).
Row-by-row subtraction lowers the contribution from 1/f noise to 6,000 DN (right).
}\label{fig:longDarkPhot}
\end{figure*}

Figure \ref{fig:longDarkPhot} shows an example dark read pair from frame 55 minus frame 54.
In the raw data, there are 4 prominent vertical stripes that are 512 pixels wide and 2048 pixels tall due to offsets in the read-out amplifiers.
These amplifier offsets are caused by drifts in the reference voltage that are reset once per frame and vary from amplifier to amplifier.
Additionally, there are obvious correlations along the fast-read (X) direction due to 1/f noise that appear as horizontal stripes in the image.

The ramps-to-slope pipeline \texttt{ncdhas} or the STScI JWST pipeline will apply reference pixel correction, bad pixel masks and non-linearity correction before fitting the groups up the ramp to a line.
In this work, we use \texttt{ncdhas}.
The reference pixels are comprised of a 4 pixel boundary along the edges of the detector, which are not connected to detector photodiodes and can therefore be used to measure common-mode electronic noise without being affected by photon counting noise.
As seen in Figure \ref{fig:longDarkPhot}, reference pixels also share the amplifier offsets, so they can be used to reduce the large offsets between amplifiers.

The reference pixels are most useful for subtracting the offsets between amplifiers because there are 8 rows of 512 pixels or a total of 4096 reference pixels to average for each amplifier at the bottom and top of the arrays, as shown in Figure \ref{fig:detectorLayout}.
There are an additional 4 columns on the left and right side boundaries (8176 pixels) that can be used to further correct the left and right amplifier offests.
However, the 1/f noise that correlates along the fast-read (X) direction is more difficult to remove with reference pixels.
Only 4 pixels on the left and 4 pixels on the right side may be used to subtract the 1/f noise in a given row and the 4 amplifiers can exhibit different 1/f noise. 
Therefore, significant residuals remain along the fast-read (X) direction in the reference-corrected read pair shown in Figure \ref{fig:longDarkPhot}.
These side reference pixels also cannot correct for read noise correlations that happen on shorter length scales than 512 pixels (5 ms in time).

\subsection{Photometric Aperture}\label{sec:photApertureDark}

Exoplanet transit measurements are dependent on the photometric stability within an aperture or spectrum.
We therefore perform aperture photometry on the dark frame as if it were a weak lens image with a 70 pixel radius circular source aperture and a 75 to 100 pixel background annulus illustrated in Figure \ref{fig:longDarkPhot}.
We place 3 different source apertures \edit1{with the same radii} on the detector for comparison of different regions.
The background subtraction will largely remove the offsets affecting the whole 512$\times$2048 vertical stripe read out by one amplifier.
However, the correlated 1/f noise along the X direction will not be subtracted efficiently by an annulus.
Instead, a row-by-row line fit to the background can be more effective.

For a read pair subtraction, the ``CDS'' (correlated double sample) read noise is 13.3 e$^-$ or 7.2 DN for the A5 detector studied here.
If we assume each pixel's read noise is uncorrelated with its neighbors, the total noise within the source aperture is then $\sqrt{N_{px}} \times RN $ where $N_{px}$ is the number of pixels and $RN$ is the read noise of a single pixel.
If we also assume that the background pixels are uncorrelated with each other and the source pixels, then the total noise in background-subtracted photometry is 1310 DN.

This theoretical read noise is smaller than the photon noise expected inside an aperture during photometric stability time series applications.
To calculate the photon noise as well as convert the noise into parts per million, we need to assume a well filling fraction for the integrations.
Here, we assume a peak pixel count of 30,000 DN or about 60\% well depth.
With this detector well filling fraction and a Weak Lens plus 8 waves (WLP8) point spread function, the number of DN collected within a 70 pixel radius aperture is $S_{tot}=3.9\times 10^7$ DN during the integration.
The Poisson shot noise for photons is $\sigma_{phot} = \sqrt{S_{tot}} / \sqrt{G} = 4,400$ DN, where $G$ is the gain ($G \approx 1.8$ for the LW detectors).
In terms of fractional noise, the photon shot noise should be 113 ppm and the read noise should be 33 ppm.
Therefore, the read noise is expected to be about 1/4 the shot noise for 2 group RAPID read mode (i.e., CDS).
When averaging together 111 full frames over 60 minutes (an order-of-magnitude transit duration for most planets), the photon noise would contribute 11 ppm and the read noise would contribute 3 ppm.

In our experimental time series, however, the standard deviation in the time series in these 3 aperture source locations is 22,000 to 25,000 DN (without any attempt to correct for amplifier offsets or 1/f noise using reference pixels or otherwise).
If uncorrected, this extra read noise will contribute 560 to 630 ppm to the time series per read pair!
This read noise would be $\sim$5$\times$ the $\sim$113 ppm photon noise.
However, there are ways to significantly reduce the pre-amplifier offsets and 1/f noise.

\begin{deluxetable*}{c|cc|cc}[b!]
\tablecaption{Summary of Noise Sources in Long Dark Test}
\tabletypesize{\footnotesize}
\tablewidth{0pt}
\tablehead{
\colhead{Aperture Type} &
\multicolumn{2}{c}{WLP8} &
\multicolumn{2}{c}{Grism 0.17 $\mu$m bin}\\
\colhead{Source Aperture Size} &
\multicolumn{2}{c}{70 px radius} &
\multicolumn{2}{c}{$\Delta$X= 165 px, $\Delta$Y= 14 px}\\
\colhead{Noise Description} &
\colhead{$\sigma$} &
\colhead{$\sigma$} &
\colhead{$\sigma$} &
\colhead{$\sigma$} \\
\colhead{} &
\colhead{DN} &
\colhead{ppm} &
\colhead{DN} &
\colhead{ppm}\\}
\startdata
Ideal Photon Noise & 4,400 & 113 & 1,430 & 388 \\
Ideal Uncorrelated Read Noise & 1,300 & 33 & 350 & 94 \\
\hline
Measured Read Noise with amplifier offsets (no correction) & 23,500 & 600 & 80,930 & 21,900  \\
Measured Read Noise (reference pixel correction) & 10,800 & 280 & 9,900 & 2,680 \\
Measured Read Noise (row-by-row subtraction) & 6,000 & 150 & 3,860 & 1044 \\
Measured Read Noise (row-by-row amp-by-amp mean subtraction) & 2,900 & 75 & & \\
Measured Read Noise (smoothed 161 px kernel interpolating over aperture) & 3,300 & 85 & & \\
\hline
Measured Read Noise (reference channel subtraction) & 9,700 & 250 & & \\
Measured Read Noise (20 Component PCA subtraction) & 5,400 & 140 & & \\
Measured Read Noise (5 Component PCA on each amplifier) & 3,200 & 81 & & \\
\hline
Measured Read Noise (variance-weighted summation)	& & & 3,060 & 830 \\
Measured Read Noise (covariance-weighted summation)	& & & 850 & 230 \\
\hline
Measured Read Noise (GRISMC row-by-row subtraction) & & & 683 & 205 \\
\enddata
\tablecomments{
We evaluate the theoretical and measured noise sources from 54 read pairs contained in a long dark integration.
The ppm units and photon uncertainties are calculated by assuming a peak 60\% well depth or an aperture sum of 3.9$\times 10^7$ DN during an integration for the weak lens PSF and 3.7$\times 10^6$ DN during and integration for the grism PSF.
The noise is reported for the aperture/spectral extraction (in DN) or relative to the photon signal (in ppm).
The integration is assumed to be a bright source with 2 groups of 1 integration each up the ramp (Correlated Double Sampling or CDS mode).
For the grism, statistics are shown for a representative 0.165~$\mu$m wide (165 pixel) wavelength bin centered on 3.34~$\mu$m with a spatial extent of 14 pixels in the Y direction.
\textcolor{magenta}{The PCA-based methods and reference channel subtraction are discussed in Appendix \ref{sec:moreSophisticatedOneOverFAppendix}.}}
\label{tab:noiseSummary1overf}
\end{deluxetable*}

These large correlations in read noise can be reduced with the reference pixels, which are designed to track electronic noise without the presence of background photon noise.
We use the \texttt{ncdhas} pipeline to apply reference pixel correction, which dramatically reduces the amplifier offsets and also reduces 1/f noise correlations along the fast read (X) direction.
The standard deviation of the flux for the 3 source apertures considered drops down to levels from 10,000 DN to 11,500 DN (depending on the aperture), but still falls short of the expectation of 1,310 DN if the read noise is uncorrelated.

The background annular aperture can affect the precision of a lightcurve, but there is a balance between photon noise, background noise and 1/f noise.
The aperture radius studied here (70~px) has an area (15,400~px$^2$) that is larger than the area of background annulus (13,700~px).
If background and photon noise were the only factor, the background annular aperture could be larger to increase the precision.
However, a large background becomes decreasingly correlated with the source pixels and 1/f noise makes using a background annulus much larger than 100 pixels less helpful for foreground-limited exoplanet system observations.

\subsubsection{Row-by-Row Subtraction}\label{sec:photRowByRow}
\edit1{It is important to subtract the values from either background pixels or reference pixels that are read out as close in time to the science target pixels as possible, where the correlations are strong.
Therefore, the column-by-column background subtraction shown in Figure \ref{fig:ampOffsetsCV3GrismSlope} along the slow-read direction is not very good at removing 1/f noise, even though it efficiently removes amplifier offsets.
Instead, pixels of the same row (ie. along the fast-read direction) are preferable to reduce 1/f noise.}

The 8 reference pixels in each row (as shown in Figure \ref{fig:detectorLayout}) are not sufficient to remove all 1/f noise in each row because the 8 pixels have their own read noise and they do not fully encapsulate the row's 1/f behavior.
Therefore, we perform another correction step that uses more pixels.
We find the median of each group up the ramp along the fast read (X) direction outside the source aperture and subtract this from every row.
We call this a row-by-row background subtraction.
The algorithm to perform row-by-row background subtraction is as follows:
\begin{enumerate}[noitemsep]
	\item \edit1{Choose one group of an exposure}\label{it:groupSelect}
	\begin{enumerate}
		\item Remove the pre-amplifier offsets within each amplifier by subtracting the median reference pixel\label{it:preAmpRemoval}
		\item Select a row of the group image\label{it:rowSelection}
		\begin{enumerate}
			\item Select all pixels outside the source aperture and outside other bright sources
			\item Calculate the median pixel count of the selected pixels
			\item Subtract this median pixel value from all pixels in the row including the source aperture\label{it:rowSubtraction}
		\end{enumerate}
		\item Repeat Step \ref{it:rowSelection} for all rows in the group\label{it:repeatRows}
	\end{enumerate}
	\item Repeat \edit1{Step \ref{it:groupSelect}} for all groups in an exposure \edit1{because each group will have its own pre-amplifier offsets.}\label{it:repeatGroups}
\end{enumerate}

\edit1{With row-by-row subtraction,} the standard deviation of the time series drops to 6000 DN, but is still larger than the expectation of 1310 DN if all noise were uncorrelated.
\edit1{
Row-by-row subtraction is shown in Figure \ref{fig:longDarkPhot}, both for an example image as well as a time series.
We note that Steps \ref{it:preAmpRemoval} through \ref{it:repeatRows} nearly commutes with the slope fit for an integration (within 1 ppm), so one can also apply Steps \ref{it:preAmpRemoval} through \ref{it:repeatRows} on a slope image rather than group-by-group as in Step \ref{it:repeatGroups}.}

\subsubsection{Row-by-Row Amp-by-Amp Subtraction}\label{sec:indAmpAvg}
A modification to the above strategy is to find the median of each amplifier individually, which we call row-by-row amp-by-amp subtraction.
This takes care of any differences between amplifiers.
The algorithm is the same as in Section \ref{sec:photRowByRow}, with a modification to Step \ref{it:rowSelection} to select the pixels of a row that correspond to one amplifier (512 px per amplifier for FULL or STRIPE subarrays) and to repeat this for all four amplifiers.

Row-by-row amp-by-amp subtraction dramatically decreases the errors as compared to reference pixels alone.
The standard deviations of the background-subtracted photometry range from 2,540 to 3,300 DN or 65 to 85 ppm across detectors, as listed in Table \ref{tab:noiseSummary1overf}.
Therefore, this simple approach can dramatically decrease the 1/f noise to well below the photon limit.
The row-by-row amp-by-amp subtraction is possible for weak lens photometry but not GRISMR spectroscopy, \edit1{which is} dispersed parallel to detector rows (the fast-read direction).
This is because the spectrum will span across the entirety of one or more amplifiers, as shown in Figure \ref{fig:detectorLayout}, so there are no source-free pixels in one or more amplifiers to calculate the median 1/f noise from.
\edit1{We note that row-by-row amp-by-amp subtraction can be difficult in dense stellar fields.
If there are bright sources along the same rows as a source and also within the same amplifier, it can make finding clean background regions in a 512 px wide amplifier challenging.
In this case, one may have to revert to the row-by-row subtraction discussed above in Section \ref{sec:photRowByRow}.}

\subsubsection{More sophisticated 1/f noise reduction techniques}
We also consider more sophisticated ways to reduce read noise than the median row subtraction discussed in Section \ref{sec:photRowByRow} and \ref{sec:indAmpAvg}.
These include a smoothing kernel, using an amplifier as a ``reference channel'' and principal component analysis.
None of these techniques yielded better results than row-by-row amp-by-amp subtraction in Section \ref{sec:indAmpAvg}, but the ideas are discussed in Appendix \ref{sec:moreSophisticatedOneOverFAppendix} to be used as a reference for future work to reduce 1/f noise.

\subsection{Grism Extraction}\label{sec:GrismDarkExtraction}

\begin{figure*}[!hbtp]
\centering
\includegraphics[width=.32\columnwidth]{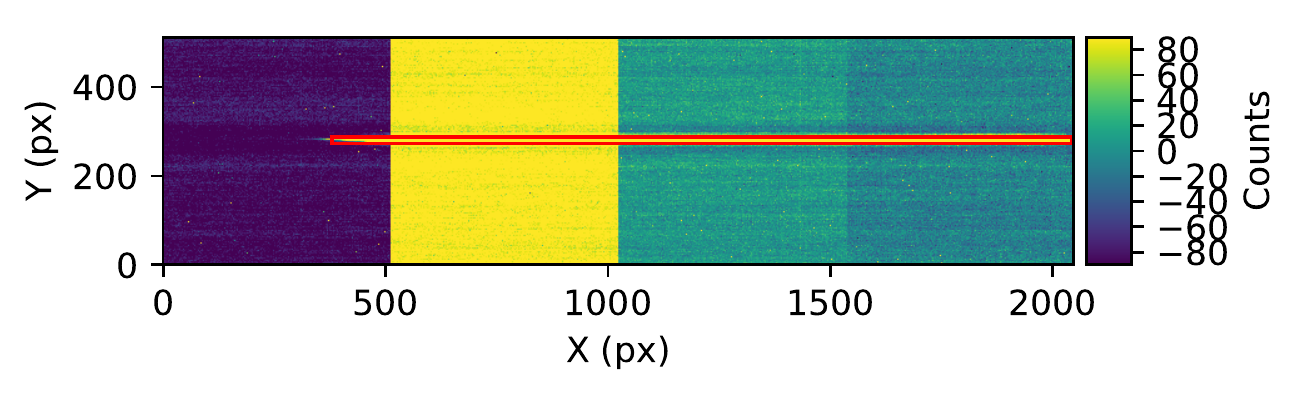}
\includegraphics[width=.32\columnwidth]{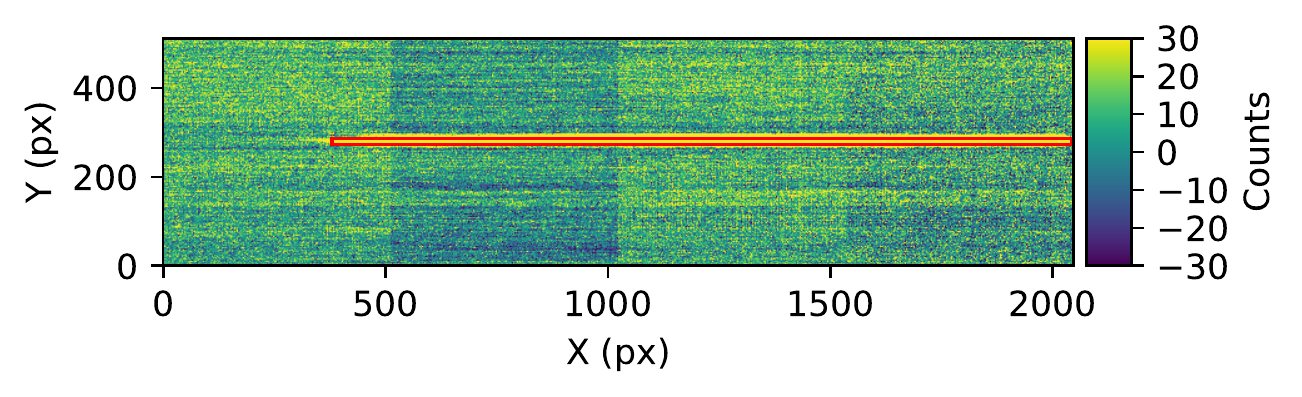}
\includegraphics[width=.32\columnwidth]{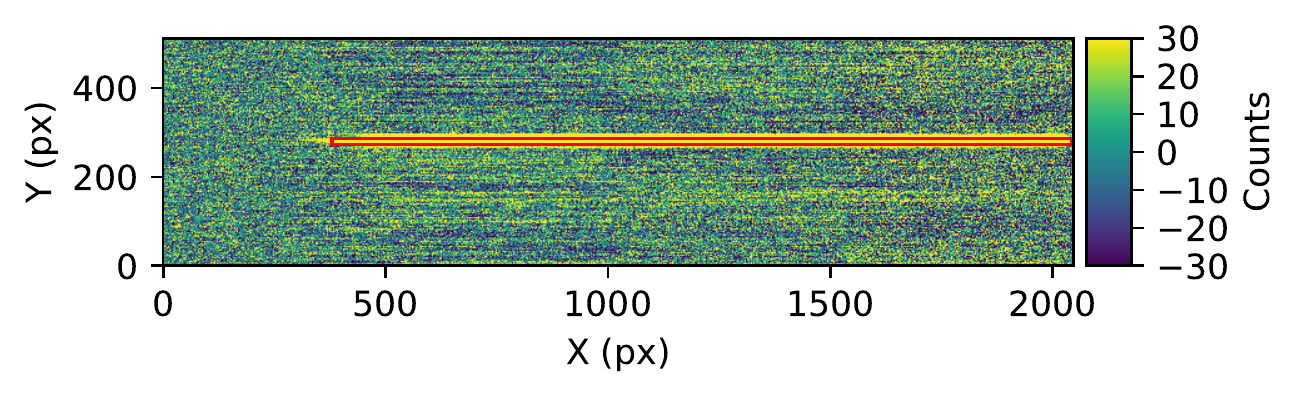}
\includegraphics[width=.32\columnwidth]{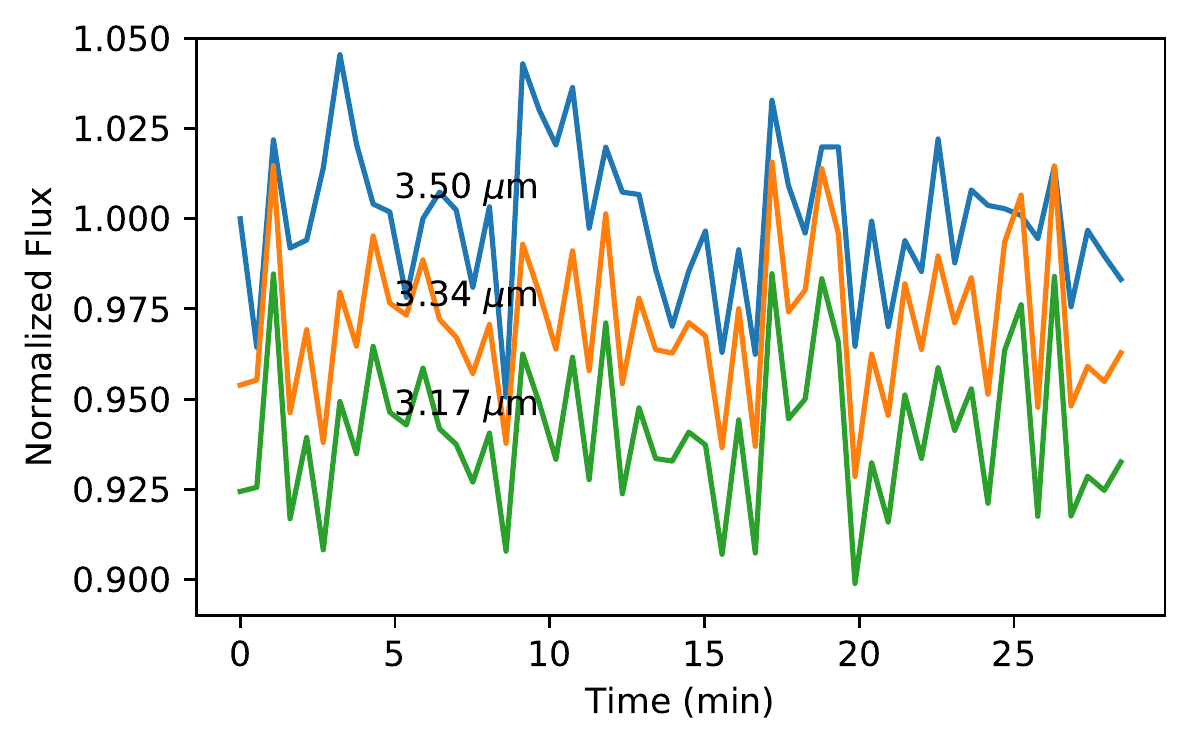}
\includegraphics[width=.32\columnwidth]{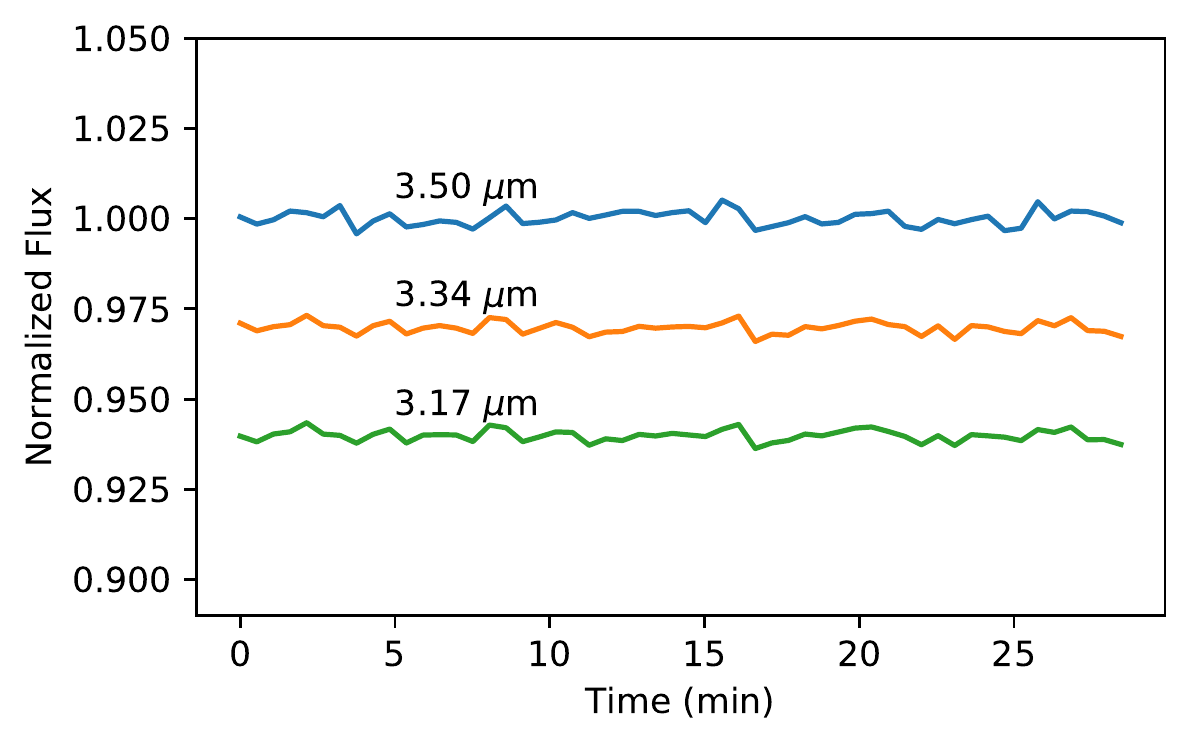}
\includegraphics[width=.32\columnwidth]{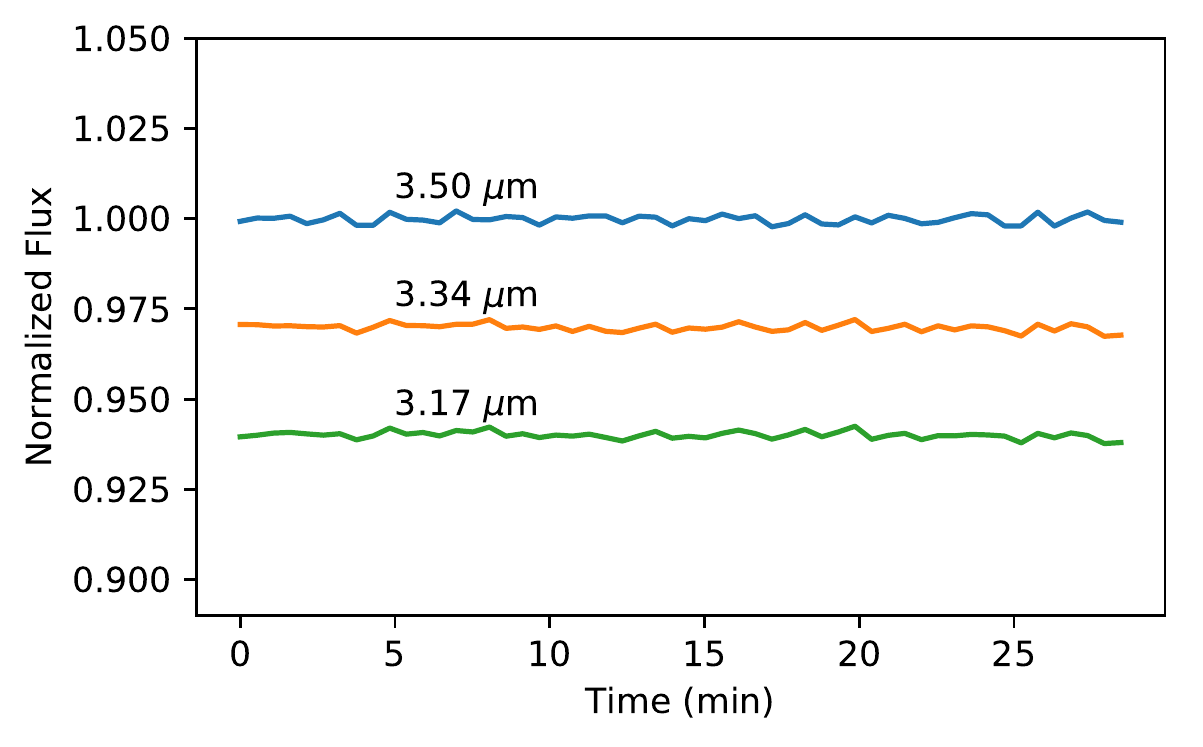}
\caption{{\it Top Row:} The read pairs have significant pre-amp offsets and 1/f noise (left), which are reduced by reference pixels (middle) and column-by-column, row-by-row subtraction (right).
The spectrum (outlined in red aperture box) is dispersed along the fast read (horizontal) direction, so it is particularly susceptible to 1/f noise.
We extract the full spectrum and create normalized time series for three example wavelength bins with aperture sizes with $\Delta$X=165~px (0.165~$\mu$m) and $\Delta$Y=14~px \edit1{(0.88\arcsec)}.
{\it Bottom Row:} The standard deviation of the time series dramatically decreases from 21,900 ppm to 2,680 ppm with reference pixels and 1,044 ppm with column-by-column, row-by-row subtraction for the 3.34~$\mu$m bin.
1/f noise still dominates over the theoretical photon noise limit of 388 ppm for the 3.34~$\mu$m bin.
Here, the simulated grism image for an F322W2 filter spans 3 amplifier channels (each 512 pixels wide) and part of the leftmost amplifier, so only reference pixels and horizontal pixels with X $<$ 300 can be used to reduce the 1/f noise.
}\label{fig:longDarkGrism}
\end{figure*}

The current NIRCam grism time series mode, which \edit1{uses} the GRISMR \edit1{element} in the pupil wheel, is particularly susceptible to 1/f noise because the dispersion direction is parallel to the detector's fast-read direction as shown in Figure \ref{fig:detectorLayout}.
This is also visible in Figure \ref{fig:longDarkGrism} where the grism apertures (in red) are parallel to the 1/f noise correlations.
The background subtraction along the spatial direction (vertically) will efficiently remove pre-amplifier offsets from frame to frame and amplifier to amplifier.
However, there are only limited background regions along the horizontal direction available to subtract out 1/f noise.
This will result in increased noise on the right three amplifiers for the F322W2 filter (2.4 to 4.0~$\mu$m over $\sim$1600 px) and the left two amplifiers for the F444W filter (3.9 to 5.0~$\mu$m over $\sim$ 1100 px), in the native ``raw'' detector pixel coordinate system used in this paper.

We give some illustrative numbers for a 0.165 $\mu$m wide (R$\sim$20) bin of a grism spectrum centered on 3.34 $\mu$m as listed in Table \ref{tab:noiseSummary1overf}.
Without reference pixel subtraction, the grism apertures have standard deviations of 80,930 DN.
Reference pixel subtraction drops this value to 9,900 DN.

As in Section \ref{sec:photRowByRow}, we perform row-by-row subtraction to further reduce the noise to 3,860 DN.
We use the pixels from X=5 to X=301 for the row-by-row subtraction to ensure there is essentially no source flux being subtracted.
Note this is only possible when there are no nearby sources at X=5 to X=301.
For a well depth of 60\%, there are $3.7 \times 10^6$ DN of source counts in the 0.17 $\mu$m bin, which amounts to a photon error of 1,430 DN assuming a gain of 1.8.
Therefore, 1/f noise can potentially dominate the error budget for broadband grism time series.
Expressed in ppm, the read noise after processing with row-by-row subtraction is 1,044~ppm as compared to a photon noise of 390~ppm for a CDS subtraction (2 groups in ramp).

Given the high 1/f read noise, it is important to maximize the number of samples up the ramp.
The 1/f read noise averages down like $\sqrt{N}$ statistics after background subtraction (i.e., each read can be assumed independent and identically distributed).
Therefore, for the example bin size of 0.17~$\mu$m, there should be at least 7 reads to average per integration to ensure that the read noise is less than the photon noise.
For RAPID and BRIGHT1 modes, there should therefore be 7 groups and for BRIGHT2, there should be at least 4 groups (of 2 frames each).
When deciding between BRIGHT1 and BRIGHT2, we recommend BRIGHT2 because it co-adds reads together rather than skip frames.
Using $>7$ groups for RAPID and $>$ 4 groups for BRIGHT2 would make the photon noise about equal to the read noise for this aperture size but ideally the read noise should be a small fraction of the photon noise.

\subsubsection{Using GRISMC Element}\label{sec:grismcChange}

One way to mitigate 1/f noise is to use a different grism element (GRISMC) in the pupil wheel, which disperses the spectra along the column (vertical) direction.
The NIRISS and NIRSpec instruments' spectroscopic modes are designed this way so that the dispersion direction is approximately {\it perpendicular} to the fast-read direction.
This allows spatial background subtraction along the fast-read direction to mitigate 1/f noise.
\edit1{GRISMC is being discussed as a science enhancement for a future JWST cycle, most likely beyond Cycle 1.}

The tradeoff with the GRISMC pupil element is that vertical subarrays will only be read with one amplifier (WINDOW mode), and thus will be about 4 to 4.5 times longer than four amplifiers (STRIPE mode), depending on line overheads.
The overheads come into play the most for small subarrays.
A 2048 $\times$ 64 STRIPE subarray has a frame time equal to 0.34061 s whereas a 64 $\times$ 2048 WINDOW mode has a frame time equal to 1.558 s.
Therefore, the maximum brightness observable without saturation with a GRISMC pupil element and 64 $\times$ 2048 WINDOW mode would be $K \approx 5.7$ compared to the GRISMR 2048 $\times$ 64 STRIPE mode with $K \approx 4.0$.
For this reason, the GRISMC is not currently available for time series observations.
If it is enabled as a new mode, GRISMC observations could have substantially reduced 1/f noise for the majority of known transiting planets, which are fainter than $K \approx 5.7$.

\begin{figure*}[!hbtp]
\centering
\includegraphics[width=.25\columnwidth]{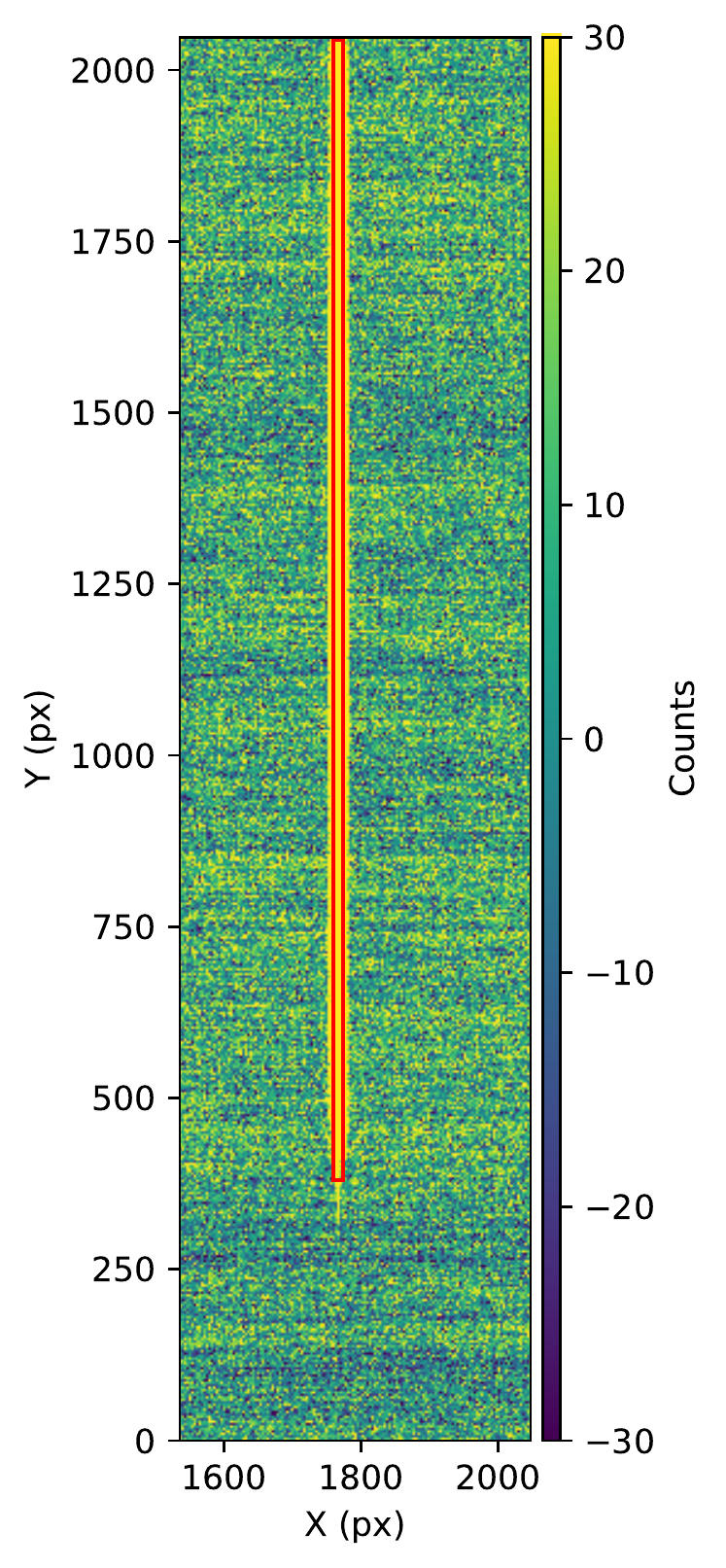}
\includegraphics[width=.25\columnwidth]{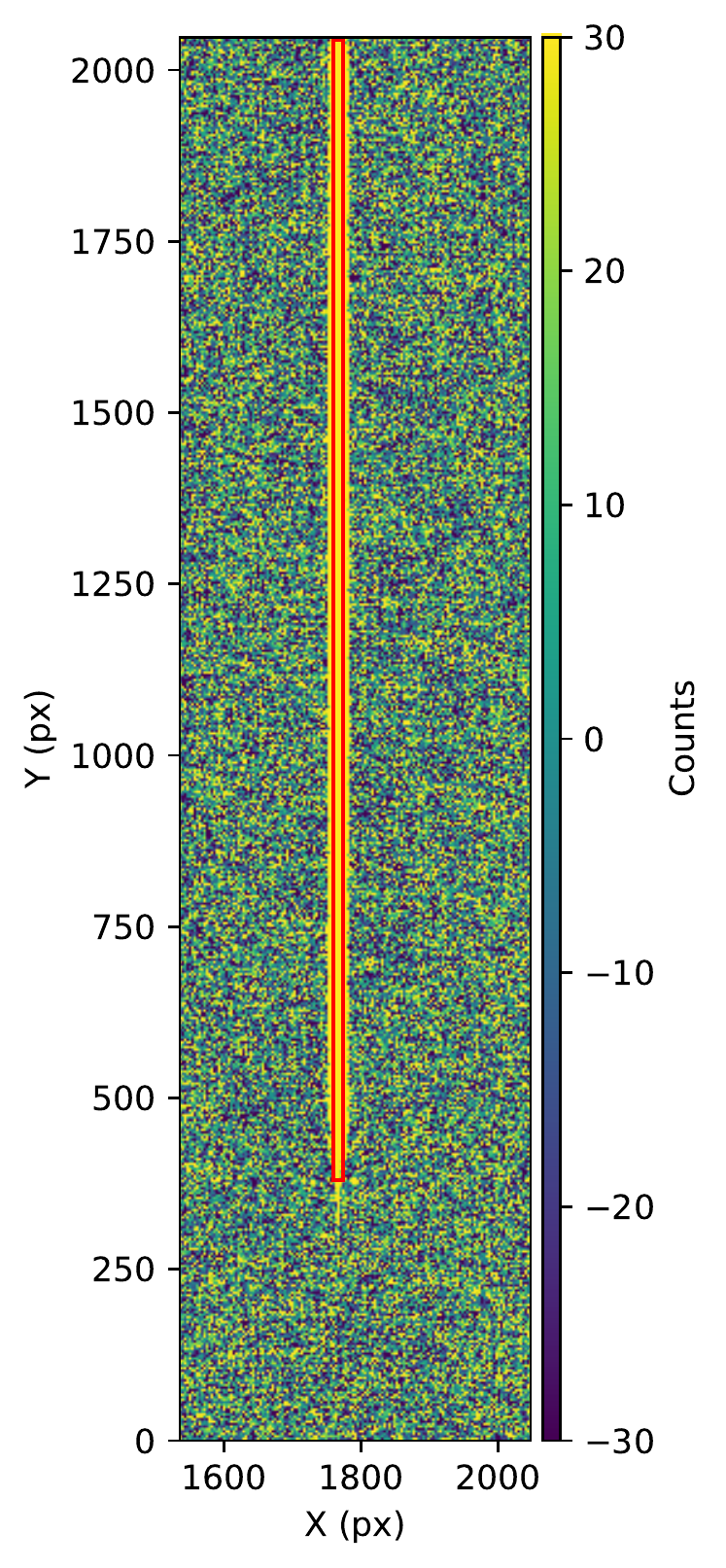}
\includegraphics[width=.36\columnwidth]{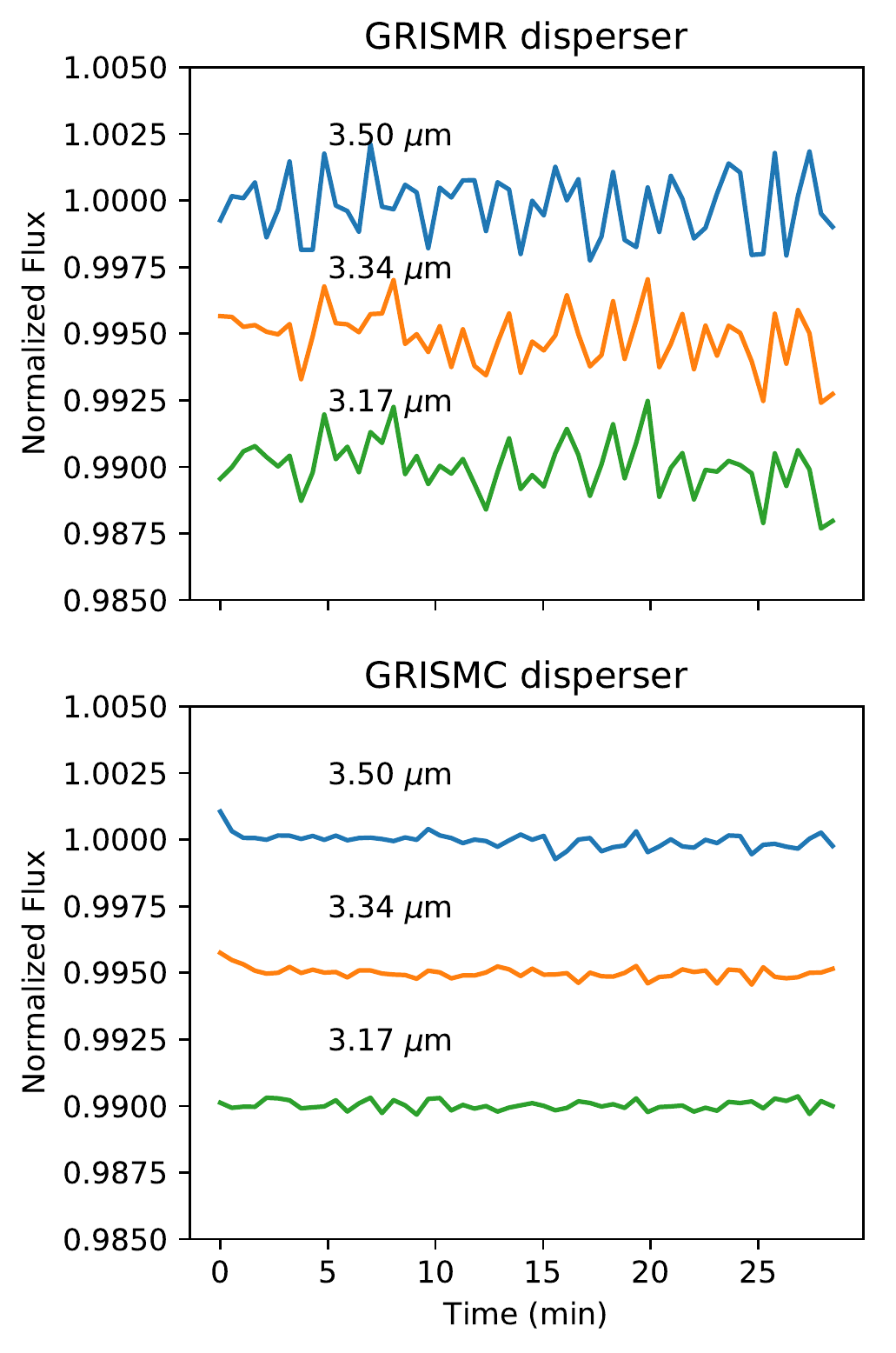}
\caption{{\it Left Panel:} The GRISMC element can be used to disperse the spectrum along the column direction so that it is approximately perpendicular to the fast-read direction.
{\it Middle Panel:} With the GRISMC disperser, row-by-row background subtraction (outside the source aperture) and within an amplifier efficiently reduces the 1/f noise.
{\it Right panel:} The simulated time series with the GRISMC disperser dramatically lowers the detector noise in the time series (205 ppm) as compared to the GRISMR disperser (1,044 ppm) because of the reduced 1/f noise.
In this extraction, the background subtraction aperture was optimized to reduce the 1/f noise.
}\label{fig:GRISMRvsGRISMC}
\end{figure*}

Figure \ref{fig:GRISMRvsGRISMC} demonstrates the advantage of the GRISMC dispersing element in NIRCam's long wavelength pupil wheel.
Before background subtraction (left panel), prominent 1/f noise can be \edit1{seen} as horizontal lines and bands approximately perpendicular to the dispersion direction.
We show an example row-by-row linear background subtraction using the pixels from X=1537 to X=1760 and X=1774 to X=2044.
This efficiently reduces the 1/f noise along the rows, but leaves a small residual nearby the source.
\edit1{The residuals are difficult to see by eye in the image, but are apparent in the covariance matrix in Appendix \ref{sec:covMatrixRowByRow}}.

Background subtraction can alternatively be performed on short pixel distances ($\lesssim 70$ px), where the source flux is low ($\gtrsim$ 10 px spatial pixels from the source center) with the GRISMC element.
As seen in Figure \ref{fig:pixelTSeriesPSpec}, the 1/f noise begins to appear for more than $\sim$ 70 clock cycles corresponding to distances more than about 70 pixels along a row and frequencies below 1400 Hz.
Subtracting out 1/f noise at 70 clock cycle timescales is not possible with the GRISMR element, where the spectrum extends more than 1100 px along the fast-read direction for the F444W filter.
Furthermore, the background subtraction with the GRISMC element can be performed within one amplifier rather than extrapolating one \edit1{amplifier's} 1/f noise behavior from other amplifiers as was done with the GRISMR element.
We found that a background from 7 px to 30 px from the source efficiently removed 1/f noise down to 683 DN (205 ppm) in a \edit1{0.165}~$\mu$m-wide wavelength bin as shown in Figure \ref{fig:GRISMRvsGRISMC} (right).
The GRISMC therefore allows a standard background subtraction and summation source extraction methods, which efficiently reduce 1/f noise to below the photon noise level.

\subsubsection{Small Spatial Apertures}\label{sec:smallSpatialAp}

Given the high level of correlation between pixels, another way to reduce 1/f noise is with a small aperture in the spatial (Y) direction.
The spatial point spread is tightly concentrated with a full width half maximum (FWHM) of 1.3 to 2.6 pixels depending on wavelength, so smaller apertures are possible.
For example, a 14 px (spatial Y-direction) tall aperture results in a measured read noise of 3,860 DN (1044 ppm) for the same 0.17$\mu$m wide bin centered at 3.34$\mu$m whereas a 3 px tall aperture results in a measured read noise of 1,070 DN (290 ppm), which is below the expected photon noise.
However, this will sacrifice photons, so there must be a balance between reducing 1/f noise while capturing as many photons as possible.
Small spatial apertures will also be more susceptible to pointing jitter or wavefront variations that manifest as variable aperture losses.
Therefore, a weighting scheme expanded from optimal extraction \citep{horne1986optimalE} is preferred.

\subsubsection{Optimal Co-Variance Weighted Flux}\label{sec:optimalCovWeights}

Here, we provide a modification to the ``conventional'' optimal extraction formula from \citet{horne1986optimalE}, that uses the covariance matrix of correlated pixels.
The goal is to find a scheme that estimates the total flux by summing the value in each pixel $j$.
We start by calculating a profile function $P_j$ in the spatial direction that describes the fraction of the total flux in pixel $j$.
The profile is normalized so that
\begin{equation*}
\sum_{j=1}^{m} P_j = 1.
\end{equation*}
The summation for NIRCam would be along the Y (spatial) direction for a fixed wavelength.
\edit1{There is a small tilt of the grism spectrum of $0.11\deg$ (ie 1:480), so summing along the Y direction will slightly mix together different wavelengths at the sub-pixel level.
The exoplanet science applications we focus on here are concentrated on measuring broad molecular features, where the maximum possible resolution of the instrument is not as critical.
Therefore, we allow for a small amount of wavelength mixing by studying summations along the Y direction instead of a more complicated interpolated model.}
The technique below can also be modified to include summation along the spatial and spectral directions.

Any pixel can be used to estimate the total flux by dividing by the profile.
\begin{equation*}
f_j = \frac{S_j}{P_j},
\end{equation*}
where $S_j$ is the background-subtracted flux.
Therefore, the flux total is a weighted average of each pixel's estimate of the total flux $f_j$.
The goal becomes a weighting scheme to determine which weights $w_j$ minimize the variance in the total background estimate:
\begin{equation*}
f_{opt} = \frac{\sum_{j=1}^{m} w_j \frac{S_j}{P_j}}{\sum_{j=1}^{m} w_j}.
\end{equation*}
The weights $w_j$ can take on any value while still giving an unbiased estimator of the flux \citep{horne1986optimalE}.
The weights are not normalized like the profile, $P_j$.

If one assumes that all pixels are independent thereby having zero correlation, the optimal weights (ones that minimize the variance in $f_{opt}$) are
\begin{equation}\label{eq:varianceweights}
w_{j,var} = \frac{1}{\mathrm{Var}(\frac{S_j}{P_j})} = \frac{1}{\frac{1}{P_j^2} \mathrm{Var}(S_j)} = \frac{P_j^2}{V_j},
\end{equation}
where Var() is the variance or the expectation of the squared deviation from the mean and $V_j =$Var$(S_j$) \citep{horne1986optimalE}.

Exoplanet observations, where high signal to noise is needed to separate the planet signal from the stellar signal, are generally in the foreground limit (where the variance approaches 
\edit1{the} number of photons), so $V_j = N_{phot,j}$ in this limit.
\edit1{The profile is also proportional to the number of photons from the source, so}
\begin{equation*}
P_j = k N_{phot,j},
\end{equation*}
\edit1{where $k$ is a proportionality constant.}
\begin{equation*}
k = \frac{1}{\sum_{i=1}^{m} N_{phot,i}}.
\end{equation*}
\edit1{In this case, the weights become:}
\begin{equation}\label{eq:profileWeights}
w_{j,sum} = \frac{k^2 N_{phot,j}^2}{N_{phot,j}} = k^2 N_{phot,j} = k P_j
\end{equation}
Therefore, using this traditional formula \edit1{(Equation \ref{eq:varianceweights})} for bright sources will provide nearly the same extraction as straight summation,
\begin{equation}\label{eq:StraightSummation}
f_{sum} = \sum_{j=1}^{m} S_j.
\end{equation} 
In other words, in the limit where the read noise and background noise goes to zero, conventional variance weighting approaches straight summation.

However, in the presence of noise correlations, the formulae become modified.
The optimal weights for correlated data are the sums of the inverse covariance matrix along one dimension  \citep{schmelling1995averagingCorrelatedData},
\begin{equation}\label{eq:covarianceweights}
w_{j,cov} = \sum_{i=0}^{m} C_{ij}^{-1},
\end{equation}
where $C_{ij}$ is the covariance (Cov()) of the variable $S_i/P_i$ with $S_j/P_j$,
\begin{equation}
C_{ij} = \mathrm{Cov}\left(\frac{S_i}{P_i},\frac{S_j}{P_j}\right) = \frac{1}{P_i P_j} \mathrm{Cov}(S_i,S_j).
\end{equation}

The variance in the total flux is
\begin{equation}
\mathrm{Var}(f_{opt}) = \frac{1}{\sum_{j=1}^m w_j}
\end{equation}
for both conventional optimal extraction and covariance weighted extraction.
\edit1{In Section \ref{sec:empCovarianceMatrix}, we compare the results of conventional variance weights (Equation \ref{eq:varianceweights}), profile weights (Equation \ref{eq:profileWeights}) that result in straight summation and covariance weights (Equation \ref{eq:covarianceweights}).}
\begin{figure}[!hbtp]
\centering
\includegraphics[width=.49\columnwidth]{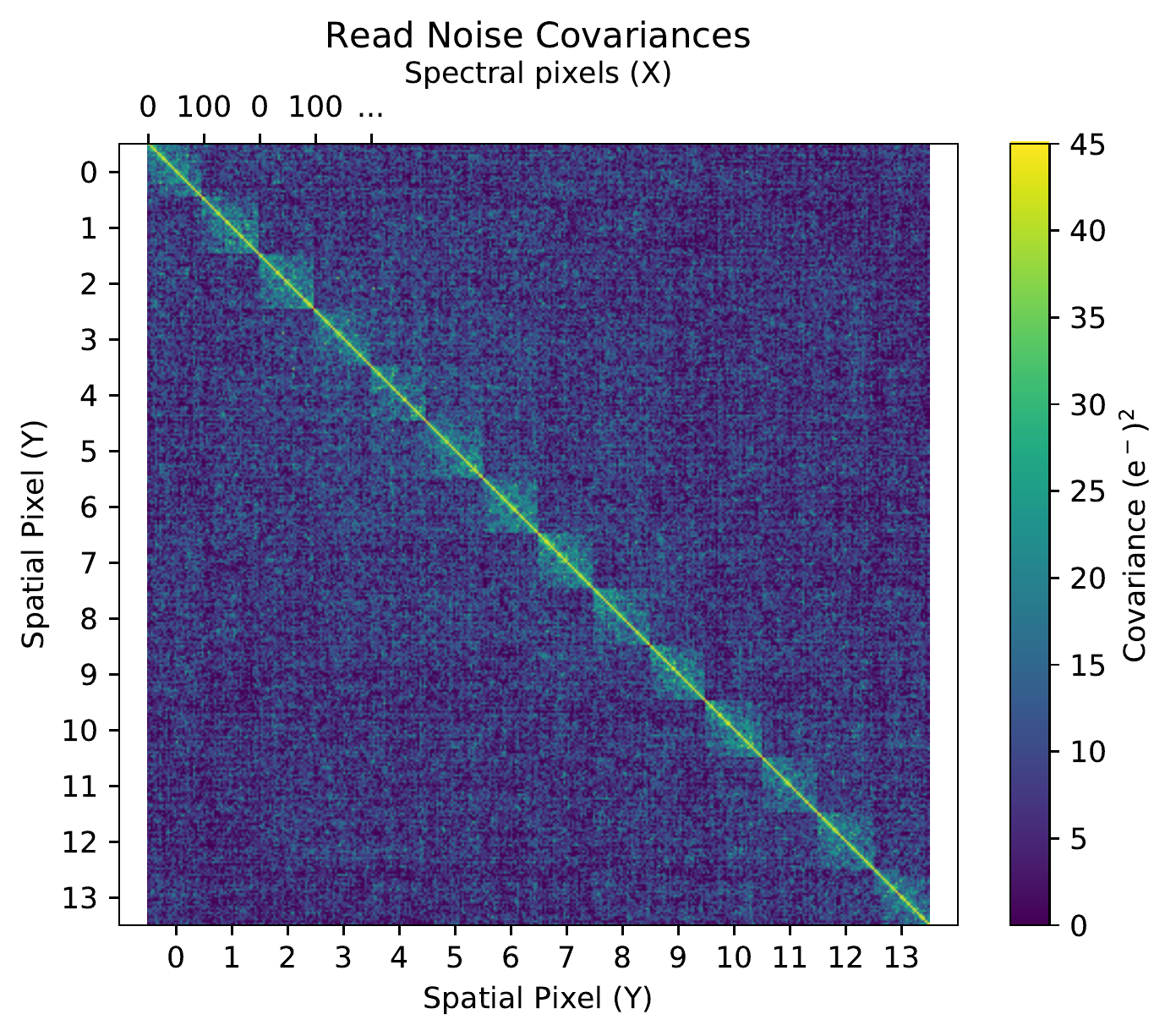}
\includegraphics[width=.49\columnwidth]{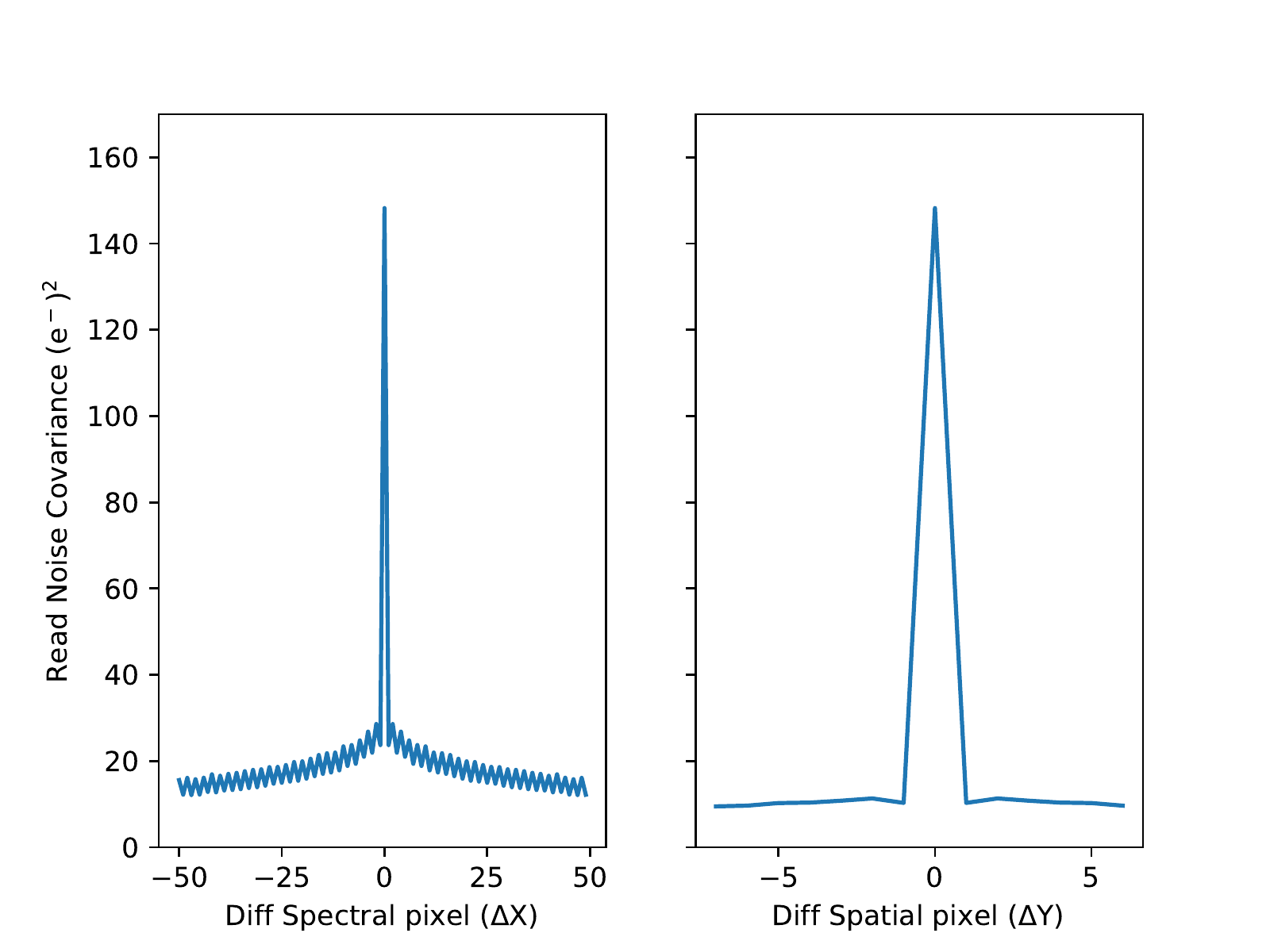}
\caption{{\it Left:} The read noise shows significant correlations between pixels -- unlike the traditional assumption that each pixel's noise is statistically independent.
The data \edit1{were} computed from a long dark exposure of 54 read pairs where reference pixel corrections, column-by-column subtraction and row-by-row subtraction have been performed.
The detector's diagonal elements are slightly below the published detector read noise ($13^2 = 169 (e^-)^2$) due to the row-by-row and column-by-column subtraction step.
{\it Right:} The spectral (X) and spatial (Y) directions from the matrix are averaged to build up a higher signal to noise covariance profile.
Spectrally, the pixels have a covariance of $\sim$20~$(e^-)^2$ expected from 1/f noise. Spatially, they have a somewhat smaller covariance of $\sim$10$(e^-)^2$.}\label{fig:covSpecSpatial}
\end{figure}

\subsubsection{Empirical Covariance Matrix}\label{sec:empCovarianceMatrix}
The ideal weighting of the spectrum from Equation \ref{eq:covarianceweights} depends on the covariance matrix $C_{ij}$.
We use a long dark exposure from the JWST Optical Telescope Element and Integrated Science Instrument Module (OTIS) test performed at NASA's Johnson Space Center on August 23, 2017 to estimate the read noise in $C_{ij}$.
We process the data by performing CDS read pairs of long dark data (108 frames) with reference pixel corrections, column-by-column median subtraction and row-by-row median subtraction.
These steps will be performed on grism time series before extraction on an observation of a bright source with 2 groups per integration in RAPID read mode (1 frame per group).
The resulting images from the column-by-column and row-by-row subtraction are similar to Figure \ref{fig:longDarkGrism} (right).
We consider a wavelength bin that is 100 pixels in the dispersion direction (0.1$\mu$m) and 14 pixels in the spatial direction.

Figure \ref{fig:covSpecSpatial} shows the covariance matrix of this wavelength bin, taken from the median covariance matrix of 20 regions of the detector, where each region is 100 px in the dispersion direction (X) and 14 px in the spatial direction (Y).
The diagonal elements $\sim 150 (e^-)^2$ are smaller than published NIRCam detector performance tables ($\sigma^2 \approx 13^2 (e^-)^2=169 (e^-)^2$ for CDS mode).
However, if we skip the row-by-row and column-by-column subtraction step, they are consistent with $\sigma^2 \approx 13^2 (e^-)^2=169 (e^-)^2$.
It is clear that 1/f noise causes covariance most prominently along the spectral direction (the fast-read direction), with a covariance near 20~$(e^-)^2$.
Surprisingly, there is also a covariance between spatial pixels (the slow read direction) of $\sim$10 $(e^-)^2$.
The average covariance profile in the dispersion direction shows jaggedness to it in Figure \ref{fig:covSpecSpatial} where the covariance rises and falls every other pixel with a peak-to-peak amplitude of $\sim$4~$(e^-)^2$.
This is the even/odd column offset discussed as noise contribution \ref{it:evenOddOffsets}.
Note that the read noise covariance matrix described here is for CDS mode, equivalent to when there are 2 groups per integration.
The covariance matrix elements will decrease by a factor of approximately $\sqrt{2/N_G}$ where $N_G$ is the number of groups.

The empirical covariance matrix from Figure \ref{fig:covSpecSpatial} is used to estimate the optimal pixel weights using Equation \ref{eq:covarianceweights}.
We reduce the noise and jaggedness of the empirical covariance matrix in Figure \ref{fig:covSpecSpatial} by averaging the value along the diagonal, across all spectral pixels and across the spatial pixels with the same value of $i -j$.
Here, the profile is the value from \texttt{pynrc} \citep{leisenring2020pynrc0p8dev}, which uses \texttt{webbpsf} \citep{perrin2014webbpsf}.
We assume that the photon and background noise is $\sqrt{N_{phot}}$ and that they are completely uncorrelated from one pixel to the next, so the covariance matrix is the sum, in quadrature, of the empirical read noise covariance matrix and a diagonal photon noise matrix.
To calculate $N_{photon}$ we assume a source is bright enough that the maximum well depth in the spectrum image is 60\% of the well depth.

\begin{figure}[!hbtp]
\centering
\includegraphics[width=.66\columnwidth]{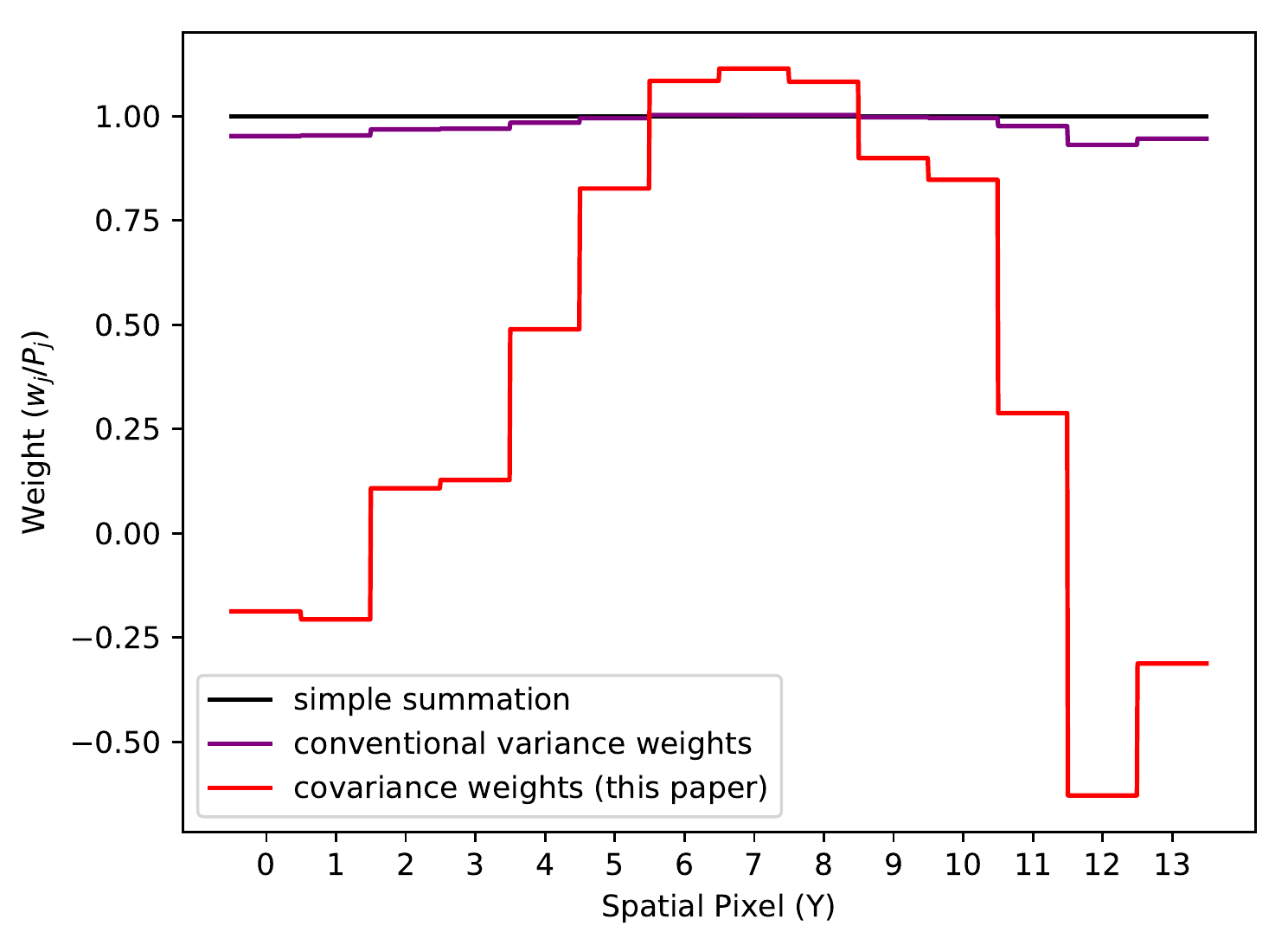}
\includegraphics[width=.32\columnwidth]{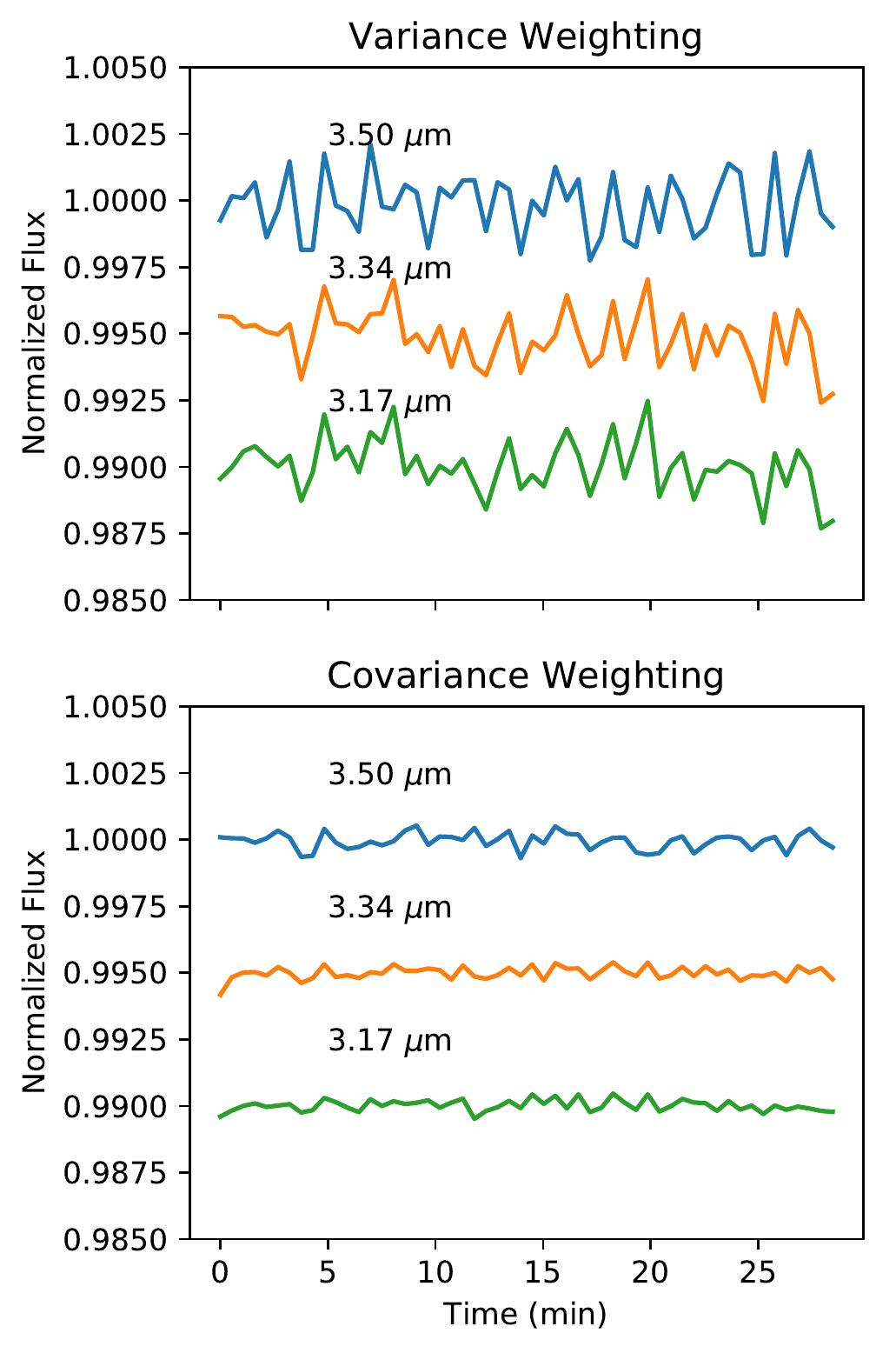}
\caption{ {\it Left:} Conventional variance weighting (black line) provides nearly the same weights as simple summation (purple line) for targets that are significantly brighter than the background flux, such as exoplanet host stars.
However, covariance weighting from Equation \ref{eq:covarianceweights} (red) strongly weights the core and negatively weights the wings to reduce the contribution from correlated 1/f noise.
Here, we assume a maximum well depth in the image of 60\% and a worst-case scenario of 2 groups RAPID (1 frame per group) in the integration.
{\it Right:} Covariance weighting will dramatically lower the scatter in the simulated spectrum plus dark frame time series compared to variance weighting.}\label{fig:covWeightedSummation}
\end{figure}

The resulting optimal weighting function $w_j/P_j$ is plotted in Figure \ref{fig:covWeightedSummation}.
This optimal weighting significantly increases the contributions from the inner pixels compared to variance weights.
The optimal weighting scheme also negatively weights some pixels outside the core of the PSF to reduce correlated noise.

Conventional optimal variance weighting slightly lowers the contribution of read noise to the overall error budget.
We test this conclusion with a series of grism images comprised of a constant A0V star spectrum added to the CDS read pairs, but do not add photon noise to focus on the read noise contribution.
We do background subtraction along the dispersion and spatial directions, as would be done in a spectroscopic extraction pipeline.
We perform sum extraction over a 14 spatial pixels and then combine together 165 spectral pixels into a 0.165 $\mu$m wavelength bin.
The standard deviation of sum extraction (black weights in Figure \ref{fig:covWeightedSummation}) is 3,860 DN, which is 2.7$\times$ the read noise, as shown in Table \ref{tab:noiseSummary1overf}.
The standard deviation of the extracted flux drops to 3,060 DN if conventional variance weighting (purple weights in Figure \ref{fig:covWeightedSummation}) is used.

Optimal covariance weighting {\it significantly} lowers the contribution of read noise to the overall error budget.
We use Equation \ref{eq:covarianceweights} (red weights in Figure \ref{fig:covWeightedSummation}) to reduce the read noise.
For the covariance weights, we assume constant off-diagonal matrix elements of 15 (e$^-)^2$ and do not include the spectral covariance.
The resulting standard deviation of 850 DN is now 60\% of the photon noise, but still above ideal read noise limit of 350 DN, due to remaining pixel correlations.
Thus, we recommend optimal covariance weighting when extracting grism time series to dramatically decrease the 1/f noise contribution.
As the number of groups of the ramp is increased, the read noise should drop as 1/$\sqrt{N_G}$ so the covariance weighting is most important when $N_G$ is small.

The methodology in calculating $P_j$ and Cov($S_i,S_j$) can significantly affect the precision on the time series.
Typically $P_j$ and the photon noise contribution to Cov($S_i,S_j$) can be found by fitting smooth functions (polynomials) along the dispersion direction.
If $P_j$ is calculated from each integration individually, it will inherit the 1/f noise variations of each integration.
Therefore, we hold $P_j$ constant along the time series with a value saved from the middle image at all wavelengths.
Alternatively, $P_j$ could be calculated from a PSF library or theoretically from \texttt{webbpsf} \citep{perrin2014webbpsf}.
As we discussed in Section \ref{sec:smallSpatialAp}, one can also use a very small spatial aperture to reduce the contributions from 1/f noise.
However, the optimal covariance approach allows one to use a wider aperture that will be less sensitive to pointing jitter and wavefront variations.

\subsection{The impact of correlated read noise on time series}

\begin{figure}[!hbtp]
\centering
\includegraphics[width=.49\columnwidth]{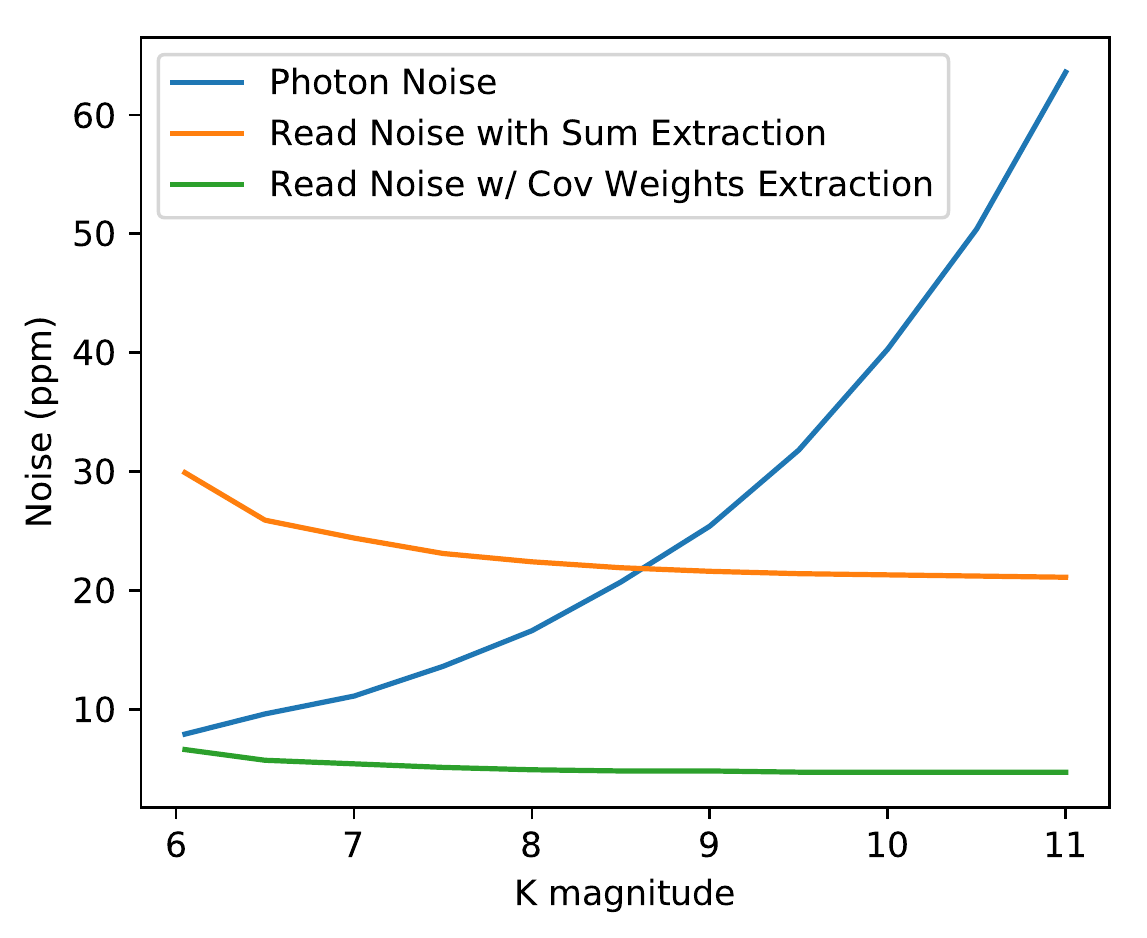}
\caption{Bright sources, where only a handful of groups per integration are possible without saturation, can be dominated by read noise if simple sum extraction is used.
A covariance-weighted extraction can substantially decrease the contribution of correlated read noise to below the photon limit.
The errors here are for a grism time series wavelength bin that is 0.165~$\mu$m wide centered on 3.34~$\mu$m for a G-type host star and a 3 hour transit duration.
We assume a 2048$\times$256 subarray and 4 amplifiers (No. of output channels=4).
}\label{fig:genericSourceErrorContrib}
\end{figure}

So far, we have discussed the magnitude of correlated read noise on an integration-by-integration basis, but here we explore how it will impact science targets.
We calculate the expected noise in the transit depth or eclipse depth of a generic exoplanet.
For this calculation, we assume a planet orbits a G-type 5800 K star with a 3 hour transit duration, is observed using a 2048$\times$256 STRIPE subarray and is read out in RAPID mode with 1 frame per group and 4 amplifiers (No. of output channels=4).
We consider a range of $K$ magnitudes from 6 to 11.
We then calculate the maximum number of groups up the ramp that can be commanded while keeping the peak pixel count below 50\% well depth to safely avoid severe non-linearities.
\edit1{Higher percentages of well depth, such as 80\% or 90\%, are possible to fit more groups in and thus increase the signal to 1/f noise ratio.
The tradeoff is that detector non-linearities may go uncorrected by the pipeline or there could be reciprocity failure.
Here, we assume a conservative 50\% value to also ensure that inaccuracies in the JWST NIRCam throughput or a source's absolute calibration do not risk saturating the brightest pixels.
}
From there, we calculate the noise due to photon counting statistics (i.e. $\sqrt{N_{phot}}$) and compare it to the read noise.
As in Section \ref{sec:GrismDarkExtraction}, we focus on a 0.17~$\mu$m wide bin centered at 3.34~$\mu$m

Figure \ref{fig:genericSourceErrorContrib} shows the photon noise as a function of $K$ Vega magnitude.
The photon noise scales as $\sqrt{f}$, where $f$ is the flux because it is \edit1{governed by Poisson statistics}.
The read noise drops slowly with $K$ magnitude because the time to reset between integrations takes a smaller and smaller fraction of the total exposure as the number of groups per integration grows.
We show the read noise assuming sum extraction using Equation~\ref{eq:StraightSummation} (1,044 ppm for CDS integration) and covariance-weighted extraction using Equation~\ref{eq:covarianceweights} (230 ppm for CDS integration).
If a sum extraction is used, the read noise will dominate for $K \lesssim 8.7$.
However, covariance weights can ensure that read noise is smaller than photon shot noise for $K > 6.0$.

Some of the most interesting targets will be nearby bright sources, where the photon noise is small and the feature size is small.
For context, the features size of atmospheric signatures on archetypical planets are expected to be $\sim$300~ppm for hot Jupiters and Neptunes but only 10 ppm for cool Super-Earths like K2-3~b if they have high metallicity atmospheres \citep{greene2016jwst_trans}.
Therefore, observations of transiting cool super-Earths will require careful treatment of the correlated read noise, such as the covariance weighting scheme, to achieve the precision needed to probe their atmospheres.

We note that Figure \ref{fig:genericSourceErrorContrib} shows a somewhat counter-intuitive result: read noise dominates for bright source and becomes less important for faint sources.
The reason is that for exoplanet transit observations, the detector well depth is filled to approximately the same level ($\sim 32,000$ DN or 50\% well depth per pixel) regardless if the system's magnitude is 6 or 11.
For the bright source with a few groups up the ramp, the read noise will be a larger fraction of 32,000 DN than for a faint source with many groups up the ramp.

\section{Random Telegraph Noise and RC pixels}\label{RTNandRC}

\begin{figure*}
\gridline{\fig{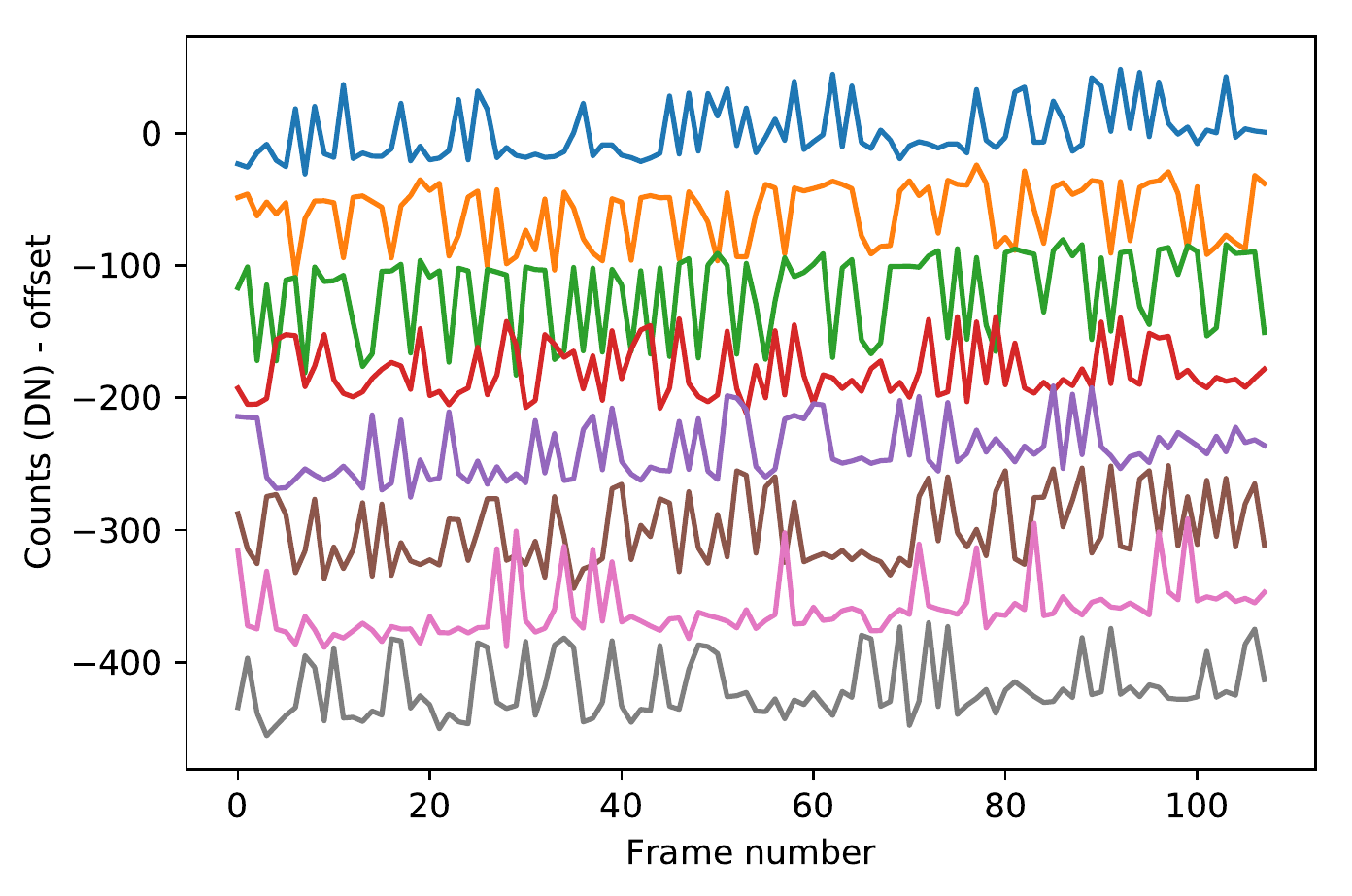}{0.45\textwidth}{RTN Pixels} 
\fig{{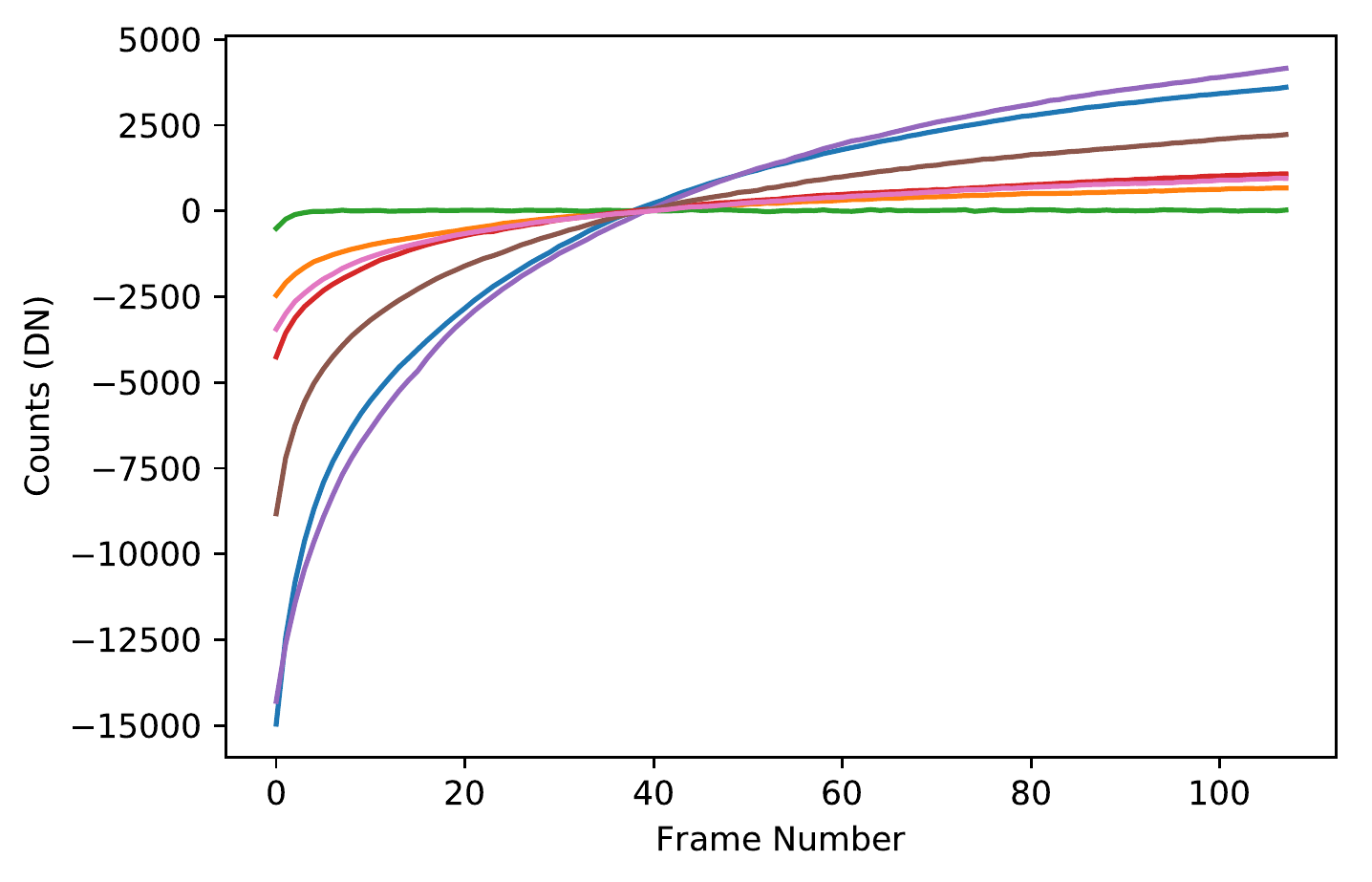}}{0.45\textwidth}{RC pixels}}
\caption{Pixels sometimes exhibit unusual behaviors that are not related to counting photon flux.
Examples are shown for RTN and RC pixels.
}\label{fig:RTNandRC}
\end{figure*}

Several pixels exhibit RC (exponential) behavior \citep{rowlands2012FGSpixelClassification} and others exhibit unpredictable jumps in their DC levels.
The jumps are called Random Telegraph Noise \citep[RTN][]{bacon2005burstNoise,rauscher2014newBeterDetectors} and some examples are shown in Figure~\ref{fig:RTNandRC}.
RTN is also referred to as burst noise and popcorn noise \citep{bacon2005burstNoise}.
If RTN occur on a frequency similar to the frame rate, they can be more pronounced \citep{smith2012H2RGreadNoisekHzframeRates}.
These are rare events (400-1000 pixels out of 4$\times 10^6$) so they are unlikely to \edit1{land} in a grism or in-focus aperture.

For a Weak Lens photometric aperture (such as a 70 pix radius aperture), there will be of order 2-4 pixels exhibiting RTN.
If these are not identified, the RTN can contribute $\sim$200 counts in 4$\times 10^7$ or 5 ppm changes.
However, the RTN pixels shown in Figure \ref{fig:RTNandRC} have large enough jumps that they can be identified as noisy pixels.
\edit1{If these RTN are directly in the middle of the source, the photon levels approach $\sim$3,000 DN per group, so the photon shot noise is $\sim$40 DN.
In these pixels, the photon noise can make identifying RTN more difficult.
Additionally, if there are only 2 groups per integration, the RTN cannot be identified from a single integration and they would have to be identified at the exposure level.
RTN are not currently flagged in the pipeline as do-not-use, although they are marked as having increased noise.
There is ongoing work to improve the reference files to include flags for RTN pixels.
}

There are also RC pixels (as shown in Figure~\ref{fig:RTNandRC}) that have exponential curves like a resister-capacitor circuit.
These pixels consistently show RC behavior and are flagged as ``Do-not-use'' for the pipeline so they may be skipped in the \edit1{aperture} summation.
\edit1{In this case, the spectrum will have sharp dips in flux relative to nearby bins.
Exoplanet transmission and emission spectra are differential measurements that essentially compare the flux in-eclipse with out-of-eclipse so the dips will not change the percentage difference.
A spectroscopic pipeline can also add corrections to wavelength bins that contain do-no-use pixels by assuming a smooth spatial profile or a theoretical PSF. 
}
There are 19 RC pixels in a 1100$\times$20~pixel F444W aperture and 32 RC pixels in a 1600$\times$20~pixel F322W2 aperture.

\section{Conclusions}\label{sec:Conclusion}
JWST time series modes will be a valuable tool to characterize the atmospheres of exoplanets.
In order to reap the benefits of NIRCam's grism time series mode, however, it is critical to understand the various random and systematic error sources that will impact the science of transiting planets.
We studied the random errors that can impact NIRCam photometric and grism time series modes in this paper that add to the photon counting error.

We list the known sources of random error that can impact time series stability in Section~\ref{sec:knownEffects}.
We conclude the following about these random sources of error:
\begin{itemize}[noitemsep]
	\item \textbf{1/f Noise is a major contributor to noise, especially in the grism time series.}
	It is important to subtract along the fast-read (horizontal) direction to reduce the correlated noise.
	Fortunately, 1/f noise averages like $1/\sqrt{N_{int}} $ from integration to integration after background subtraction.
	We found that subtracting the median value from each row in each amplifier efficiently reduces the 1/f noise.
	This will be possible for imaging time series but not grism spectroscopy, where one must use the outer background regions in the spectral dispersion direction to reduce 1/f noise.
	For a small number of groups up the ramp ($<7$ groups), the 1/f noise will be larger than the photon noise using sum extraction along a 14 px tall spatial extraction region and a 165 px wide spectral (0.165 $\mu$m) wavelength bin.
	A covariance weighting scheme described in Section \ref{sec:optimalCovWeights} can significantly lower the read noise from $\sim$1000 ppm to $\sim$250 ppm per double correlated sampling by positively weighting the central pixels.
	\edit1{In addition, the GRISMC element can disperse the spectra perpendicular to the 1/f noise correlations so that they can be naturally subtracted out with spatial background subtraction.
	A GRISMC mode is being discussed as a future science enhancement.}
	\item \textbf{Background Subtraction Efficiently Removes Most Other Effects.}
	We find that background subtraction efficiently removes many random detector effects including:
		\subitem Pre-Amplifier Reset Offsets
		\subitem Amplifier Discontinuities
		\subitem Even-odd Offsets \\
	These effects all become smaller than the 1/f noise with column-by-column (ie. spatial) background subtraction.
	\item \textbf{RTN and RC pixels only affect a small fraction of pixels.} 
	We find that if RTN are ignored by a pipeline, they could contribute about 5 ppm per integration, but the value will be less if the jumps are detected by a pipeline.
	The RC pixels affect about 0.1\% of pixels but are marked as do-no-use in the pipeline so they should not affect time series stability when ignored in aperture summation.
\end{itemize}

These random error sources are only one class of noise sources to consider and tend to average down with the number of integrations.
In Paper II, we will discuss the systematic noise that can arise from other time-variable behavior such as pointing drifts and detector charge trapping.

\acknowledgments

\section*{acknowledgements}
Funding for the E. Schlawin is provided by the JWST project at NASA Goddard Space Flight Center. 
Thanks to Rafia Bushra for the aperture radius exploration that led to several insights about 1/f noise.
Thanks to Nikolay Nikolov and John Stansberry for helpful discussions about 1/f noise.
\edit1{Thanks to Doug Kelly for planning many of the time series tests we used to assess stability.
We also thank the anonymous referee for helpful and prompt comments to improve this work.}

\software{
\edit1{\texttt{astropy} package \citep{astropy2013},
 \texttt{pynrc} \citep{leisenring2020pynrc0p8dev},
\texttt{webbpsf} \citep{perrin2014webbpsf},
\texttt{numpy} \citep{vanderWalt2011numpy},
\texttt{scipy} \citep{virtanen2020scipy},
\texttt{pysynphot} \citep{lim2015pysynphot},
\texttt{ncdhas},
\texttt{photutils} \citep{bradley2016photutilsv0p3},
\texttt{matplotlib} \citep{Hunter2007matplotlib}
}}

\facilities{\edit1{JWST(NIRCam), NASA Goddard Space Environment Simulator, NASA Johnson Chamber A}}

\appendix

\section{A Primer on Detector Operation}\label{sec:detectorPrimer}

\subsection{Photosensitive Operation}

\begin{figure}[!hbtp]
\centering
\includegraphics[width=.99\columnwidth]{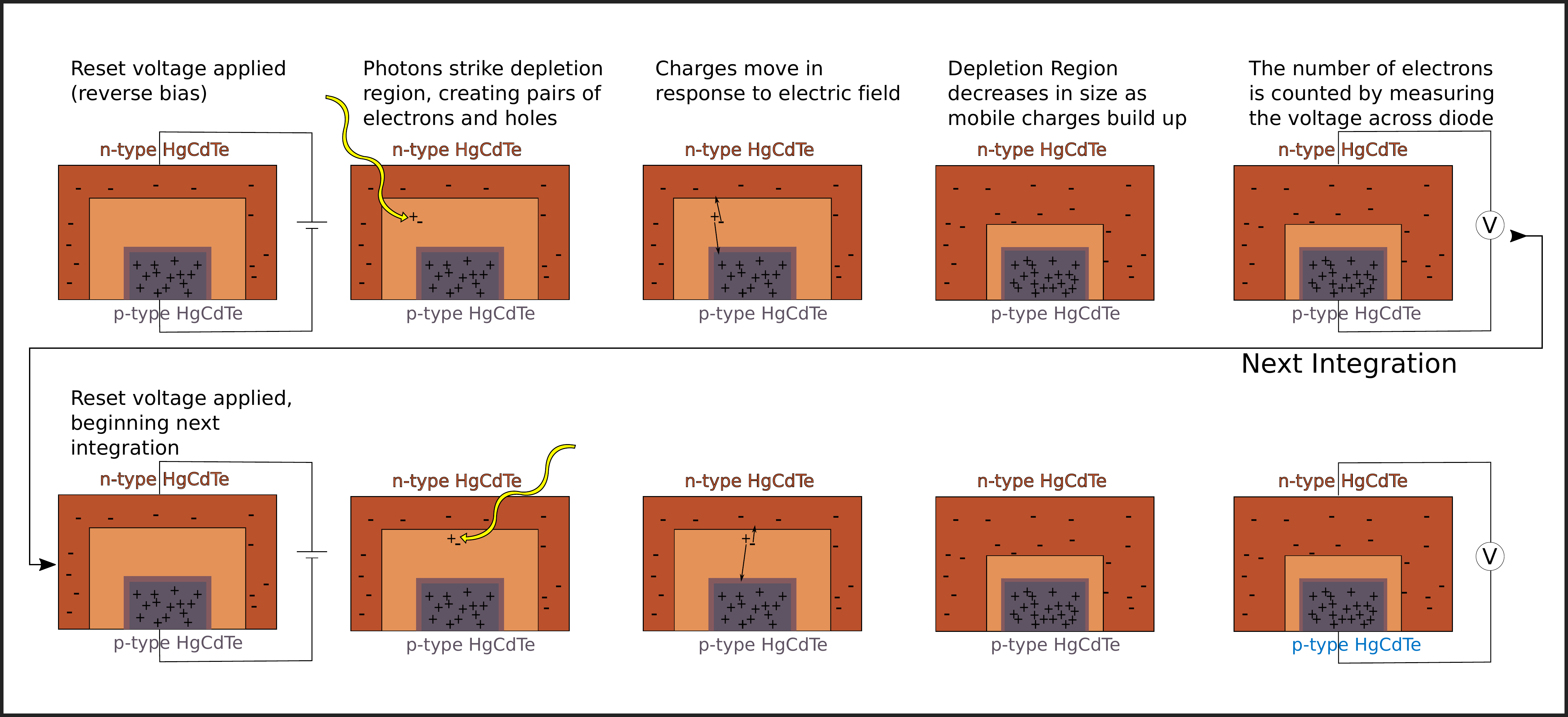}
\caption{A cartoon schematic of a H2RG photodiode detector inspired by \citet{smith2008imgPersistence} and \citet{tulloch2018persistenceH2RG}.
The orange and blue purple colors depict the N-type and P-type semiconductors, respectively.
The undepleted zones of the semiconductor (darker colors with - and + signs) contain mobile charge carriers whereas the depletion region (lighter color without - and +) has no mobile carriers and all electrons are in the valence state.
The depletion region is widest after the detector resets and shrinks as the detector is illuminated with photons that create electron-hole pairs.
The electron-hole pairs migrate across the depletion region towards the undepleted zones due to an electric field in the depletion zone.
The voltage can be measured non-destructively (for many reads) as the well fills.}\label{fig:npSchematic}
\end{figure}

It is helpful to review the basic operation of a NIR detector \citep[e.g.][]{rieke2007irDetectorReview} in order to understand the systematics affecting time-series data.
Figure~\ref{fig:npSchematic} shows a schematic of an NP semiconductor junction for a H2RG detector \edit1{inspired by} \citet{smith2008imgPersistence}.
The crystalline HgCdTe structure has 4 valence electrons that form covalent bonds in a tetrahedral structure.
By varying the relative amount of Hg, Cd and Te (doping), it is possible to increase or decrease the number of electrons to create mobile charge carriers that move along the crystaline lattice.
The N-type semiconductor is a negatively-doped HgCdTe crystalline structure that has negative charge carriers (electrons).
The P-type semiconductor is a positively-doped HgCdTe crystalline structure that has positive charge carriers (vacant holes in the lattice structure than can move like electrons).

Here, we discuss how the N-type and P-type semiconductors operate as a photodiode.
The junction of these two HgCdTe materials enables electrons to flow from the N-type material to the P-type material where it can complete a valence shell in the semiconductor material.
This produces a negative charge on the P-type side.
Conversely, the positively charged holes from the P-type material flow to the N-type material to complete the valence shell and this produces a net positive charge on the N-type side.
The resulting voltage difference is called the contact potential.

In the region where electrons have completed the valence shells in the P-type material and holes in the N-type material, the valence shells are complete.
This region spanning across the two semiconductor materials is called the depletion region (pictured in Figure~\ref{fig:npSchematic}), because it is depleted of mobile charge carriers.
The positive net charge on the N-type material and negative charge in the P-type material causes a voltage (potential) difference across the detector.
It should be noted that the ``depletion region'' is depleted in {\it mobile charge carriers}, but is {\it not} depleted in net charge because it has a net positive charge in the N-type depletion region and negative charge in the P-type depletion region.

The NP junction in Figure~\ref{fig:npSchematic} is ``reverse-biased'' with a positive potential on the N-type material and negative potential on the P-type.
The positive potential on the N-type material attracts the negative charge carriers (electrons) whereas the negative potential on the P-type material attracts the positive charge carriers (holes).
The reverse bias thus reduces the number of mobile charge carriers and increases the size of the depletion region.
The increased depletion region size also increases the well depth or capacity of electrons that the detector may collect before saturation.
The reverse bias is applied at each reset at the beginning of a detector's integration.
Figure~\ref{fig:npSchematic} shows the initial reverse voltage that is applied during a pixel reset. 
The physical size of the N-type HgCdTe semiconductor is much larger than the P-type HgCdTe because the density of carriers is lower in the N-type semi-conductor.

Note that Figure~\ref{fig:npSchematic} shows the mobile charge carriers only.
Textbook NP junction schematics often show the net charge and thus symbolize the regions with charge carriers as empty and the depletion region with ``+'' symbols on the N-type semiconductor and ``-'' symbols on the P-type \citep[e.g.][]{halliday2004physicsText}.
The schematic in Figure~\ref{fig:npSchematic} instead shows the mobile charge carriers as ``+'' and ``-'' because it is easier to visualize the charge trapping effect and motion of mobile charges this way \citep[e.g.][]{smith2008imgPersistence}.

When an incoming photon strikes the N-type depletion region, it will excite an electron from the valence to conduction band which, creates an electron-hole pair \citep{rieke2007irDetectorReview}.
The depletion region has a net electric field pointed toward the P-type side, which controls the direction electrons and holes will migrate.
The electron moves to the extremity of the depletion zone in the N-type material whereas the positive hole generated crosses the NP junction and adds a mobile charge to the P-type side.
The incoming photon has the effect of decreasing the size of the depletion region.
A second photon will excite another electron-hole pair so that a hole moves to the P-type side.
This continues and decreases the size of the depletion region.
The diffusion of charge carriers across the junction changes the voltage across the NP junction, which can be measured with a sensitive amplifier circuit.

As the size of the depletion region decreases, the detector becomes decreasingly sensitive to new photons.
Thus, the NIRCam HgCdTe detectors are never strictly linear.
Eventually, with enough photons, the depletion region is too small to create electron-hole pairs with incoming photons.
This is where the detector approaches saturation and no longer changes voltage with more light.

\subsection{Readout Circuit}\label{sec:readout}

The accumulated charge from absorbed photons in the detector is measured with a multiplexor (MUX) that forms a readout integrated circuit (ROIC) \citep{rieke2007irDetectorReview}.
This readout circuitry enables the signal to be measured non-destructively up the ramp in so-called ``multiaccum" mode.

A sensitive amplifier converts the measured voltage potential (as compared with the bias level after reset) into a signal.
The electronic circuit that digitizes the voltages is tuned to ensure the 16 bits (65,536 possible Data Numbers) measure the voltage across the NP junction from the value after reset (\edit1{and before most} photons are detected) to the saturation voltage where the maximum number of photons are collected and no more electron-hole pairs can be produced in the depletion region.
The minimum voltage is several thousands of counts above zero to ensure that fluctuations in the bias signal do not go below 0.
The 65,536 DN level is tuned to be just above the HgCdTe \edit1{maximum} well capacity to measure the full well capacity of a pixel.

\section{More sophisticated 1/f noise reduction techniques}\label{sec:moreSophisticatedOneOverFAppendix}
Here we discuss some more sophisticated 1/f noise reduction techniques.
These were applied to the same time series explored in Section \ref{sec:longDarks}
These methods were not as effective as the simpler row-by-row amp-by-amp method from Section \ref{sec:indAmpAvg}, but are worth discussing in the search for better ways to reduce 1/f noise.

\subsection{Smoothed Kernel Subtraction}\label{sec:smoothKernelSub}
We start by applying a smoothing kernel to each group (1 read for RAPID mode) up the ramp of the raw data.
The kernel is a uniform normalized kernel 1 pixel in height and 161 pixels along a row.
This enables subtraction of 1/f noise that has length scales shorter than the line time (ie., timescales shorter than 512 clock cycles or 5.12 ms), which the median subtraction cannot remove.
Care must be taken not to apply the kernel through the source aperture, which would subtract and distort any signal from a star.
The kernel should be wider than the source aperture diameter and interpolate through the source points.

We use the \texttt{astropy.convolution.convolve()} routine, which calculates a convolution of the kernel by the image.
\texttt{astropy.convolution.convolve()} ignores all NaN values in the mask and renormalizes the kernel by the sum of the remaining kernel points.
This efficiently interpolates over bad pixels and can also interpolate over the source aperture by creating a circle of NaN values around the source.
This is similar to the row-by-row method where the pixels are selected outside the source aperture with no bright sources but it operates on timescales shorter than 5.12 ms.
Our smoothing kernel is a flat boxcar normalized to 1.0.
The resulting time series has a standard deviation of 2,800 DN to 3,700 ppm.
This is slightly larger than the much simpler row-by-row amp-by-amp subtraction in Section \ref{sec:indAmpAvg}.

\begin{figure*}[!hbtp]
\centering
\includegraphics[width=.32\columnwidth]{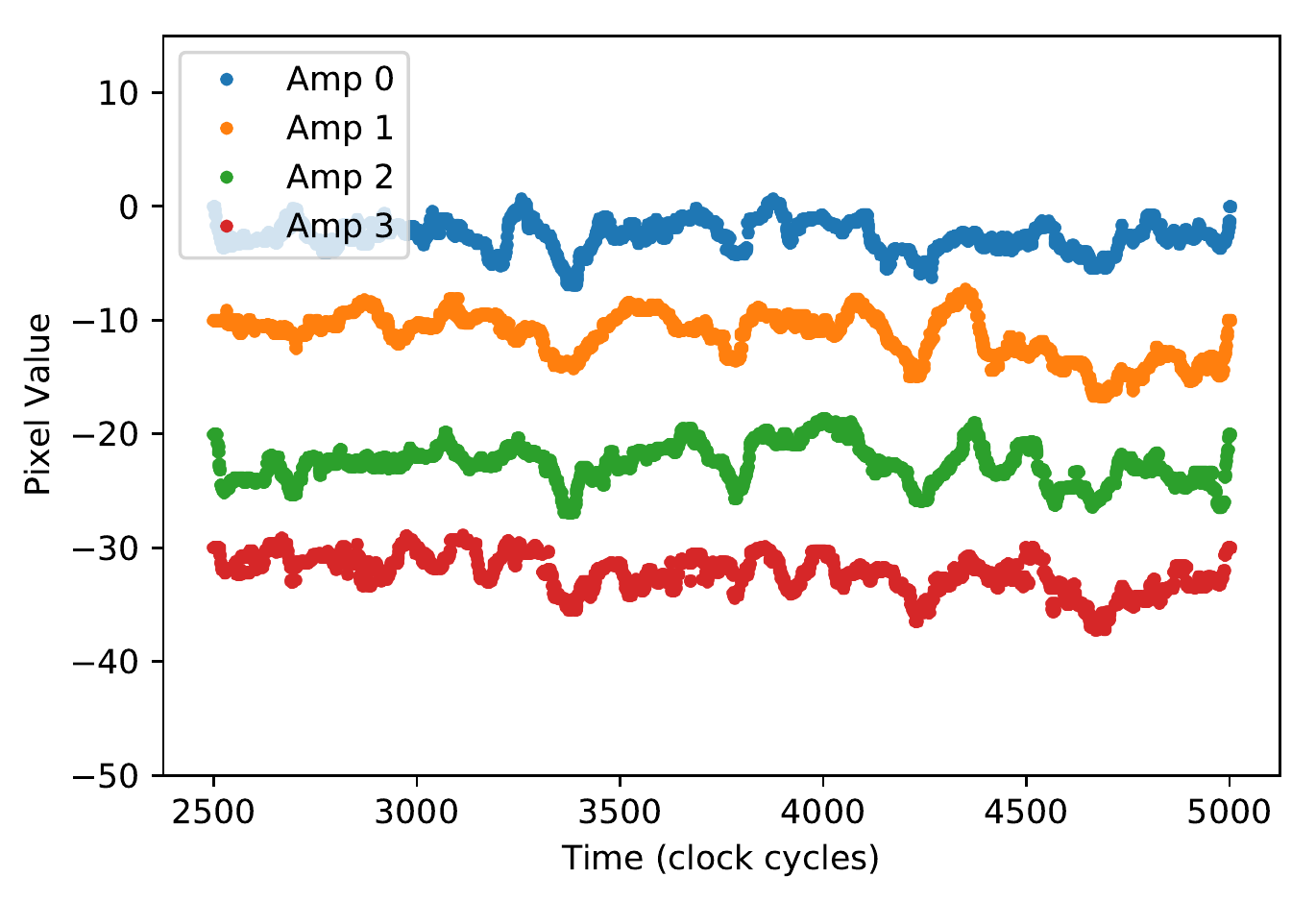}
\includegraphics[width=.32\columnwidth]{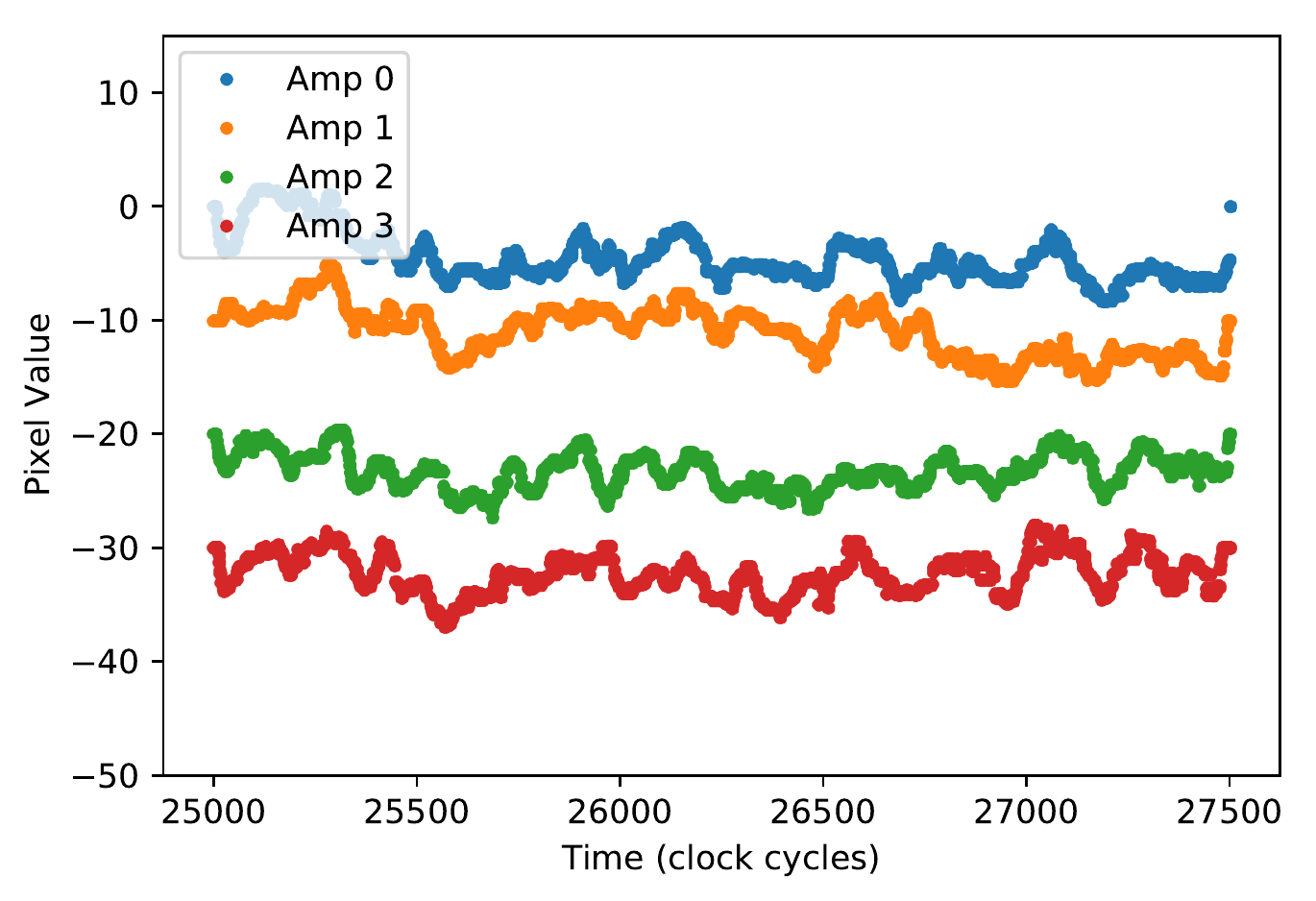}
\caption{The time series for all 4 amplifiers share common-mode behaviors for some intervals but not always (left and right plots).
This is visible in a short segment of the time series of all pixels for an example group up-the-ramp 53 in a dark ramp where the bias has already been subtracted.
The pixels in each time series are smoothed with a 51 pixel-wide median filter to show the low frequency noise.
The left panel shows the a time near the beginning of a frame, where amplifiers 1 and 2 share some common-mode read noise.
The right panel shows another part of the time series where amplifiers 1 and 2 have more disparate noise signatures.
}\label{fig:darkPixelTimeSeries}
\end{figure*}

\subsection{Reference Channel Subtraction}
We note that the 4 channels can sometimes exhibit the same 1/f correlation behavior, as shown in Figure \ref{fig:darkPixelTimeSeries} \edit1{because they share a common reference voltage.}
Another option to remove 1/f noise is to subtract the 3 out of 4 output amplifiers that have the source on them by the 1 output amplifier without the target.
We subtract the first amplifier's time series (pixels X=1 through X=512) from all amplifiers by shifting and flipping the 2048$\times$256 image sub-section to the corners of the 4 amplifiers at (X=1, X=513, X=1025 and X=1537).
Of course, this means that this amplifier is perfectly subtracted and no noise or signal remains in this 1/4 of the image.
The subtraction step should increase the read noise on an individual pixel (for the remaining 3 output amplifiers) by $\sim \sqrt{2}$ because the errors from the subtracted pixel should add in quadrature to the original pixels.
This increase in read noise on a per-pixel level might be acceptable if it decreases the correlated errors along the rows.
On the other hand, if each amplifier exhibits its own individual correlated noise behavior, the reference amplifier will fail to subtract out all correlated noise and may introduce more noise.

The net effect of this method is to increase the noise over a row-by-row median subtraction.
The standard deviations in the time series range from 8,500 DN to 10,800 DN for this reference amplifier method as opposed to the 5,500 DN to 6,600 DN for a row-by-row median.
This means the correlations within an amplifier are not efficiently subtracted and may be propagated to the remaining amplifiers.

\subsection{PCA Correction}

\begin{figure*}[!hbtp]
\centering
\includegraphics[width=.4\columnwidth]{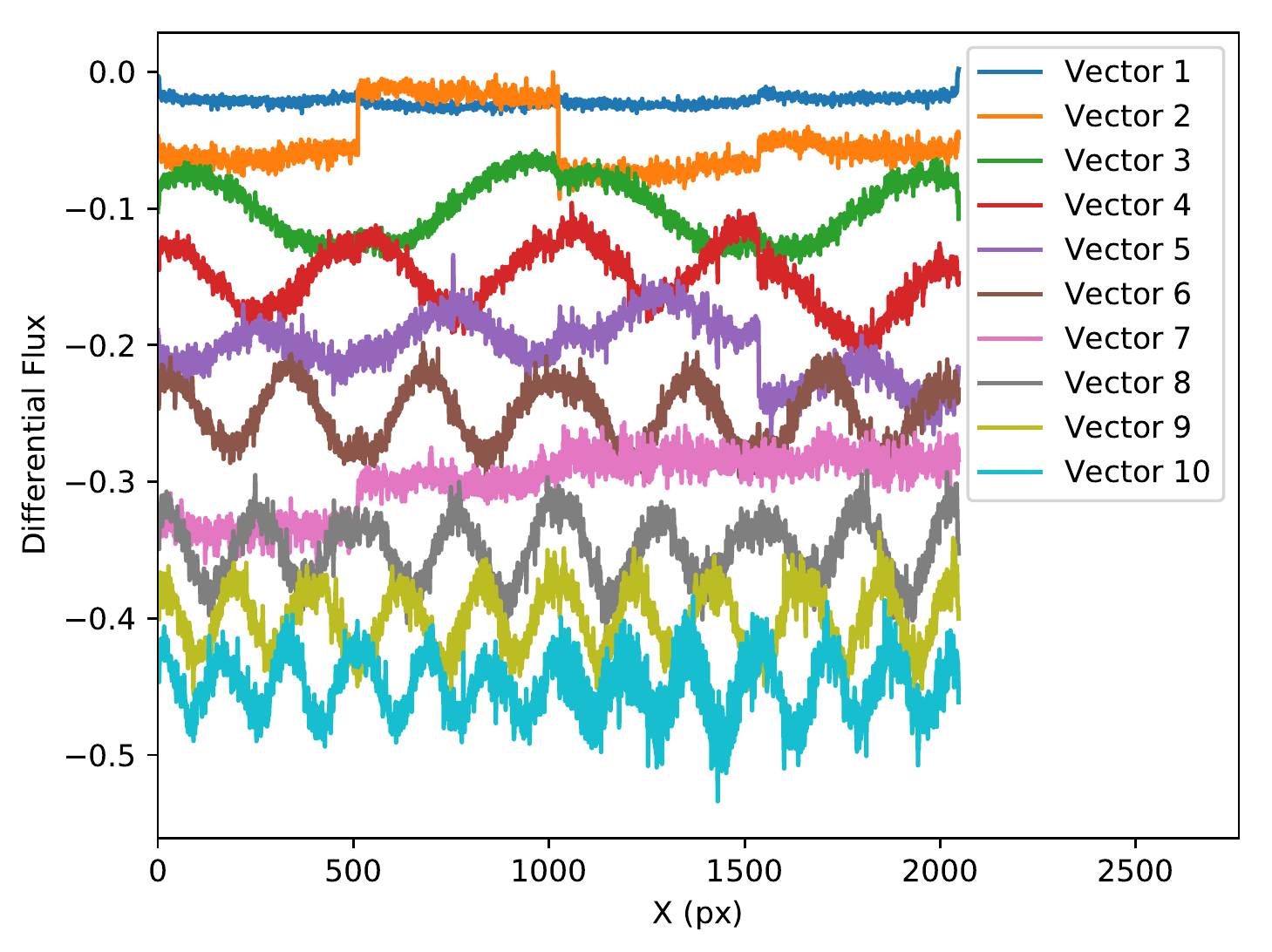}
\includegraphics[width=.4\columnwidth]{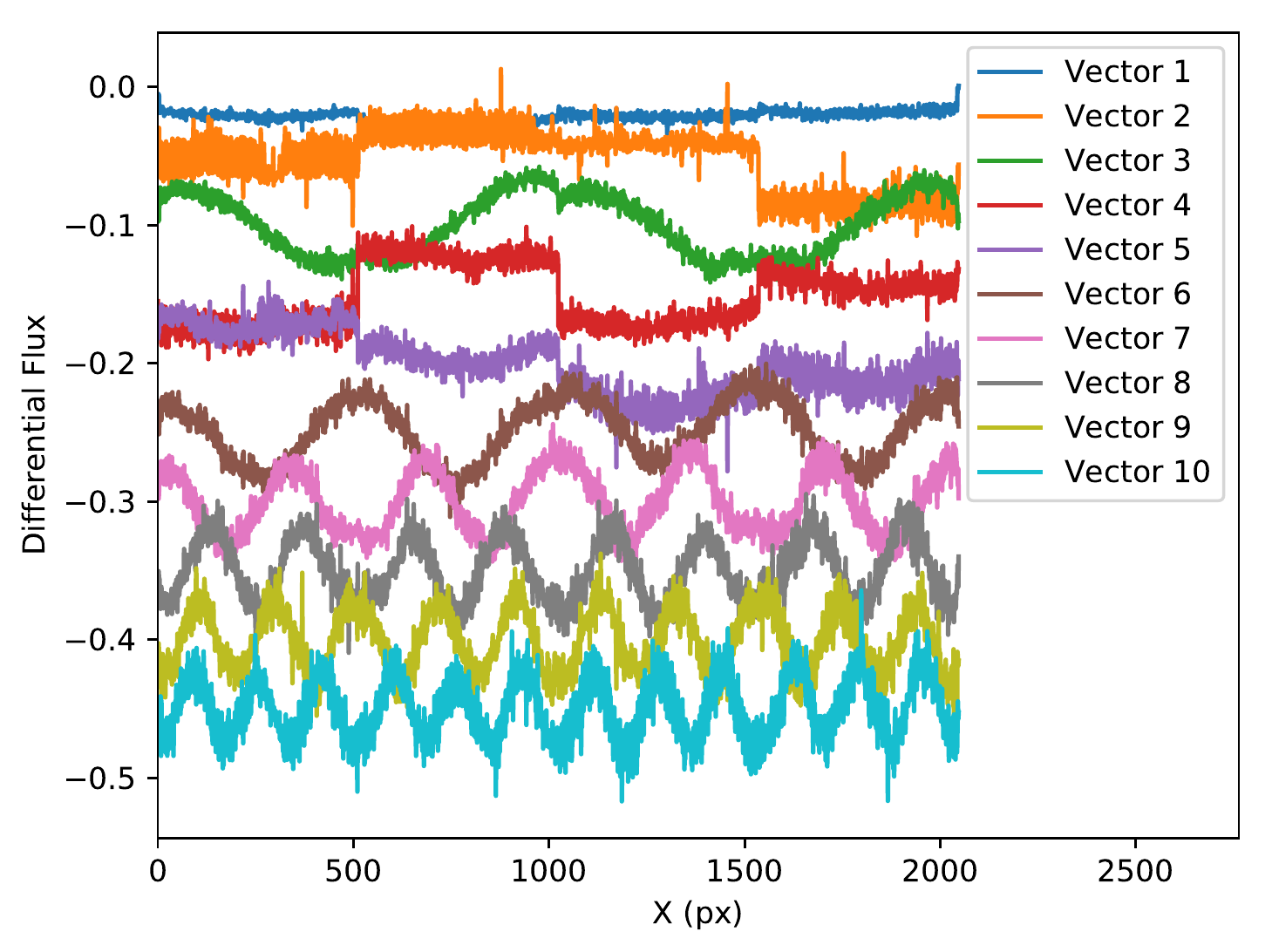}
\caption{The first 10 principal component eigenvectors identified from group 54 (left) and group 55 (right).
Pre-amplifier reset values are seen in the first principal component, and change from one group (one read in RAPID) to another.
These are residual values are not fully subtracted by reference pixels.
The principal components generally increase in frequency with principal component number.
In other words, the highest frequency principal components explain the least amount of the variance as would be expected for 1/f noise.
}\label{fig:pcaEigenvectors}
\end{figure*}

Another approach to treating the 1/f noise is dimensionality reduction with principal component analysis (PCA).
Here, we assume that the counts for each pixel along a column is a random variable that can be correlated with pixels in other columns.
Each row represents a different noise instantiation.
First, we apply reference pixel correction to remove the pre-amplifier resets.
Next, we calculate the mean value of each pixel in the 108 group stack and subtract all groups by this mean image.
This removes all bias levels and focuses on the time variability of the pixels.
Here, we assume the dark current is negligible.
Next, we mask all bad pixels by setting all pixels with an absolute value larger than 200 DN and replace them with 0 DN to remove bad pixels that can drive the principal components.
The first 10 principal components are plotted in Figure \ref{fig:pcaEigenvectors} for two example groups (54 and 55).

The first 10 principal components explain 14\% of the total variance in each group up the ramp.
The principal components from one group to another are different, but the ordering of these components generally follows a pattern of increasing frequency with higher order principal components (smaller eigenvalues).
This is expected for 1/f noise where the lowest frequencies have most of the power.
Since the first principal components are different from one group to another, we calculated the principal components for each group independently.
We then create a noise model image which is the matrix multiplication of the eigenvectors by the principal components for 10 components.
This model is subtracted from each group to create a 1/f-corrected image.

The PCA subtraction is applied to each group up the ramp and then these groups are subtracted in pairs to create a time series.
We use 10 principal components as a starting point for this method.
The final time series has a measured standard deviation of 5,300 to 5,500 DN within the background-subtracted aperture.
This is comparable to the row and column subtraction standard deviation of 5,500 to 6,500 DN.
Adding 10 more principal components for a total of 20 eigenvectors only reduces the time series noise to 5,100 to 5,400 for the 3 apertures.

\begin{figure*}[!hbtp]
\centering
\includegraphics[width=.4\columnwidth]{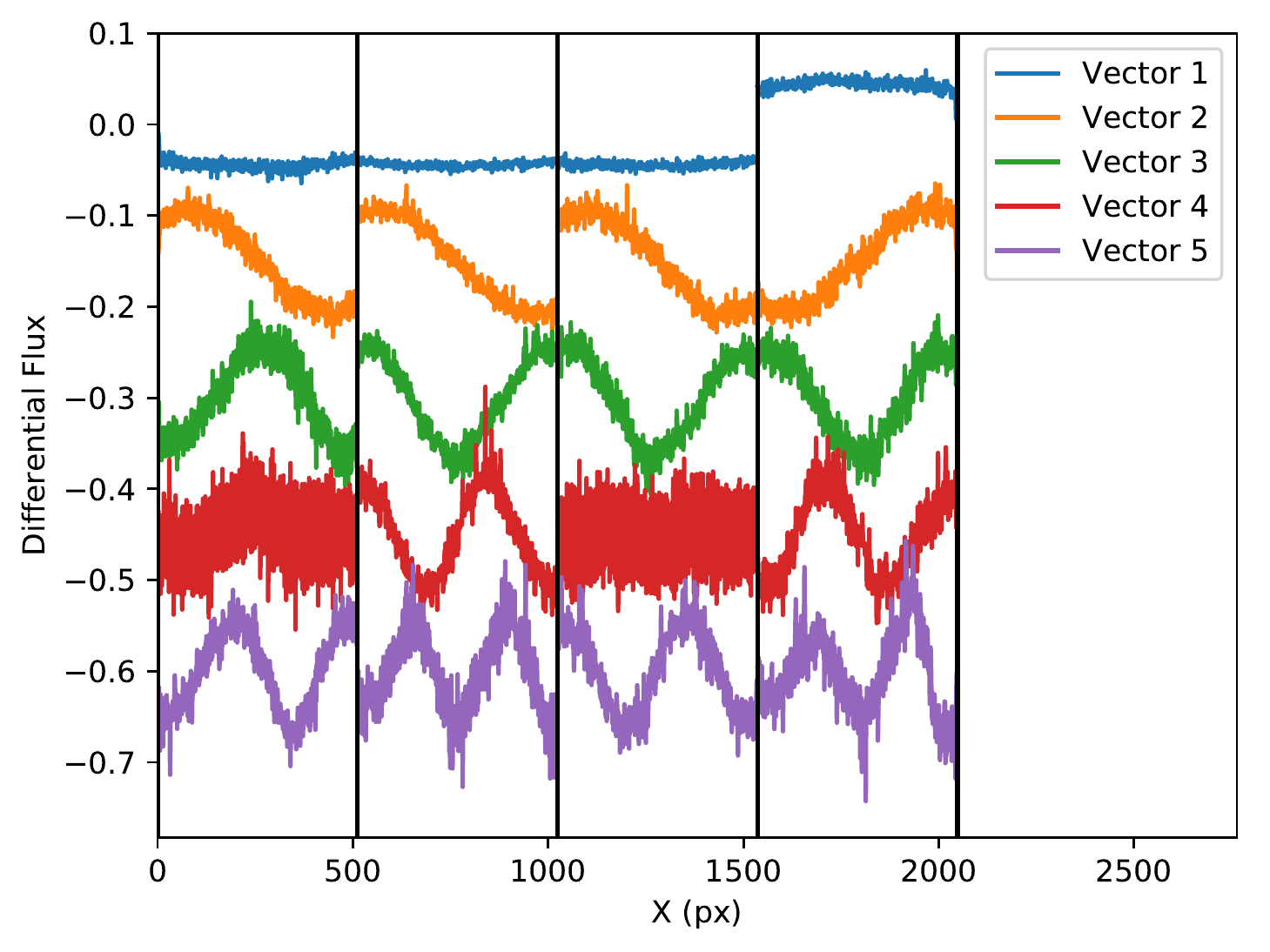}
\includegraphics[width=.4\columnwidth]{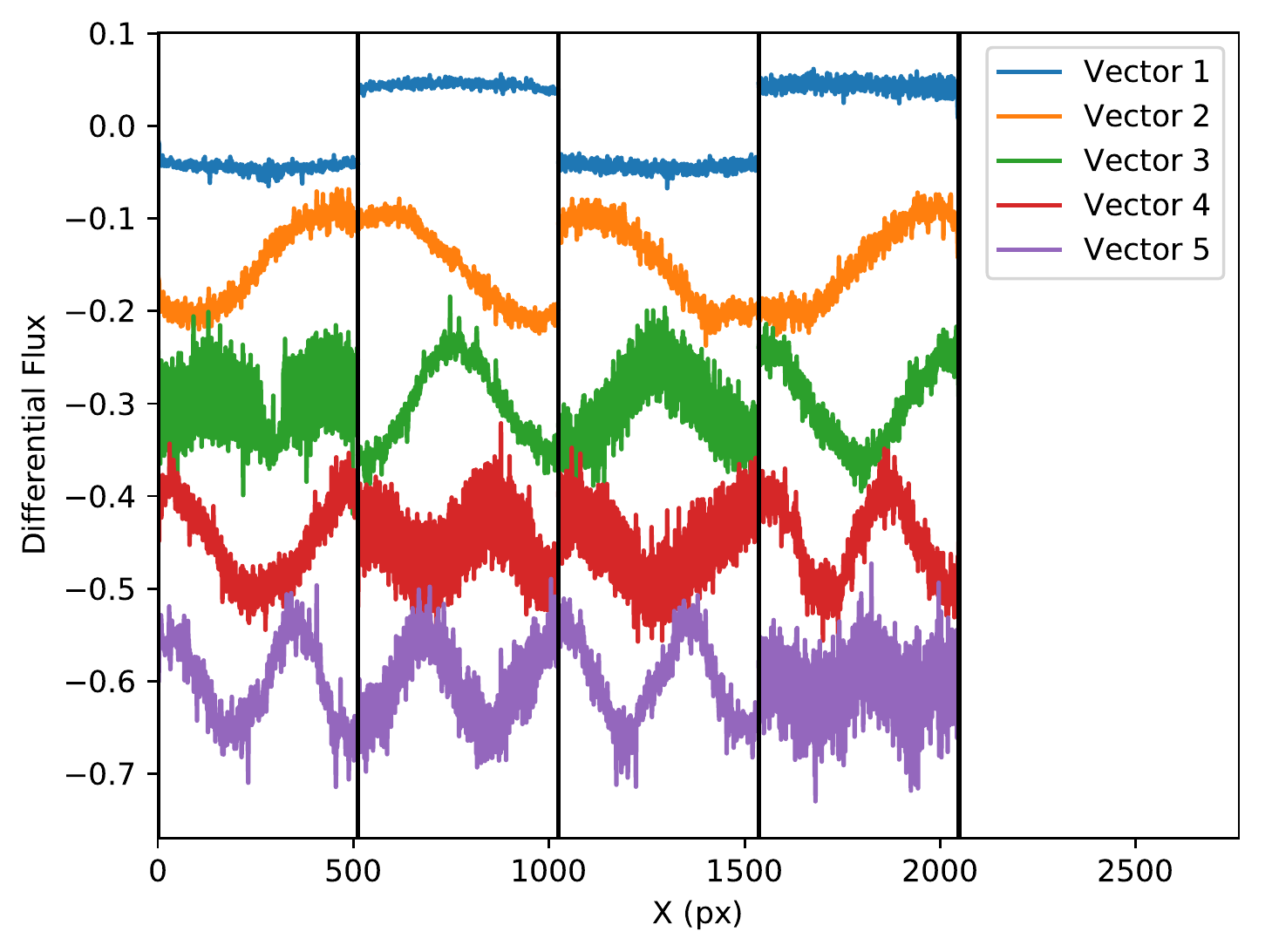}
\caption{The first 5 principal component eigenvectors identified from group 54 (left) and group 55 (right) where each amplifier has been analyzed independently.
The amplifier boundaries are demarcated by solid vertical lines.
}\label{fig:pcaEigenvectorsIndAmp}
\end{figure*}

In another approach, we treat each amplifier separately so that the principal components are calculated independently.
This approach means that a noise signature in 3 amplifiers but not the fourth does not propagate to fourth amplifier when subtracting the PCA-based model.
This individual-amplifier model has four times the number of variables as the combined PCA approach, so we fit with only the first 5 principal components in each amplifier.
The resulting eigenvectors are shown in Figure \ref{fig:pcaEigenvectorsIndAmp}.

If we treat each amplifier separately, the PCA model is more effective in reducing 1/f noise.
We begin by keeping 5 principal components within each amplifier and subtracting these from each amplifier separately.
The resulting standard deviation of the time series is 2,500 to 3,820 DN.
This improvement is beyond what is possible on a real source because the principal components include pixels over the source aperture.
To simulate a real photometry scenario, it is necessary to interpolate over the source aperture points as done in Section \ref{sec:smoothKernelSub} for the smoothing kernel.
However, we stop the PCA at this point because it does not do as well as the simple row-by-row amplifier-by-amplifier mean subtraction performed in Section~\ref{sec:indAmpAvg}.

\section{Covariance Matrix with Row-by-Row Subtraction}\label{sec:covMatrixRowByRow}

\begin{figure*}[!hbtp]
\centering
\includegraphics[width=.6\columnwidth]{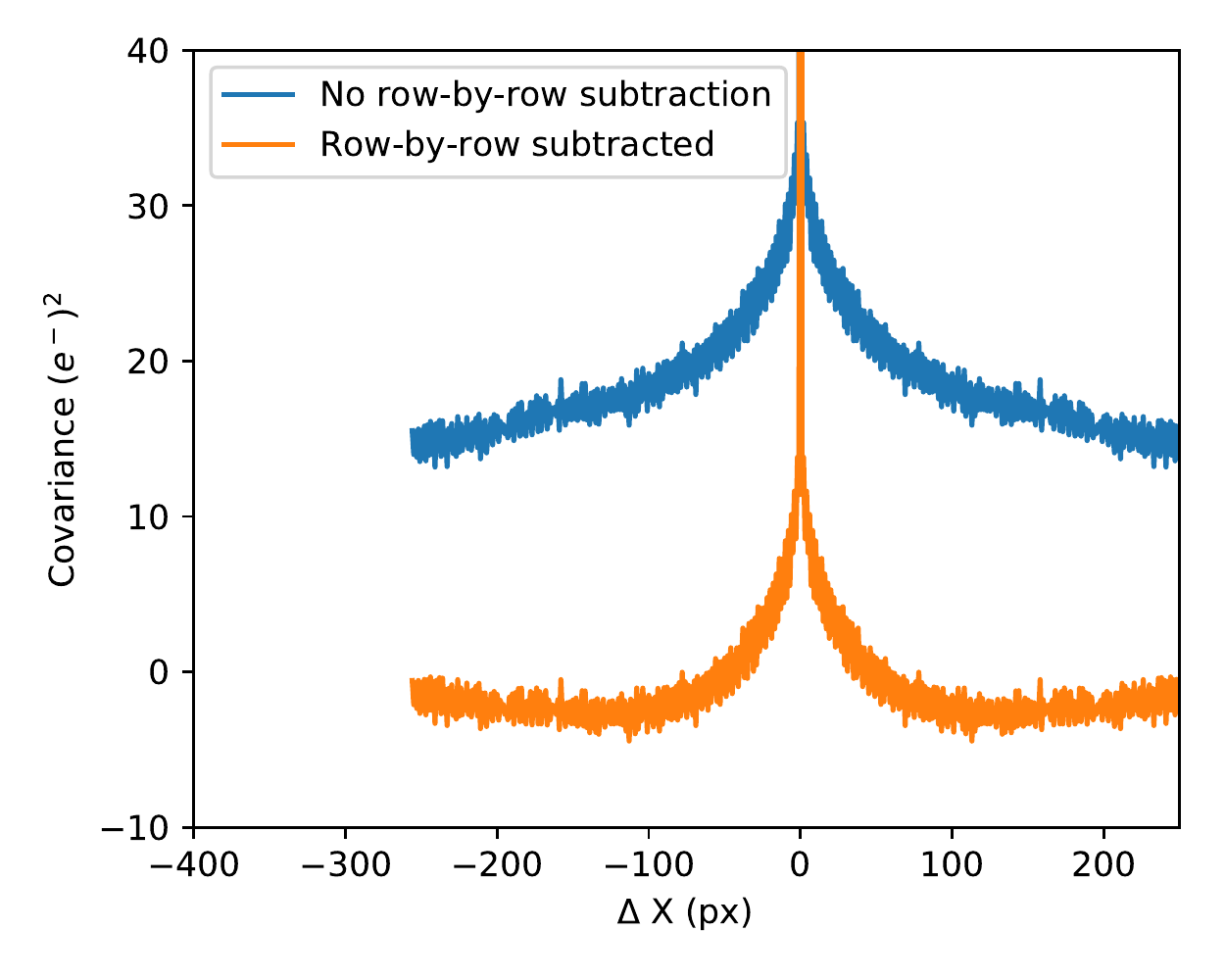}
\caption{\edit1{1/f noise causes read noise covariance between pixels, but can be reduced in GRISMC spectroscopy considerably with row-by-row subtraction.
Before any row-by-row subtraction (blue line), all pixels within an amplifier are positive correlated.
These can be reduced with row-by-row linear subtraction using all pixels within a row that are outside the source and within an amplifier.
The resulting images have smaller overall covariance (orange line), but still have correlations for closely spaced pixels ($|\Delta X| < 50 px$).
Therefore, the optimal background subtraction will use a tighter background subtraction region ($|\Delta X| < 50 px$) to reduce 1/f noise.}
}\label{fig:covAfterRowbyRow}
\end{figure*}

\edit1{Here, we examine the covariance matrix profile for GRISMC images to show the effect of row-by-row median subtraction within an amplifier.
This uses the same image and simulated grism spectrum as in Section \ref{sec:grismcChange}.
We calculate the covariance matrix for the pixels in amplifier 4, which is labeled in Figure \ref{fig:detectorLayout}.
For the covariance, we treat each row in the image as an independent noise instance and find the covariance between pixels within that row.
We exclude all pixels that include the bright source region and then average all covariance matrix elements C$_{ij}$ that have the same value for $i -j$ to create a differential covariance profile.
As shown in Figure \ref{fig:covAfterRowbyRow}, the covariance level is 15-30 $(e^-)^2$ due to 1/f noise and rises for nearby pixels (as $\Delta X$ is reduced).}

\edit1{Row-by-row linear subtraction for all pixels outside the source substantially lowers the covariance, and in some cases causes a mild 3-4 $(e^-)^2$ anti-covariance.
However, there still remains positive covariance between pixels closer than 50 px (or 50 clock cycles).
Therefore, a background region that includes pixels with $|\Delta X| < 50$ px will decrease the 1/f noise residuals.
We find that a tighter background region of 7 px to 30 px optimally lowers the 1/f noise for the simulated time series.
}

\section{Dark Current Variations}\label{sec:darkCurrentVariations}
\edit1{In Section \ref{sec:longDarks}, we use a long dark integration with 108 groups to simulate a time series of 54 integrations with 2 groups each by subtracting sequential pairs of reads.
This simulation does not take into account the effect of resets that would occur at the start of every integration in a real exposure that has 2 groups and 54 integrations.
In reality, the resets can affect time series because the dark current is non-linear with time.
Here, we quantify the magnitude of this effect and how it could affect exoplanet time series.}

\begin{figure*}[!hbtp]
\centering
\includegraphics[width=.6\columnwidth]{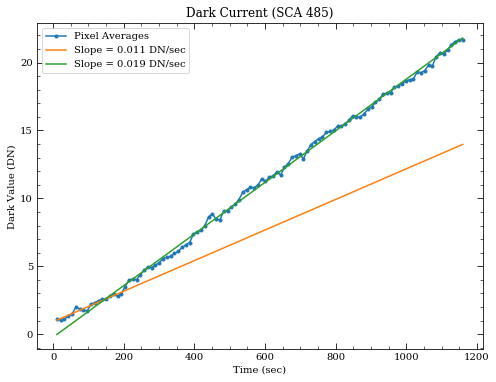}
\caption{\edit1{The dark current for the NIRCam detectors is non-linear.
After about 150 sec, the slope of the dark current changes, so a long integration from a dark is not equivalent to a series of shorter integrations.
However, the dark current is a small signal, so the difference in the dark current is only about $\sim$100 ppm of the photon signal of a bright source before doing the background subtraction, and much smaller afterwards.}
}\label{fig:darkCurrent}
\end{figure*}

\edit1{Figure \ref{fig:darkCurrent} shows the dark current measured from CV3 dark exposures composed of 108 groups up one integration in RAPID mode at a temperature of 37.0 K.
In the beginning of an integration, the slope is about 0.011 DN/s but  increases to 0.019 DN/s at around 150 sec following detector reset.
Thus, a simulation that breaks up a long integration (108 groups) into many shorter integrations will have dark current variation that would not be present for a long exposure that has 54 integrations with 2 groups each.
We expect the dark current to come out with background subtraction.
However, we calculate the magnitude of the dark current to see if any residuals in the background subtraction could affect the time series.
The difference in signal between the early and late dark currents is 0.008 DN/s, so the total DN difference in a 21.5 sec integration is 397 DN or 107 ppm in our archetype grism wavelength bin that is 165 spectral pixels and 14 spatial pixels.
Therefore, the dark current difference is only expected to be 107 ppm before subtraction and much smaller after background subtraction.
 }

\bibliographystyle{apj}
\bibliography{this_biblio}

\begin{thebibliography}{}
\expandafter\ifx\csname adsurllinklabel\endcsname\relax
  \def\adsurllinklabel{[LINK]}\fi
\expandafter\ifx\csname adsadsurllinklabel\endcsname\relax
  \def\adsadsurllinklabel{[ADS]}\fi
\expandafter\ifx\csname natexlab\endcsname\relax\def\natexlab#1{#1}\fi

\bibitem[{{Astropy Collaboration} {et~al.}(2013){Astropy Collaboration},
  {Robitaille}, {Tollerud}, {Greenfield}, {Droettboom}, {Bray}, {Aldcroft},
  {Davis}, {Ginsburg}, {Price-Whelan}, {Kerzendorf}, {Conley}, {Crighton},
  {Barbary}, {Muna}, {Ferguson}, {Grollier}, {Parikh}, {Nair}, {Unther},
  {Deil}, {Woillez}, {Conseil}, {Kramer}, {Turner}, {Singer}, {Fox}, {Weaver},
  {Zabalza}, {Edwards}, {Azalee Bostroem}, {Burke}, {Casey}, {Crawford},
  {Dencheva}, {Ely}, {Jenness}, {Labrie}, {Lim}, {Pierfederici}, {Pontzen},
  {Ptak}, {Refsdal}, {Servillat}, \& {Streicher}}]{astropy2013}
{Astropy Collaboration}, {Robitaille}, T.~P., {Tollerud}, E.~J., {et~al.} 2013,
  \aap, 558, A33
 \href{http://adsabs.harvard.edu/abs/2013A%26A...558A..33A}{\adsadsurllinklabel}

\bibitem[{{Bacon} {et~al.}(2005){Bacon}, {McMurtry}, {Pipher}, {Forrest}, \&
  {Garnett}}]{bacon2005burstNoise}
{Bacon}, C.~M., {McMurtry}, C.~W., {Pipher}, J.~L., {Forrest}, W.~J., \&
  {Garnett}, J.~D. 2005, in Society of Photo-Optical Instrumentation Engineers
  (SPIE) Conference Series, Vol. 5902, \procspie, ed. T.~J. {Grycewicz} \&
  C.~J. {Marshall}, 116--127
 \href{https://ui.adsabs.harvard.edu/abs/2005SPIE.5902..116B}{\adsadsurllinklabel}

\bibitem[{{Barstow} {et~al.}(2015){Barstow}, {Aigrain}, {Irwin}, {Kendrew}, \&
  {Fletcher}}]{barstow2015jwstSystematics}
{Barstow}, J.~K., {Aigrain}, S., {Irwin}, P.~G.~J., {Kendrew}, S., \&
  {Fletcher}, L.~N. 2015, \mnras, 448, 2546
 \href{http://adsabs.harvard.edu/abs/2015MNRAS.448.2546B}{\adsadsurllinklabel}

\bibitem[{{Barstow} \& {Irwin}(2016)}]{barstow2016trappist1habitable}
{Barstow}, J.~K., \& {Irwin}, P.~G.~J. 2016, \mnras, 461, L92
 \href{http://adsabs.harvard.edu/abs/2016MNRAS.461L..92B}{\adsadsurllinklabel}

\bibitem[{{Beichman} {et~al.}(2014){Beichman}, {Benneke}, {Knutson}, {Smith},
  {Lagage}, {Dressing}, {Latham}, {Lunine}, {Birkmann}, {Ferruit}, {Giardino},
  {Kempton}, {Carey}, {Krick}, {Deroo}, {Mandell}, {Ressler}, {Shporer},
  {Swain}, {Vasisht}, {Ricker}, {Bouwman}, {Crossfield}, {Greene}, {Howell},
  {Christiansen}, {Ciardi}, {Clampin}, {Greenhouse}, {Sozzetti}, {Goudfrooij},
  {Hines}, {Keyes}, {Lee}, {McCullough}, {Robberto}, {Stansberry}, {Valenti},
  {Rieke}, {Rieke}, {Fortney}, {Bean}, {Kreidberg}, {Ehrenreich}, {Deming},
  {Albert}, {Doyon}, \& {Sing}}]{beichman2014pasp}
{Beichman}, C., {Benneke}, B., {Knutson}, H., {et~al.} 2014, \pasp, 126, 1134
 \href{http://adsabs.harvard.edu/abs/2014PASP..126.1134B}{\adsadsurllinklabel}

\bibitem[{{Berta} {et~al.}(2012){Berta}, {Charbonneau}, {D{\'e}sert},
  {Miller-Ricci Kempton}, {McCullough}, {Burke}, {Fortney}, {Irwin}, {Nutzman},
  \& {Homeier}}]{berta2012flat_gj1214}
{Berta}, Z.~K., {Charbonneau}, D., {D{\'e}sert}, J.-M., {et~al.} 2012, \apj,
  747, 35
 \href{http://adsabs.harvard.edu/abs/2012ApJ...747...35B}{\adsadsurllinklabel}

\bibitem[{Bradley {et~al.}(2016)Bradley, Sipocz, Robitaille, Tollerud,
  Vin{\'\i}cius, Deil, Barbary, G{\"u}nther, Cara, Droettboom, Bostroem, Bray,
  Bratholm, Pickering, Craig, Barentsen, Pascual, adonath, Greco, Kerzendorf,
  StuartLittlefair, Ferreira, D'Eugenio, \& Weaver}]{bradley2016photutilsv0p3}
Bradley, L., Sipocz, B., Robitaille, T., {et~al.} 2016, astropy/photutils:
  v0.3, doi:10.5281/zenodo.164986
 \href{https://doi.org/10.5281/zenodo.164986}{\adsurllinklabel}

\bibitem[{{Deming} {et~al.}(2013){Deming}, {Wilkins}, {McCullough}, {Burrows},
  {Fortney}, {Agol}, {Dobbs-Dixon}, {Madhusudhan}, {Crouzet}, {Desert},
  {Gilliland}, {Haynes}, {Knutson}, {Line}, {Magic}, {Mandell}, {Ranjan},
  {Charbonneau}, {Clampin}, {Seager}, \& {Showman}}]{deming13}
{Deming}, D., {Wilkins}, A., {McCullough}, P., {et~al.} 2013, \apj, 774, 95
 \href{http://adsabs.harvard.edu/abs/2013ApJ...774...95D}{\adsadsurllinklabel}

\bibitem[{{Gardner} {et~al.}(2006){Gardner}, {Mather}, {Clampin}, {Doyon},
  {Greenhouse}, {Hammel}, {Hutchings}, {Jakobsen}, {Lilly}, {Long}, {Lunine},
  {McCaughrean}, {Mountain}, {Nella}, {Rieke}, {Rieke}, {Rix}, {Smith},
  {Sonneborn}, {Stiavelli}, {Stockman}, {Windhorst}, \&
  {Wright}}]{gardner2006SSRv}
{Gardner}, J.~P., {Mather}, J.~C., {Clampin}, M., {et~al.} 2006, \ssr, 123, 485
 \href{http://adsabs.harvard.edu/abs/2006SSRv..123..485G}{\adsadsurllinklabel}

\bibitem[{{Gillon} {et~al.}(2016){Gillon}, {Jehin}, {Lederer}, {Delrez}, {de
  Wit}, {Burdanov}, {Van Grootel}, {Burgasser}, {Triaud}, {Opitom}, {Demory},
  {Sahu}, {Bardalez Gagliuffi}, {Magain}, \&
  {Queloz}}]{gillon2016trappist1Discovery}
{Gillon}, M., {Jehin}, E., {Lederer}, S.~M., {et~al.} 2016, \nat, 533, 221
 \href{http://adsabs.harvard.edu/abs/2016Natur.533..221G}{\adsadsurllinklabel}

\bibitem[{{Gillon} {et~al.}(2017){Gillon}, {Triaud}, {Demory}, {Jehin}, {Agol},
  {Deck}, {Lederer}, {de Wit}, {Burdanov}, {Ingalls}, {Bolmont}, {Leconte},
  {Raymond}, {Selsis}, {Turbet}, {Barkaoui}, {Burgasser}, {Burleigh}, {Carey},
  {Chaushev}, {Copperwheat}, {Delrez}, {Fernandes}, {Holdsworth}, {Kotze}, {Van
  Grootel}, {Almleaky}, {Benkhaldoun}, {Magain}, \&
  {Queloz}}]{gillon2017trappist-1sevenp}
{Gillon}, M., {Triaud}, A.~H.~M.~J., {Demory}, B.-O., {et~al.} 2017, \nat, 542,
  456
 \href{http://adsabs.harvard.edu/abs/2017Natur.542..456G}{\adsadsurllinklabel}

\bibitem[{{Greene} {et~al.}(2016){Greene}, {Line}, {Montero}, {Fortney},
  {Lustig-Yaeger}, \& {Luther}}]{greene2016jwst_trans}
{Greene}, T.~P., {Line}, M.~R., {Montero}, C., {et~al.} 2016, \apj, 817, 17
 \href{http://adsabs.harvard.edu/abs/2016ApJ...817...17G}{\adsadsurllinklabel}

\bibitem[{{Greene} {et~al.}(2017){Greene}, {Kelly}, {Stansberry}, {Leisenring},
  {Egami}, {Schlawin}, {Chu}, {Hodapp}, \& {Rieke}}]{greene2017jatisNIRCam}
{Greene}, T.~P., {Kelly}, D.~M., {Stansberry}, J., {et~al.} 2017, Journal of
  Astronomical Telescopes, Instruments, and Systems, 3, 035001
 \href{http://adsabs.harvard.edu/abs/2017JATIS...3c5001G}{\adsadsurllinklabel}

\bibitem[{{Halliday} {et~al.}(2004){Halliday}, {Resnick}, \&
  {Walker}}]{halliday2004physicsText}
{Halliday}, D., {Resnick}, R., \& {Walker}, J. 2004, {Fundamentals of Physics,
  Part 5 (Chapters 38-44)} (WILEY), 248
 \href{http://adsabs.harvard.edu/abs/2004fpp5.book.....H}{\adsadsurllinklabel}

\bibitem[{{Horne}(1986)}]{horne1986optimalE}
{Horne}, K. 1986, \pasp, 98, 609
 \href{http://adsabs.harvard.edu/abs/1986PASP...98..609H}{\adsadsurllinklabel}

\bibitem[{{Howe} {et~al.}(2017){Howe}, {Burrows}, \&
  {Deming}}]{howe2017informationJWST}
{Howe}, A.~R., {Burrows}, A., \& {Deming}, D. 2017, \apj, 835, 96
 \href{http://adsabs.harvard.edu/abs/2017ApJ...835...96H}{\adsadsurllinklabel}

\bibitem[{Hunter(2007)}]{Hunter2007matplotlib}
Hunter, J.~D. 2007, Computing In Science \& Engineering, 9, 90


\bibitem[{{Kreidberg} {et~al.}(2014){Kreidberg}, {Bean}, {D{\'e}sert}, {Line},
  {Fortney}, {Madhusudhan}, {Stevenson}, {Showman}, {Charbonneau},
  {McCullough}, {Seager}, {Burrows}, {Henry}, {Williamson}, {Kataria}, \&
  {Homeier}}]{kreidberg2014wasp43}
{Kreidberg}, L., {Bean}, J.~L., {D{\'e}sert}, J.-M., {et~al.} 2014, \apjl, 793,
  L27
 \href{http://adsabs.harvard.edu/abs/2014ApJ...793L..27K}{\adsadsurllinklabel}

\bibitem[{{Krissansen-Totton} {et~al.}(2018){Krissansen-Totton}, {Garland},
  {Irwin}, \& {Catling}}]{krissansen-totton2018trappist1eJWST}
{Krissansen-Totton}, J., {Garland}, R., {Irwin}, P., \& {Catling}, D.~C. 2018,
  \aj, 156, 114
 \href{http://adsabs.harvard.edu/abs/2018AJ....156..114K}{\adsadsurllinklabel}

\bibitem[{{Leisenring}(2019)}]{leisenring2020pynrc0p8dev}
{Leisenring}, J. 2019, pynrc
 \href{https://pynrc.readthedocs.io/en/latest/}{\adsurllinklabel}

\bibitem[{{Lim} {et~al.}(2015){Lim}, {Diaz}, \& {Laidler}}]{lim2015pysynphot}
{Lim}, P.~L., {Diaz}, R.~I., \& {Laidler}, V. 2015, PySynphot User's Guide,
  \url{http://ssb.stsci.edu/pysynphot/docs/}


\bibitem[{{Loose} {et~al.}(2006){Loose}, {Beletic}, {Blackwell}, {Hall}, \&
  {Jacobsen}}]{loose2006sidecarAsic}
{Loose}, M., {Beletic}, J.~W., {Blackwell}, J., {Hall}, D., \& {Jacobsen}, S.
  2006, Astrophysics and Space Science Library, Vol. 336, {SIDECAR ASIC -
  Control Electronics on a Chip} (Springer), 699--706
 \href{https://ui.adsabs.harvard.edu/abs/2006ASSL..336..699L}{\adsadsurllinklabel}

\bibitem[{{Loose} {et~al.}(2003){Loose}, {Farris}, {Garnett}, {Hall}, \&
  {Kozlowski}}]{loose2003H2RGs}
{Loose}, M., {Farris}, M.~C., {Garnett}, J.~D., {Hall}, D. N.~B., \&
  {Kozlowski}, L.~J. 2003, in Society of Photo-Optical Instrumentation
  Engineers (SPIE) Conference Series, Vol. 4850, \procspie, ed. J.~C. {Mather},
  867--879
 \href{https://ui.adsabs.harvard.edu/abs/2003SPIE.4850..867L}{\adsadsurllinklabel}

\bibitem[{{Lustig-Yaeger} {et~al.}(2019){Lustig-Yaeger}, {Meadows}, \&
  {Lincowski}}]{lustig-yaeger2019detectabilityTRAPPIST-1}
{Lustig-Yaeger}, J., {Meadows}, V.~S., \& {Lincowski}, A.~P. 2019, \aj, 158, 27
 \href{https://ui.adsabs.harvard.edu/abs/2019AJ....158...27L}{\adsadsurllinklabel}

\bibitem[{{Matsuo} {et~al.}(2019){Matsuo}, {Greene}, {Johnson}, {Mcmurray},
  {Roellig}, \& {Ennico}}]{matuso2019siAsDetectorStability}
{Matsuo}, T., {Greene}, T.~P., {Johnson}, R.~R., {et~al.} 2019, \pasp, 131,
  124502
 \href{https://ui.adsabs.harvard.edu/abs/2019PASP..131l4502M}{\adsadsurllinklabel}

\bibitem[{{Morales-Calder{\'o}n} {et~al.}(2006){Morales-Calder{\'o}n},
  {Stauffer}, {Kirkpatrick}, {Carey}, {Gelino}, {Barrado y Navascu{\'e}s},
  {Rebull}, {Lowrance}, {Marley}, {Charbonneau}, {Patten}, {Megeath}, \&
  {Buzasi}}]{moralesCalderon2006LdwarfsWeatherIPC}
{Morales-Calder{\'o}n}, M., {Stauffer}, J.~R., {Kirkpatrick}, J.~D., {et~al.}
  2006, \apj, 653, 1454
 \href{http://adsabs.harvard.edu/abs/2006ApJ...653.1454M}{\adsadsurllinklabel}

\bibitem[{{Perrin} {et~al.}(2014){Perrin}, {Sivaramakrishnan}, {Lajoie},
  {Elliott}, {Pueyo}, {Ravindranath}, \& {Albert}}]{perrin2014webbpsf}
{Perrin}, M.~D., {Sivaramakrishnan}, A., {Lajoie}, C.-P., {et~al.} 2014, in
  \procspie, Vol. 9143, Space Telescopes and Instrumentation 2014: Optical,
  Infrared, and Millimeter Wave, 91433X
 \href{http://adsabs.harvard.edu/abs/2014SPIE.9143E..3XP}{\adsadsurllinklabel}

\bibitem[{{Rauscher} {et~al.}(2011){Rauscher}, {Arendt}, {Fixen}, {Lander},
  {Lindler}, {Loose}, {Moseley}, {Wilson}, \&
  {Xenophontos}}]{rauscher2011irsSquared}
{Rauscher}, B.~J., {Arendt}, R.~G., {Fixen}, D.~J., {et~al.} 2011, in
  \procspie, Vol. 8155, 81550C
 \href{https://ui.adsabs.harvard.edu/abs/2011SPIE.8155E..0CR}{\adsadsurllinklabel}

\bibitem[{{Rauscher} {et~al.}(2012){Rauscher}, {Stahle}, {Hill}, {Greenhouse},
  {Beletic}, {Babu}, {Blake}, {Cleveland}, {Cofie}, {Eegholm}, {Engelbracht},
  {Hall}, {Hoffman}, {Jeffers}, {Jhabvala}, {Kimble}, {Kohn}, {Kopp}, {Lee},
  {Leidecker}, {Lindler}, {McMurray}, {Misselt}, {Mott}, {Ohl}, {Pipher},
  {Piquette}, {Polis}, {Pontius}, {Rieke}, {Smith}, {Tennant}, {Wang}, {Wen},
  {Willmer}, \& {Zandian}}]{rauscher2012degradation}
{Rauscher}, B.~J., {Stahle}, C., {Hill}, R.~J., {et~al.} 2012, AIP Advances, 2,
  021901
 \href{https://ui.adsabs.harvard.edu/abs/2012AIPA....2b1901R}{\adsadsurllinklabel}

\bibitem[{{Rauscher} {et~al.}(2014){Rauscher}, {Boehm}, {Cagiano}, {Delo},
  {Foltz}, {Greenhouse}, {Hickey}, {Hill}, {Kan}, {Lindler}, {Mott},
  {Waczynski}, \& {Wen}}]{rauscher2014newBeterDetectors}
{Rauscher}, B.~J., {Boehm}, N., {Cagiano}, S., {et~al.} 2014, \pasp, 126, 739
 \href{https://ui.adsabs.harvard.edu/abs/2014PASP..126..739R}{\adsadsurllinklabel}

\bibitem[{{Rieke}(2007)}]{rieke2007irDetectorReview}
{Rieke}, G.~H. 2007, \araa, 45, 77
 \href{http://adsabs.harvard.edu/abs/2007ARA%26A..45...77R}{\adsadsurllinklabel}

\bibitem[{{Rieke} {et~al.}(2005){Rieke}, {Kelly}, \&
  {Horner}}]{rieke2005nircamSPIE}
{Rieke}, M.~J., {Kelly}, D., \& {Horner}, S. 2005, in SPIE Conf Series, Vol.
  5904, Cryogenic Optical Systems and Instruments XI, ed. J.~B. {Heaney} \&
  L.~G. {Burriesci}, 1--8
 \href{http://adsabs.harvard.edu/abs/2005SPIE.5904....1R}{\adsadsurllinklabel}

\bibitem[{{Robberto}(2014)}]{robberto2014refPixPreAmp}
{Robberto}, M. 2014, JWST-STScI-003852, SM-12
 \href{https://stsci.edu/files/live/sites/www/files/home/jwst/documentation/technical-documents/_documents/JWST-STScI-003852.pdf}{\adsurllinklabel}

\bibitem[{{Rowlands} {et~al.}(2012){Rowlands}, {Warner}, {Berndt}, {Albert}, \&
  {Chayer}}]{rowlands2012FGSpixelClassification}
{Rowlands}, N., {Warner}, G., {Berndt}, C., {Albert}, L., \& {Chayer}, P. 2012,
  in Society of Photo-Optical Instrumentation Engineers (SPIE) Conference
  Series, Vol. 8453, \procspie, 845313
 \href{https://ui.adsabs.harvard.edu/abs/2012SPIE.8453E..13R}{\adsadsurllinklabel}

\bibitem[{{Schlawin} {et~al.}(2018){Schlawin}, {Greene}, {Line}, {Fortney}, \&
  {Rieke}}]{schlawin2018JWSTforecasts}
{Schlawin}, E., {Greene}, T.~P., {Line}, M., {Fortney}, J.~J., \& {Rieke}, M.
  2018, \aj, 156, 40
 \href{http://adsabs.harvard.edu/abs/2018AJ....156...40S}{\adsadsurllinklabel}

\bibitem[{{Schlawin} {et~al.}(2017){Schlawin}, {Rieke}, {Leisenring}, {Walker},
  {Fraine}, {Kelly}, {Misselt}, {Greene}, {Line}, {Lewis}, \&
  {Stansberry}}]{schlawin2017dhs}
{Schlawin}, E., {Rieke}, M., {Leisenring}, J., {et~al.} 2017, \pasp, 129,
  015001
 \href{http://adsabs.harvard.edu/abs/2017PASP..129a5001S}{\adsadsurllinklabel}

\bibitem[{{Schmelling}(1995)}]{schmelling1995averagingCorrelatedData}
{Schmelling}, M. 1995, \physscr, 51, 676
 \href{https://ui.adsabs.harvard.edu/abs/1995PhyS...51..676S}{\adsadsurllinklabel}

\bibitem[{{Sing} {et~al.}(2016){Sing}, {Fortney}, {Nikolov}, {Wakeford},
  {Kataria}, {Evans}, {Aigrain}, {Ballester}, {Burrows}, {Deming},
  {D{\'e}sert}, {Gibson}, {Henry}, {Huitson}, {Knutson}, {Etangs}, {Pont},
  {Showman}, {Vidal-Madjar}, {Williamson}, \& {Wilson}}]{sing2016continuum}
{Sing}, D.~K., {Fortney}, J.~J., {Nikolov}, N., {et~al.} 2016, \nat, 529, 59
 \href{http://adsabs.harvard.edu/abs/2016Natur.529...59S}{\adsadsurllinklabel}

\bibitem[{{Smith} \& {Hale}(2012)}]{smith2012H2RGreadNoisekHzframeRates}
{Smith}, R.~M., \& {Hale}, D. 2012, in Society of Photo-Optical Instrumentation
  Engineers (SPIE) Conference Series, Vol. 8453, \procspie, 84530Y
 \href{https://ui.adsabs.harvard.edu/abs/2012SPIE.8453E..0YS}{\adsadsurllinklabel}

\bibitem[{{Smith} {et~al.}(2008){Smith}, {Zavodny}, {Rahmer}, \&
  {Bonati}}]{smith2008imgPersistence}
{Smith}, R.~M., {Zavodny}, M., {Rahmer}, G., \& {Bonati}, M. 2008, in High
  Energy, Optical, and Infrared Detectors for Astronomy III, Vol. 7021, 70210J
 \href{https://ui.adsabs.harvard.edu/#abs/2008SPIE.7021E..0JS}{\adsadsurllinklabel}

\bibitem[{{STScI}(2017)}]{jdoxNIRCamDetectors}
{STScI}. 2017, {NIRCam Detector Overview}, JDox website, Baltimore, MD
 \href{https://jwst-docs.stsci.edu/near-infrared-camera/nircam-instrumentation/nircam-detector-overview}{\adsurllinklabel}

\bibitem[{{STScI}(2020{\natexlab{a}})}]{apt2020p2}
---. 2020{\natexlab{a}}, APT
 \href{https://www.stsci.edu/scientific-community/software/astronomers-proposal-tool-apt}{\adsurllinklabel}

\bibitem[{{STScI}(2020{\natexlab{b}})}]{jdoxNIRCamDetectorPatterns}
---. 2020{\natexlab{b}}, {NIRCam Detector Readout Patterns}, JDox website,
  Baltimore, MD
 \href{https://jwst-docs.stsci.edu/near-infrared-camera/nircam-instrumentation/nircam-detector-overview/nircam-detector-readout-patterns}{\adsurllinklabel}

\bibitem[{{Tulloch}(2018)}]{tulloch2018persistenceH2RG}
{Tulloch}, S. 2018, ArXiv e-prints, arXiv:1807.05217
 \href{http://adsabs.harvard.edu/abs/2018arXiv180705217T}{\adsadsurllinklabel}

\bibitem[{{van der Walt} {et~al.}(2011){van der Walt}, {Colbert}, \&
  {Varoquaux}}]{vanderWalt2011numpy}
{van der Walt}, S., {Colbert}, S.~C., \& {Varoquaux}, G. 2011, Computing in
  Science and Engineering, 13, 22
 \href{https://ui.adsabs.harvard.edu/abs/2011CSE....13b..22V}{\adsadsurllinklabel}

\bibitem[{{Vanderburg} \& {Johnson}(2014)}]{vanderburg2014twoWheeledKeplerPhot}
{Vanderburg}, A., \& {Johnson}, J.~A. 2014, \pasp, 126, 948
 \href{https://ui.adsabs.harvard.edu/abs/2014PASP..126..948V}{\adsadsurllinklabel}

\bibitem[{{Virtanen} {et~al.}(2020){Virtanen}, {Gommers}, {Oliphant},
  {Haberland}, {Reddy}, {Cournapeau}, {Burovski}, {Peterson}, {Weckesser},
  {Bright}, {van der Walt}, {Brett}, {Wilson}, {Jarrod Millman}, {Mayorov},
  {Nelson}, {Jones}, {Kern}, {Larson}, {Carey}, {Polat}, {Feng}, {Moore}, {Vand
  erPlas}, {Laxalde}, {Perktold}, {Cimrman}, {Henriksen}, {Quintero}, {Harris},
  {Archibald}, {Ribeiro}, {Pedregosa}, {van Mulbregt}, \&
  {Contributors}}]{virtanen2020scipy}
{Virtanen}, P., {Gommers}, R., {Oliphant}, T.~E., {et~al.} 2020, Nature
  Methods, 17, 261
 \href{https://rdcu.be/b08Wh}{\adsurllinklabel}

\bibitem[{{Wakeford} {et~al.}(2017){Wakeford}, {Sing}, {Kataria}, {Deming},
  {Nikolov}, {Lopez}, {Tremblin}, {Amundsen}, {Lewis}, {Mandell}, {Fortney},
  {Knutson}, {Benneke}, \& {Evans}}]{wakeford2017hatp26}
{Wakeford}, H.~R., {Sing}, D.~K., {Kataria}, T., {et~al.} 2017, Science, 356,
  628
 \href{http://adsabs.harvard.edu/abs/2017Sci...356..628W}{\adsadsurllinklabel}

\bibitem[{{Zhou} {et~al.}(2017){Zhou}, {Apai}, {Lew}, \&
  {Schneider}}]{zhou2017chargeTrap}
{Zhou}, Y., {Apai}, D., {Lew}, B.~W.~P., \& {Schneider}, G. 2017, \aj, 153, 243
 \href{http://adsabs.harvard.edu/abs/2017AJ....153..243Z}{\adsadsurllinklabel}

\end{thebibliography}



\end{document}